\documentclass[a4paper,12pt]{article}

\usepackage{algorithm}
\usepackage{amsfonts, amsmath, amssymb, amsthm}
\usepackage{bbm, bm, mathrsfs, mathtools, wasysym}
\usepackage{array, booktabs, colortbl, multicol, multirow}
\usepackage{caption, float, subcaption}
\usepackage{color, xcolor}
\usepackage{enumerate}
\usepackage[T1]{fontenc}
\usepackage[utf8]{inputenc}
\usepackage[left = 2.5cm, right = 2.5cm, top = 2.5cm, bottom = 2.5cm]{geometry}
\usepackage{graphicx}
\usepackage{hyperref}
\usepackage[authoryear,longnamesfirst,round]{natbib}
\usepackage{setspace}
\usepackage{xr}

\newtheorem{thm}{Theorem}
\newtheorem{prop}{Proposition}
\newtheorem{lem}{Lemma}

\newtheorem{asm}{Assumption}

\newenvironment{asm2}[3]{%
\par\medskip
  \phantomsection
  \label{#3}%
  \noindent\textbf{Assumption {#1}*}\textmd{ ({#2}).}\quad\itshape
}{%
  \normalfont\par\medskip
}

\newcommand{\assref}[2]{{\hyperref[#2]{{#1}*}}}

\begin{document}    

\title{ \large{\textsc{Performance of Empirical Risk Minimization \\ for Principal Component Regression}}} 
\author{
    \normalsize{Christian Brownlees}$^{\dag}$ \and 
    \normalsize{Gu\dh mundur Stef\'an Gu\dh mundsson}$^{\ddag,*}$ \and 
    Yaping Wang$^{\dag}$
} 

\long\def\symbolfootnote[#1]#2{\begingroup\def\thefootnote{\fnsymbol{footnote}}\footnote[#1]{#2}\endgroup}
      
\maketitle

\begin{abstract}
This paper studies the predictive performance of empirical risk minimization for principal component regression.
Our analysis accommodates the leading eigenvalues of the predictor covariance matrix growing either 
linearly or sublinearly with the number of predictors.
Additionally, we allow for both light-tailed and heavy-tailed data. 
Our main result establishes that, under appropriate conditions, empirical risk minimization for principal component regression 
is consistent for prediction and achieves near-optimal performance.


\end{abstract}

\symbolfootnote[0]{\\
    \noindent
    $^{\dag}$ Department of Economics and Business, Universitat Pompeu Fabra and Barcelona GSE;\\
    e-mail: \texttt{christian.brownlees@upf.edu}, \texttt{yaping.wang@upf.edu}.\\
    $^{\ddag}$ ACE and Department of Economics and Business Economics, Aarhus University; \\
    e-mail: \texttt{gsgudmundsson@econ.au.dk}.\\
    $^*$ Corresponding author. 
}

\newpage
\doublespacing

\section{Introduction}

Principal component regression (PCR) is a regression methodology with a long and well-established tradition 
that can be traced back to at least \citet{Hotelling:1957} and \citet{Kendall:1957}. 
In a nutshell, PCR is used to forecast a prediction target of interest 
on the basis of the principal components extracted from a potentially large set of predictors.
PCR is closely related to diffusion index forecasting models, 
which are one of the leading forecasting methodologies in macroeconomics \citep{StockWatson2012b}.

In this paper, we study the properties of PCR from a learning theory perspective. 
Our central contribution is establishing nonasymptotic prediction performance guarantees for PCR. 
Our main theorems may be interpreted as nonasymptotic analogues of the classic asymptotic results 
on the prediction properties of PCR obtained in \citet{Stock:Watson:2002}. 
An important feature of our analysis is that we consider both the case in which the largest eigenvalues 
of the predictor covariance matrix grow linearly with the number of predictors 
and the case in which they grow sublinearly. 
Moreover, we allow for both light-tailed and heavy-tailed data. 
Our framework builds upon assumptions and proof strategies developed in the factor model 
literature \citep{Bai:Ng:2002,Bai:2003,Fan:Liao:Mincheva:2011,Fan:Liao:Mincheva:2013,Onatski:2012}.

PCR may be described as a two-step procedure.
Let $ \mathcal D_T = \{ (Y_t, \bm X_t')' \}_{t=1}^T $ be a stationary sequence of zero-mean random vectors 
taking values in $\mathcal Y \times \mathcal X \subset \mathbb R \times \mathbb R^p$.
The goal is to forecast the prediction target $Y_t$ using the 
$p$-dimensional vector of predictors $\bm X_t=(X_{1\,t},\ldots,X_{p\,t})'$.
The first step of PCR consists in computing the $T \times K$ principal components matrix 
$\widehat{\mathbf P} = (\widehat{\bm P}_1, \ldots, \widehat{\bm P}_T)'$ associated with 
the $T \times p$ predictor matrix $\mathbf X = (\bm X_1, \ldots, \bm X_T)'$, for some appropriate choice of $K$.
This may be defined as the solution of the constrained least squares problem 
\[
	( \widehat{\mathbf B} , \widehat{\mathbf P} ) 
    = \operatorname*{argmin}_{ \substack{\mathbf B \in \mathbb R^{p \times K} \\ 
                                \mathbf P \in \mathbb R^{T \times K}} } 
    \| \mathbf X - \mathbf P \mathbf B' \|_F^2 
	~~\text{s.t.}~{1\over T} \mathbf P' \mathbf P = \bm I_K, 
    {1 \over p} \mathbf B' \mathbf B \text{ is diagonal},
\]
where $\| \cdot \|_F$ denotes the Frobenius norm.
It is well known that $\widehat {\mathbf P}$ is given by $\sqrt{T}$ 
times the first $K$ eigenvectors of the matrix $\mathbf X \mathbf X'$. 
It is useful to remark here that this allows us to express the vector of predictors 
$\bm X_t$ as a linear combination of the $K$ principal components $\widehat{\bm P}_t$ 
plus a residual vector $\widehat{\bm u}$, that is
\begin{equation}\label{eqn:pcr:decomp}
	\bm X_t = \widehat{\mathbf B} \widehat{\bm P}_t + \hat{\bm u}_t ~,
\end{equation}
where $\hat{\bm u}_t = \bm X_t - \widehat{\mathbf B} \widehat{\bm P}_t$.
The second step of PCR consists in computing the $K \times 1$ least squares coefficient vector 
$\hat{\bm \vartheta}$ associated with the regression of the $T\times 1$ target variable vector 
$\bm Y = (Y_1, \ldots, Y_T)'$ on the principal components matrix $\widehat{\bm P}$.
This is the solution to the least squares problem
\[
	\hat{\bm \vartheta} = \operatorname*{argmin}_{\bm \vartheta \in \mathbb R^K }  
    \| \bm Y - \widehat{\mathbf P} {\bm \vartheta} \|^2_2 ~,
\]
where $\| \cdot \|_2$ denotes the Euclidean norm. 
It is straightforward to check that $\hat{\bm \vartheta} = \widehat{\mathbf P}' \bm Y / T$.
We remark that in many empirical forecasting applications it is common to augment PCR 
with lags of the dependent variable. 
We conjecture that the analysis and results of this paper can be extended to such settings,
but leave the formal analysis for future research.

PCR can be formulated as a regularized empirical risk minimization procedure.\footnote{%
We remark that in this paper we use the expression 
``empirical risk minimization for principal component regression'' only for simplicity.
A more appropriate name for the procedure we study would be 
``regularized empirical risk minimization based on principal component analysis''.}
Consider the class of linear prediction rules $\bm \theta' \bm X_t$,
indexed by $\bm \theta \in \mathbb R^p$.
Then PCR can be cast as the regularized empirical risk minimization problem given by
\begin{equation}\label{eqn:pcr:cols}
	\hat{\bm \theta} \in \operatorname*{argmin}_{\bm \theta \in \mathbb R^p } R_T(\bm \theta)
	~~\text{ s.t. }~\widehat{\mathbf V}_R' \bm \theta = \bm 0 ~,
\end{equation}
where  
\[
	R_T(\bm \theta) = {1 \over T} \sum_{t=1}^T ( Y_t - \bm \theta' \bm X_t )^2 
    = {1 \over T} \| \bm Y - {\mathbf X} {\bm \theta} \|^2_2 ~,
\]
is the empirical risk and  
\[
	\widehat{\mathbf V}_R  = (\hat{\bm v}_{K+1}, \ldots, \hat{\bm v}_p ) ~,
\]
where $\hat{\bm v}_i$ is the eigenvector associated with the $i$-th largest eigenvalue 
of the sample covariance matrix of the predictors $\widehat{\bm \Sigma} = {\mathbf X' \mathbf X}/T$.
We emphasize that the minimizer in \eqref{eqn:pcr:cols} is not necessarily unique, 
and if more than one prediction rule achieves the minimum we may pick one arbitrarily.
In what follows we refer to $\hat{\bm \theta}$ as a (regularized) empirical risk minimizer (ERM).
It is straightforward to verify that $\hat{\bm \theta}'\bm X_t = \hat{\bm \vartheta}'\widehat{\bm P}_t$,
which implies that the PCR predictions can be equivalently expressed 
as linear predictions based on constrained least squares.

One of the classical objectives in learning theory is to establish predictive performance guarantees 
of statistical learning procedures relative to a suitable benchmark.
Define the out-of-sample principal components 
$\widehat{\bm P}_{T+1} = \widehat{\bm \Lambda}_K^{-1/2} \widehat{\mathbf V}_K' \bm X_{T+1}$
with $\widehat{\mathbf V}_K = (\hat{\bm v}_1, \ldots, \hat{\bm v}_K)$ and
$\widehat{\bm \Lambda}_K = \operatorname{diag}(\hat \lambda_1, \ldots, \hat \lambda_K)$ 
where $\hat \lambda_1, \ldots, \hat \lambda_K$ are the $K$ largest eigenvalues of $\widehat{\bm \Sigma}$.
We can then define the predictive performance of PCR as its conditional risk, that is
\begin{equation}\label{eqn:risk:erm}
	R( \hat{ \bm \vartheta} )  
	= \mathbb E \left[ ( Y_{T+1} - \hat{\bm \vartheta}' \widehat{\bm P}_{T+1} )^2 \mid \mathcal D_T \right] ~.
\end{equation}
The performance measure in \eqref{eqn:risk:erm} can be interpreted as the risk of the ERM obtained from 
the ``training'' sample $\mathcal D_T$ evaluated on the ``validation'' observation $(Y_{T+1}, \bm X_{T+1})'$.
Define the optimal risk for the class of linear prediction rules as
\[
	R^* = \min_{\bm \theta \in \mathbb R^p} \mathbb E \left[ ( Y_{T+1} - \bm \theta' \bm X_{T+1} )^2 \right] ~.
\]
Our aim is to find a pair $(B_T(p,K),\delta_T)$ such that $B_T(p,K) \rightarrow 0$ and $\delta_T\rightarrow 0$ 
as $T \rightarrow \infty$ for which 
\begin{equation}\label{eqn:risk:regret}
	 R( \hat{ \bm \vartheta } ) \leq R^* + B_T(p,K) ~,
\end{equation}
holds with probability at least  $1 - \delta_T$ for any $T$ sufficiently large.
The inequality in \eqref{eqn:risk:regret} is commonly referred to as an \emph{oracle inequality}.
Oracle inequalities such as \eqref{eqn:risk:regret} provide nonasymptotic guarantees 
on the performance of empirical risk minimization and 
imply that the ERM performs asymptotically as well as the best linear predictor.

%We remark that the performance measure in \eqref{eqn:risk:erm} allows us to keep our analysis close
%to the bulk of the contributions in the learning theory literature (which typically focus on the analysis of i.i.d.~data) and facilitates comparisons.
%We remark that \citet{brownlees-gudmundsson-2025} and \citet{Brownlees:LlorensTerrazas:2021} consider 
%alternative performance measures such as the conditional out-of-sample average risk of the ERM, 
%which has a more attractive interpretation for time series applications. 
%It turns out that these alternative measures lead to essentially the same theoretical analysis 
%at the expense of introducing additional notation.
%We therefore focus on the performance measure defined in \eqref{eqn:risk:regret} for clarity.

This paper is related to various strands of the literature.
First, the literature on diffusion index forecasting models, which includes
\citet{Stock:Watson:2002}, \citet{StockWatson2002b}, \citet{BaiNg2006}, \citet{BoivinNg2006} 
and \citet{fan-masini-medeiros-2023}. 
In particular, our work is close in spirit to the important contribution of \citet{fan-masini-medeiros-2023} 
that studies the properties of a large class of high-dimensional models, which includes factor models.
In contrast to \citet{fan-masini-medeiros-2023},
we allow the growth rate of the largest eigenvalues of the covariance matrix of the predictors 
to range from sublinear to linear. 
Specifically, our largest eigenvalues are proportional to $p^\alpha$ for $\alpha \in (0,1]$, 
whereas \citet{fan-masini-medeiros-2023} focus on the case of $\alpha = 1$.
Moreover, we derive explicit conditions under which principal component regression 
attains the oracle rate $K/T$, up to logarithmic factors. 
We further establish a matching lower bound (up to logarithmic factors), 
showing that this rate is essentially tight.
Second, our work is also related to the literature on approximate factor models, 
principal component analysis and spiked covariance models, 
which includes \citet{Fan2017}, \citet{Donoho_2018}, 
\citet{Forni:Hallin:Lippi:Reichlin:2000,Forni:Hallin:Lippi:Reichlin:2005}, \citet{Bai:Ng:2019}, 
\citet{Bai_2012, Bai_2016}, \citet{Lam2011}, \citet{Fan2016}, \citet{Gon_alves_2020}, \citet{Barigozzi2020}, 
\citet{Su2017} and \citet{Uematsu2022}.
Third, the literature on the small-ball method, which includes \citet{LecueMendelson:2016}, 
\citet{LecueMendelson:2017}, \citet{Mendelson:2018} and \citet{LecueMendelson:2018}.
Fourth, the literature on empirical risk minimization for linear regression, 
which includes \citet{BirgeMassart:1998} and \citet{Tsybakov:2003} among others.
In particular, our contribution is close to the subset of the literature that deals with dependent data, 
as in \citet{JiangTanner:2010,brownlees-gudmundsson-2025,brownlees-llorens_terrazas-2024}.
Relative to \citet{brownlees-gudmundsson-2025}, who also study the performance of a
linear regression procedure, the main novelty lies in the fact that in this paper the 
regressors (i.e.~the estimated principal components) are themselves 
subject to estimation error, which introduces a substantial 
additional layer of technical complexity into the analysis.

The rest of the paper is structured as follows.
Section \ref{sec:asm} introduces additional basic notation, assumptions and preliminary results.
Section \ref{sec:result} establishes a bound on the performance of empirical risk minimization 
for thin-tailed data, which is the main result of the paper, as well as a lower bound.
Section \ref{sec:proof} contains the proof of the main result.
Section \ref{sec:heavy} contains an extension of our main result that allows for heavy-tailed data.
Concluding remarks follow in Section \ref{sec:end}. 
All remaining proofs are given in the appendix or the online appendix.

\section{Notation, Preliminaries and Assumptions} \label{sec:asm}

In this section we lay out the assumptions required for our analysis.
We first introduce some basic notation.
For a generic vector $\bm x \in \mathbb R^d$ we define $\| \bm x \|_r$ as $( \sum_{i=1}^d |x_i|^r )^{1/r}$ 
for $ 1 \leq r < \infty$ and $\max_{i=1,\ldots,d} |x_i| $ for $r=\infty$.
For a generic random variable $X \in \mathbb R$ we define $\| X \|_{L_r}$ as $(\mathbb E[ |X|^r ] )^{1/r}$ 
for $ 1 \leq r < \infty$ and $\inf \{ a : \mathbb P( |X|> a ) = 0 \}$ for $r = \infty$.
For a positive semi-definite matrix $\mathbf M$ we use $\mathbf M^{ {1/2} }$ to denote the positive 
semi-definite square root matrix of $\mathbf M$ 
and $\mathbf M^{ -{1/2} }$ to denote the generalized inverse of $\mathbf M^{1/2}$.
For a generic matrix $\mathbf M$ we use $\| \mathbf M \|_2$ to denote the spectral norm of $\mathbf M$.

We begin with a preliminary lemma that establishes the population analogue of the principal components 
decomposition of the prediction vector $\bm X_t$ introduced in \eqref{eqn:pcr:decomp}.
\begin{lem}\label{lem:pcr}
    Let $Y_t$ be a random variable such that $\| Y_t \|_{L_2} < \infty$ and $\bm X_t$ be a zero-mean 
    $p$-dimensional random vector such that $\| X_{i\,t} \|_{L_2} < \infty$ for $i=1, \ldots, p$.
    For any $K \in \{1, \ldots, p\}$, define $\bm \Lambda_K = \operatorname{diag}(\lambda_1,\ldots,\lambda_K)$, 
    $\bm \Lambda_R = \operatorname{diag}(\lambda_{K+1}, \ldots, \lambda_p)$, 
    $\mathbf V_K = (\bm v_1, \ldots, \bm v_K)$ and $\mathbf V_R = (\bm v_{K+1},\ldots,\bm v_p)$
    where $\lambda_1, \ldots, \lambda_p$ and $\bm v_1, \ldots, \bm v_p$ denote the non-increasing sequence 
    of eigenvalues of $\bm \Sigma= \mathbb E(\bm X_t \bm X_t')$ and the corresponding sequence of eigenvectors.

    Then, $(i)$ it holds that
    \[
	    \bm X_t = \mathbf B \bm P_t + \bm u_t,
    \]
    where $\mathbf B = \mathbf V_K \bm \Lambda_K^{1/2}$, $\bm P_t = \bm \Lambda_K^{-1/2} \mathbf V_K' \bm X_t$ 
    and $\bm u_t = \mathbf V_R (\mathbf V_R' \mathbf V_R)^{-1} \mathbf V_R' \bm X_t $,
    with $\mathbf B' \mathbf B $ diagonal, $\mathbb E( \bm P_t \bm P_t' ) = \mathbf I_K $, 
    $\mathbb E( \bm u_t \bm u_t' ) = \mathbf V_R \bm \Lambda_R \mathbf V_R'$ and 
    $\mathbb E( \bm P_t \bm u_t' ) = \mathbf 0_{K \times p}$.

    $(ii)$ Let $\bm \theta^* \in \mathbb R^p$ be a vector of coefficients associated 
    with the best linear predictor of $Y_t$ based on $\bm X_t$. 
    Then, it holds that 
    \[
	    \bm X_t' \bm \theta^* = \bm P_t' \bm \vartheta^* + \bm u_t' \bm \gamma^*, 
    \]
    where $\bm \vartheta^* = \bm \Lambda_K^{1/2} \mathbf V_K' \bm \theta^*$ and 
    $\bm \gamma^* = \mathbf V_R (\mathbf V_R' \mathbf V_R)^{-1} \mathbf V_R' \bm \theta^*$.

    $(iii)$ $\bm \vartheta^*$ is unique provided that $\lambda_K > 0$, 
    and $\bm u_t' \bm \gamma^*$ is unique almost surely.
    
    $(iv)$ $\|\bm\vartheta^*\|_2 \le \| Y_t \|_{L_2}$ 
    and $\| \bm\Lambda_R^{1/2} \mathbf V_R' \bm\gamma^* \|_2 \le \| Y_t \|_{L_2}$.
\end{lem}

Part $(i)$ of the lemma states that $\bm X_t$ can be decomposed into two parts:
a common component consisting of the product of the matrix of coefficients $\mathbf B$ 
with the random vector $\bm P_t$, and the residual random vector $\bm u_t$, 
where $\bm P_t$ and $\bm u_t$ are orthogonal.
We call $\bm P_t$ the population principal components and $\bm u_t$ the idiosyncratic component.
Part $(ii)$ implies that the best linear predictor for $Y_t$ can alternatively be represented as a function 
of the population principal components $\bm P_t$ and the idiosyncratic component $\bm u_t$ 
with the coefficient vectors $\bm \vartheta^*$ and $\bm \gamma^*$.
Note that since $\bm P_t$ and $\bm u_t$ are orthogonal, $\bm \vartheta^*$ may be interpreted 
as the best linear predictor based on the population principal components $\bm P_t$.
We remark that, clearly, the matrix $\bm V_R (\bm V_R' \bm V_R)^{-1} \bm V_R'$ may be simplified to 
$\bm V_R \bm V_R'$ in the definitions of $\bm u_t$ and $\bm \gamma^*$, 
but we prefer the former as it emphasizes that this is a projection matrix.
Moreover, note that the decomposition established in part $(ii)$ holds for any vector in $\mathbb R^p$,
but for our purposes it suffices to focus on a vector of coefficients associated 
with the best linear predictor of $Y_t$ based on $\bm X_t$.
Part $(iii)$ implies that, despite the fact that the vector of coefficients of the best linear predictor 
is not unique in general, $\bm \vartheta^*$ is unique when $\lambda_K > 0$,
and $\bm u_t' \bm \gamma^*$ is unique almost surely.

We now lay out the assumptions of our analysis.
We say that the $d$-dimensional random vector $\bm U$ is sub-Gaussian with parameter $C_U > 0$ if,
for any $\varepsilon>0$ and some constant $C_0 > 0$, it holds that
\[	
    \sup_{\bm v: \| \bm v \|_2=1} \mathbb P\left( |\bm v'\bm U| > \varepsilon \right) 
    \leq C_0 \exp( -C_U \varepsilon^2 ) ~.
\]
For a univariate random variable $U$ this is equivalent to 
$\mathbb P(|U| > \varepsilon) \le C_0 \exp( -C_U \varepsilon^2 )$.

\begin{asm}[Distribution] \label{asm:dist}
    There exists a constant $C_Z$ such that $Y_t$ and $\bm Z_t = \bm \Sigma^{-1/2} \bm X_t$ 
    where $\bm \Sigma = \mathbb E(\bm X_t \bm X_t')$ are sub-Gaussian with parameter $C_Z$.
\end{asm}

Assumption \ref{asm:dist} requires the prediction target $Y_t$ and the standardized predictors 
$\bm Z_t$ to have sub-Gaussian tails, so that the tails of the data decay exponentially. 
We remark that such a condition is fairly standard in the analysis of large-dimensional factor models 
\citep{Fan:Liao:Mincheva:2011}.
In Section \ref{sec:heavy} we relax this assumption and study the performance of PCR under heavy-tailed data,
at the cost of more involved proofs.

Let $\mathcal F_{-\infty}^t$ and $\mathcal F_{t+l}^{\infty}$ be the $\sigma$-algebras generated by 
$\lbrace (Y_s, \bm X_s')': -\infty \leq s \leq t \rbrace$ and 
$\lbrace (Y_s, \bm X_s')': t + l \leq s \leq \infty\rbrace$ respectively for some $t \in \mathbb Z$ 
and define the $\beta$-mixing coefficients
\[
    \beta(l) 
    = \frac12 \sup_{\{A_i\}_{i=1}^I, \{B_j\}_{j=1}^J}
    \sum_{i=1}^I \sum_{j=1}^J \left| \mathbb P(A_i \cap B_j) - \mathbb P(A_i)\mathbb P(B_j) \right|,
\]
where the supremum is taken over all finite measurable partitions
$\{A_i\}_{i=1}^I$ and  $\{B_j\}_{j=1}^J$ such that $A_i \in \mathcal F_{-\infty}^t$ 
and $B_j \in \mathcal F_{t+l}^{\infty}$.

\begin{asm}[Dependence] \label{asm:mix}
   $(i)$ There exist constants $C_\beta > 0$ and $r_\beta \ge 1$ such that the $\beta$-mixing coefficients 
   satisfy $\beta(l) \le \exp( -C_\beta l^{r_\beta} )$. 
   $(ii)$ Let $\varepsilon_t = Y_t - \bm X_t' \bm \theta^*$ where $\bm \theta^*$ is a vector coefficients 
   associated with the best linear predictor of $Y_t$ based on $\bm X_t$.
   The process $\{ (Y_t,\bm X_t')' \}$ satisfies 
   $\mathbb E[ \varepsilon_{T+1} \bm X_{T+1} \mid \mathcal D_T ] = \bm 0$,
   $\mathbb E[ \varepsilon_{T+1}^2 \mid \mathcal D_T] = \mathbb E[ \varepsilon_t^2 ]$
   and $\mathbb E[ \bm X_{T+1} \bm X_{T+1}' \mid \mathcal D_T ] = \mathbb  E[ \bm X_t \bm X_t' ]$.   
\end{asm}

Part $(i)$ of Assumption \ref{asm:mix} states that the process $\{(Y_t, \bm X_t')'\}$ is absolutely regular 
and has geometrically decaying $\beta$-mixing coefficients. 
We recall that $\beta$-mixing implies $\alpha$-mixing, which is a fairly standard assumption in the analysis 
of large-dimensional time series models \citep{JiangTanner:2010,Fan:Liao:Mincheva:2011,Kock:Callot:2015}.
We require $\beta$-mixing to use the matrix Bernstein inequality of \citet{banna-merlevede-youssef-2016}
which allows us to establish predictive performance guarantees for PCR for heavy-tailed data.\footnote{We 
remark that Theorem \ref{thm:erm-pcr} can be derived under geometric $\alpha$-mixing. 
However, for simplicity, we prefer to rely on a single notion of dependence throughout the paper.}
Part $(ii)$ of Assumption \ref{asm:mix} imposes three conditions: 
a conditional martingale difference sequence condition on the score of the best linear predictor,
conditional homoskedasticity on the prediction error of the best linear predictor,
and a conditional homoskedasticity condition on the predictors. 
These conditions play a key role in decomposing the predictive performance (see Lemma~\ref{lem:decomp} below), 
allowing the analysis of the out-of-sample risk of the ERM to proceed analogously to the i.i.d. case 
despite the dependent data setting. 
Notably, part $(ii)$ plays no role in controlling the estimation error of PCR.

\begin{asm}[Eigenvalues] \label{asm:eigen}
    There is an integer $K \in \{1, \ldots, p\}$, a constant $\alpha \in (0,1]$
    and a sequence of non-increasing nonnegative constants $c_1, \ldots, c_p$ with $c_K > 0$
    such that $\lambda_i = c_i p^\alpha$ for $i = 1, \ldots, K$ and
    $\lambda_i = c_i$ for $i = K+1, \ldots, p$.
\end{asm}

Assumption \ref{asm:eigen} allows the first $K$ eigenvalues of the covariance matrix $\bm \Sigma$
to diverge with the number of predictors $p$, where the rate of divergence is determined by $\alpha$. 
We distinguish between different regimes that depend on the value of this constant.
We label $\alpha=1$ the strong signal regime, 
$\alpha \in (1/2,1)$ the semi-strong signal regime, and $\alpha \in (0,1/2]$ the semi-weak signal regime.
We point out that here $K$ is assumed known and that there is a large literature devoted to the estimation 
of this quantity \citep{Bai:Ng:2002,Amengual2007,Onatski_2010,Lam:Yao:2012,Ahn:Horenstein:2013,Yu_2019}.
It is important to emphasize that the assumption allows for the non-diverging eigenvalues of $\bm \Sigma$ 
to be zero, so our framework allows $\bm \Sigma$ to be singular.

\begin{asm}[Dimensions] \label{asm:npk}
    (i) There are constants $C_p \ge 1$ and $r_p \in [0, \overline r_p)$ such that 
    $p = \lfloor C_p T^{r_p} \rfloor$ where 
    $\overline r_p = r_\beta / ( 1 + (1- \alpha) r_\beta )$;
    (ii) There are constants $C_p \ge C_K \ge 1$ and $r_K \in [0, \overline r_K)$ such that 
    $K = \lfloor C_K T^{r_K} \rfloor$ where 
    $\overline r_K = r_\beta / ( 1 + r_\beta )$; and
    (iii) $r_p$ and $r_K$ additionally satisfy $r_K + r_p (1 - \alpha) < 1$ and $r_k \leq r_p$.
\end{asm}

Assumption \ref{asm:npk} permits both the predictor dimension and the number of retained principal components 
to increase with the sample size $T$.
Their admissible growth rates depend on the rate of decay of the temporal dependence, 
as measured by $r_\beta$, and on the signal strength, measured by $\alpha$. 
The weaker the dependence and the stronger the signal, the larger the numbers of allowed predictors 
and principal components.
The assumption permits the number of predictors to exceed the sample size, 
but requires the number of principal components to be smaller than $T$.

\section{Performance of Empirical Risk Minimization}\label{sec:result}

We now state our main result on the performance of PCR for thin-tailed data. 
The theorem establishes a nonasymptotic oracle inequality that bounds the excess risk of PCR 
relative to the best linear predictor based on $\bm X_t$. 

\begin{thm} \label{thm:erm-pcr}
    Suppose Assumptions \ref{asm:dist}--\ref{asm:npk} are satisfied.

    Then for every $\eta>0$ there exist constants $C_1, C_2 > 0$ such that, for all $T$ sufficiently large
    \begin{align*} 
        R( \hat{\bm \vartheta} ) - R^* 
        &\le 2 (\bm \gamma^*)' \mathbf V_R \bm \Lambda_R \mathbf V_R' \bm \gamma^* \notag \\
        &+ C_1 \left(1 + { ( p + \log(T) )^{ {r_\beta +1 \over \, r_\beta}  } \over T }
        + \sqrt{ { ( p + \log(T) )^{ {r_\beta +1 \over \, r_\beta}  } \over T }} \right)
        \frac{1}{p^{2\alpha - 1}} \frac{K \log(T)}{T} \\
        &+ C_2 \frac{K \log(T)}{T},
    \end{align*}
    holds with probability at least $1 - T^{-\eta}$.
\end{thm}

The theorem establishes a regret bound on the excess risk of PCR relative to the best linear predictor 
that can be obtained on the basis of the predictors $\bm X_t$.
The gap is made up of three terms. 
The first accounts for the approximation error of PCR and the last two account for the estimation error.
The approximation error measures the gap between the performance of the best linear predictor based on 
the population principal components $\bm P_t$ and the best linear predictor based on the predictors $\bm X_t$.
The estimation error measures the gap between the performance of PCR relative to the best linear predictor 
based on the population principal components $\bm P_t$.

A natural question that arises from Theorem \ref{thm:erm-pcr} is whether the rate for the estimation error 
is sharp. 
The following theorem establishes a lower bound for the performance of PCR under 
a Gaussian data-generating process.

\begin{thm} \label{thm:lower-bound}
    Suppose $\{ (\bm X_t', \varepsilon_t)' \}$ is an i.i.d.~sequence of Gaussian vectors such that
    $\bm X_t \sim \mathcal N(\bm 0, \bm\Sigma)$ and $\varepsilon_t \sim \mathcal N(0, \sigma^2)$, 
    with $\varepsilon_t$ independent of $\bm X_t$.
    Let there be an integer $K \in \{1, \ldots, p\}$, a constant $\alpha \in (0,1]$
    and a sequence of non-increasing strictly positive constants $c_1, \ldots, c_K$ such that 
    $\lambda_i = c_i p^\alpha$ for $i = 1, \ldots, K$ and $\lambda_i = 1$ for $i = K+1, \ldots, p$.
    Suppose moreover that it holds that 
    \[
        Y_t = \bm X_t' \bm\theta^* + \varepsilon_t, \quad t=1,\ldots,T ~,
    \]
    where $\bm\theta^* \in \operatorname{span}(\mathbf V_R)$ and thus satisfies 
    $\bm \vartheta^* = \bm \Lambda_K^{1/2} \mathbf V_K' \bm \theta^* = \bm 0$ and
    $\bm \gamma^* = \mathbf V_R \mathbf V_R' \bm \theta^* = \bm \theta^*$.
    Finally, suppose that Assumption \ref{asm:npk} is satisfied.

    Then we have that for every $\eta > 0$, and all sufficiently large $T$
    \[
        R(\hat{\bm \vartheta}) - R^*
        \ge \frac16 \left( \| \bm\gamma^* \|_2^2 + \sigma^2 \frac{K}{T} \right)
    \]
    holds with probability at least $1 - T^{-\eta} - e^{-K/9}$.
\end{thm}

The theorem establishes that, under a specific Gaussian data-generating process, 
the estimation error of PCR is bounded from below by a term of order $K/T$ with positive probability. 
It is straightforward to verify that the assumptions of the theorem are compatible with 
Assumptions~\ref{asm:dist}--\ref{asm:npk}.

A number of remarks on the performance of PCR are in order.
First, we discuss under which conditions the estimation error rate established in Theorem \ref{thm:erm-pcr} is,
in some appropriate sense, optimal.
For ease of exposition we assume that the approximation error is zero for the remainder of this paragraph.
Theorem \ref{thm:lower-bound} establishes that the lower bound on the performance of PCR is of the order $K/T$.
This coincides with the optimal rate that the least squares estimator could achieve 
if the population principal components were observed \citep{Tsybakov:2003}.
This motivates us to label the $K/T$ rate as the optimal rate. 
Inspection of Theorem \ref{thm:erm-pcr} reveals that the estimation error rate depends crucially 
on the signal strength $\alpha$. 
When $\alpha \in [1/2,1]$, the near-optimal rate $(K \log(T))/T$ is achieved provided that 
the number of predictors is not too large in the sense that
\[
    r_p \le \frac{r_\beta}{1 + 2 (1 - \alpha) r_\beta} ~.
\]
This condition is, for example, satisfied when $p$ is fixed ($r_p = 0$), and more generally, 
the admissible range of $r_p$ is increasing in both $\alpha$ and $r_\beta$. 
Notice that the condition simplifies to $r_p < r_\beta$ when $\alpha = 1$,
which is already guaranteed by Assumption \ref{asm:npk}.
When $\alpha \in (0, 1/2)$, the signal is too weak for the near-optimal rate to be achieved unless $p$ is fixed.
This is because the factor $p^{1-2\alpha}$ in the estimation error diverges with $p$, 
and the bound is strictly slower than $(K \log(T))/T$ as long as $p$ is growing. 
In this regime PCR remains consistent for prediction under Assumption~\ref{asm:npk}, 
but at a rate that is penalized by the weakness of the signal.

Second, it is insightful to provide an alternative representation of the approximation error of PCR. 
This may be equivalently expressed as
\[
	(\bm \gamma^*)' \mathbf V_R \bm \Lambda_R \mathbf V_R' \bm \gamma^* 
    = (\bm \theta^*)' \mathbf V_R \bm \Lambda_R \mathbf V_R' \bm \theta^* ~.
\]
This highlights that the approximation error of PCR is small when the projection of the best linear predictor 
$\bm \theta^*$ on the subspace spanned by the population eigenvectors $\bm v_{K+1}, \ldots, \bm v_p$ is small.
Equivalently, if the contribution of the idiosyncratic component vector $\bm u_t$ is negligible 
then PCR has a negligible approximation error.
Clearly, when $\bm \gamma^* = \bm 0$ the best linear predictor based on $\bm X_t$ coincides with the 
best linear predictor based on $\bm P_t$ and the approximation error is zero.
We remark that part $(iv)$ of Lemma \ref{lem:pcr} shows that the approximation error is bounded by a constant 
that does not depend on $K, p$ or $T$.

Finally, we note that the bound in Theorem~\ref{thm:erm-pcr} provides some theoretical support for the 
empirical finding of \citet{BoivinNg2006} that adding more predictors to a diffusion index model 
does not always improve forecast performance. 
When $\alpha \in (0, 1)$, Assumption~\ref{asm:npk} restricts the admissible growth rate of $p$ relative to $T$ 
more severely than in the strong signal setting $\alpha = 1$,
illustrating how expanding the predictor set indiscriminately can be detrimental to forecast performance.

\section{Proof of Main Result}\label{sec:proof}

The proof of Theorem \ref{thm:erm-pcr} begins with a decomposition of an upper bound of the excess risk 
of the empirical risk minimizer.
Define the approximate rotation matrix 
$\mathbf H = \widehat{\bm \Lambda}_K ^{-1/2} \widehat{\mathbf V}_K' \mathbf V_K \bm \Lambda_K ^{1/2}$ 
\citep{Bai:Ng:2002} and the infeasible PCR estimator 
$\tilde{\bm \vartheta} = \operatorname{argmin}_{\bm \vartheta \in \mathbb R^K } 
\frac1T \sum_{t=1}^T (Y_t - \bm P_t' \bm \vartheta)^2$.
The following lemma provides a useful bound for the excess risk.
\begin{lem} \label{lem:decomp}
    Suppose that $\{ (Y_t, \bm X_t')' \}_{t}$ is a stationary sequence taking values in 
    $\mathcal Y \times \mathcal X \subset \mathbb R \times \mathbb R^p$ such that 
    $\mathbb E(Y_t^2) < \infty$ and $\mathbb E(X_{i\,t}^2) < \infty$ for all $i = 1, \ldots, p$.
    Let $\varepsilon_t = Y_t - \bm X_t' \bm \theta^*$ where $\bm \theta^*$ is a vector coefficients 
    associated with the best linear predictor of $Y_t$ based on $\bm X_t$.
    Suppose $\{ (Y_t,\bm X_t')' \}$ satisfies 
    $\mathbb E[ \varepsilon_{T+1} \bm X_{T+1} \mid \mathcal D_T ] = \bm 0$,
    $\mathbb E[ \varepsilon_{T+1}^2 \mid \mathcal D_T] = \mathbb E[ \varepsilon_t^2 ]$
    and $\mathbb E[ \bm X_{T+1} \bm X_{T+1}' \mid \mathcal D_T ] = \mathbb  E[ \bm X_t \bm X_t' ]$.   
    
    Then, for all $T \geq K$ it holds that 
    \begin{align}
	& R( \hat{\bm \vartheta} ) - R^* 
        \le 2 \| \hat{\bm \vartheta} \|_2^2
        \mathbb E( \| \widehat {\bm P}_{T+1} - \mathbf H {\bm P}_{T+1} \|^2_2 \mid \mathcal D_T ) 
        + 2 \| \tilde{\bm \vartheta} - \mathbf H ' \hat{\bm \vartheta} \|^2_2 \nonumber \\
    	&\quad + 2 \| \bm \vartheta^* - \tilde{\bm \vartheta} \|^2_2 
        + 2 (\bm \gamma^*)' \mathbf V_R \bm \Lambda_R \mathbf V_R' \bm \gamma^*. \label{eqn:exriskdecomp}
    \end{align}
\end{lem}

Lemma \ref{lem:decomp} implies that in order to control the excess risk of PCR it suffices 
to provide appropriate bounds on the terms on the right hand side of \eqref{eqn:exriskdecomp}.
The following three propositions establish such bounds.

The first two propositions control the terms 
$\mathbb E( \|  \widehat {\bm P}_{T+1} - \mathbf H {\bm P}_{T+1} \|^2_2 \mid \mathcal D_T )$ and 
$\| \tilde{\bm \vartheta} - \mathbf H ' \hat{\bm \vartheta} \|^2_2 $.
We remark that these two terms capture the risk of PCR arising from the estimation of the principal components.

\begin{prop} \label{prop:p-est-oos}
    Suppose Assumptions \ref{asm:dist}--\ref{asm:npk} are satisfied. 
    Then for every $\eta > 0$ there exists a constant $C > 0$ such that,
    for all $T$ sufficiently large
    \[
        \mathbb E( \| \hat{\bm P}_{T+1} - \mathbf H \bm P_{T+1} \|_2^2 \mid \mathcal D_T)
        \le C \frac{1}{p^{2 \alpha - 1}} \frac{K \log(T)}{T},
    \]
    holds with probability at least $1 - T^{-\eta}$.
\end{prop}

\begin{prop} \label{prop:p-est-is}
    Suppose Assumptions \ref{asm:dist}--\ref{asm:npk} are satisfied. 
    Then for every $\eta>0$ there exists a constant $C > 0$ such that, for all $T$ sufficiently large
    \[
        \| \tilde{\bm \vartheta} - \mathbf H ' \hat{\bm \vartheta} \|^2_2 
        \le C \left(1 + { ( p + \log(T) )^{ {r_\beta +1 \over \, r_\beta}  } \over T }
        + \sqrt{ { ( p + \log(T) )^{ {r_\beta +1 \over \, r_\beta}  } \over T }} \right)
        \frac{1}{p^{2\alpha - 1}} \frac{K \log(T)}{T},
    \]
    holds with probability at least $1 - T^{-\eta}$.
\end{prop}

The next proposition controls the term $\| \bm \vartheta^* - \tilde{\bm \vartheta} \|^2_2$.
We remark that this term may be interpreted as the risk of the least squares estimator of PCR 
based on the population principal components.

\begin{prop} \label{prop:linreg}
    Suppose Assumptions \ref{asm:dist}--\ref{asm:npk} are satisfied.
    Then for every $\eta>0$ there exists a constant $C > 0$ such that, for all $T$ sufficiently large
    \[   
	    \| \bm \vartheta^* - \tilde{\bm \vartheta} \|^2_2 \leq C {K \log (T) \over T},
    \]
    holds with probability at least $1 - T^{-\eta}$.
\end{prop}

The claim of Theorem \ref{thm:erm-pcr} then follows from Propositions \ref{prop:p-est-oos} to \ref{prop:linreg},
along with Proposition \ref{prop:norm-y} which establishes that for every $\eta>0$ there exists a constant 
$C > 0$ such that, for all $T$ sufficiently large, $\| \hat{\bm \vartheta} \|_2^2 < C$ holds with probability
at least $1 - T^{-\eta}$, after a straightforward application of the implication rule and the union bound.

\section{Performance of Empirical Risk Minimization for Heavy-tailed Data}\label{sec:heavy}

The analysis in the preceding sections assumes that the data are light-tailed, 
in the sense that $Y_t$ and the standardized predictors $\bm Z_t$ have sub-Gaussian tails. 
In this section we relax this condition and study the performance of PCR under heavy-tailed data. 
The main result of this section establishes a bound on the predictive performance of PCR analogous 
to Theorem~\ref{thm:erm-pcr} under a polynomial tail condition.

To this end we introduce modified versions of the distributional and dimensionality assumptions.

\begin{asm2}{1}{Distribution}{asm:dist-ht}
    Let $\bm Z_t = \bm \Sigma^{-1/2} \bm X_t$.
    There exists a constant $m > 4$ and a positive constant $C_Z$ such that 
    $\mathbb E( |Y_t|^m) \le C_Z$ and $\sup_{\| \bm v \|_2 = 1} \mathbb E( |\bm v'\bm Z_t|^m) \le C_Z$.
\end{asm2}

Assumption \assref{1}{asm:dist-ht} is a fairly weak polynomial tail assumption. 
Note that we assume that all possible one-dimensional marginals of the standardized predictors $\bm Z_t$ 
have polynomial tails. 

\begin{asm2}{4}{Dimensions}{asm:npk-ht}
    Let $0 < \kappa < \min \lbrace 1 - 2 / m, (m - 4) / 2 \rbrace$ be a constant.
    $(i)$ There are constants $C_p \ge 1$ and $r_p \in [0, \overline r_p)$ such that 
    $p = \lfloor C_p T^{r_p} \rfloor$ where 
    $(1 - \alpha) \overline r_p = 1 - 2 / m - \kappa$ if $\alpha \in (0, 1)$
    and $\overline r_p = \infty$ if $\alpha = 1$;
    $(ii)$ There are constants $C_p \ge C_K \ge 1$ and $r_K \in [0, \overline r_K)$ such that 
    $K = \lfloor C_K T^{r_K} \rfloor$ where 
    $\overline r_K = 1 - 2 / m - \kappa $; and
    $(iii)$ $r_p$ and $r_K$ additionally satisfy $r_K + r_p ( 1 - \alpha) < 1$ and $r_K \le r_p$.
\end{asm2}

Assumption \assref{4}{asm:npk-ht} states the allowed growth regimes for $p$ and $K$.
In particular, the larger $m$ is, the higher the admissible growth rates.
The dependence of the growth rate of $p$ on $\alpha$ is analogous to that in the thin-tailed setting.
Note that the growth rates here are restricted by $m$, 
whereas in Assumption \ref{asm:npk} they are restricted by $r_\beta$.

We next state our result on the performance of PCR for heavy-tailed data.

\begin{thm} \label{thm:erm-pcr-ht}
    Suppose Assumptions \assref{1}{asm:dist-ht}, \ref{asm:mix}, \ref{asm:eigen} 
    and \assref{4}{asm:npk-ht} are satisfied.

    Then for every $\kappa / m > \delta > 0$ and every $m \delta / 2  > \eta > 0$
    there exist constants $C_1, C_2 > 0$ such that, for all $T$ sufficiently large
    \begin{align*} 
        R( \hat{\bm \vartheta} ) - R^*
        &\le 2 (\bm \gamma^*)' \bm V_R \bm \Lambda_R \bm V_R' \bm \gamma^* \notag \\
        &+ C_1 \left(1 + \sqrt{ \frac{p \log(T) }{T} }
        + \frac{p \log(T)^3}{T^{1 - 2/m - \delta}} \right) 
        \times \frac{1}{p^{2\alpha - 1}} \frac{K \log(T)}{T} \\
        &+ C_2 \frac{K \log(T)}{T},
    \end{align*}
    with probability at least $1 - T^{-\eta}$.
\end{thm}

A number of remarks on Theorem~\ref{thm:erm-pcr-ht} are in order. 
The lower bound established in Theorem~\ref{thm:lower-bound} continues to hold in the present setting, 
as it is derived under a Gaussian data-generating process which satisfies Assumption~\assref{1}{asm:dist-ht}
and is consistent with Assumption \assref{4}{asm:npk-ht}.
The optimal rate therefore remains $K/T$, as in the light-tailed case. 
The key difference between the bounds obtained in Theorems \ref{thm:erm-pcr-ht} and \ref{thm:erm-pcr}
is that the former incurs a penalty based on the number of moments $m$,
whereas the latter is penalized by $r_\beta$.
If we set $\delta = c / m$ for some $\kappa > c > 2 \eta$ and let $m$ and $r_\beta$ be large,
the theorems give essentially the same conclusions.\footnote{Note that the theorems respectively require 
$m$ and $r_\beta$ to be finite.}
Similar to the thin-tailed setting, 
the near-optimal rate $(K \log(T)) / T$ is obtained without further restrictions if $\alpha = 1$.
If $\alpha \in [1/2, 1)$, it is obtained when $r_p < (1 - 2/m - \delta) / (2 (1 - \alpha))$,
which is slightly stricter than what Assumption \assref{4}{asm:npk-ht} stipulates.
If $\alpha \in (0, 1/2)$ the signal is too weak for optimality unless $p$ is fixed.

\section{Conclusions}\label{sec:end}

This paper develops a nonasymptotic theory of prediction for principal component regression. 
Our main result is an oracle inequality bounding the excess risk of principal component regression 
relative to the best linear predictor. 
Under appropriate conditions, the oracle inequality shows that principal component regression achieves 
near-optimal performance up to a logarithmic factor.
Several features of the analysis merit attention. 
The framework accommodates both linearly and sublinearly growing leading eigenvalues 
of the predictor covariance matrix, indexed by the signal strength parameter $\alpha \in (0,1]$. 
Signal strength governs not only the convergence rate of the estimation error 
but also the admissible growth rate of the predictor dimension: 
stronger signals permit larger predictor sets while preserving near-optimal performance, 
whereas weaker signals impose tighter dimension restrictions. 
The analysis also covers a broad class of tail behaviours, 
encompassing both thin-tailed and heavy-tailed distributions.
Our results provide theoretical grounding for the empirical finding of \citet{BoivinNg2006} that 
indiscriminately expanding the predictor set can be detrimental to forecast accuracy.

\appendix

\section{Proofs}

\setcounter{prop}{0}
\setcounter{lem}{0}
\setcounter{thm}{0}
\setcounter{equation}{0}
\renewcommand\theprop{\thesection.\arabic{prop}}
\renewcommand\thelem{\thesection.\arabic{lem}}
\renewcommand\thethm{\thesection.\arabic{thm}}
\renewcommand\theequation{\thesection.\arabic{equation}}

We recall that $\bm \Sigma = \mathbf V_K \bm \Lambda_K \mathbf V_K' + \mathbf V_R \bm \Lambda_R \mathbf V_R'$
where $\mathbf V_K = (\bm v_1, \ldots, \bm v_K)$, 
$\mathbf V_R = (\bm v_{K+1}, \ldots, \bm v_p)$, 
$\bm \Lambda_K = \operatorname{diag}(\lambda_1, \ldots, \lambda_K)$ 
and $\bm \Lambda_R = \operatorname{diag}(\lambda_{K+1}, \ldots, \lambda_p)$.
Analogously, we have that 
$\widehat{\bm \Sigma} = \widehat{\mathbf V}_K \widehat{\bm \Lambda}_K \widehat{\mathbf V}_K' 
+ \widehat{\mathbf V}_R \widehat{\bm \Lambda}_R \widehat{\mathbf V}_R'$
where $\widehat{\mathbf V}_K = (\hat{\bm v}_1, \ldots, \hat{\bm v}_K)$, 
$\widehat{\mathbf V}_R = (\hat{\bm v}_{K+1}, \ldots, \hat{\bm v}_p)$, 
$\widehat{\bm \Lambda}_K = \operatorname{diag}(\hat \lambda_1, \ldots, \hat \lambda_K)$ 
and $\widehat{\bm \Lambda}_R = \operatorname{diag}(\hat \lambda_{K+1}, \ldots, \hat \lambda_p)$.
For a random variable $X$, we introduce the shorthand 
$\| X \|_{L_2,\mathcal D_T} = \sqrt{ \mathbb E( X^2 | \mathcal D_T ) }$ for convenience.

\begin{proof}[Proof of Lemma \ref{lem:pcr}]
    $(i)$ Note that by definition
    \[
        \mathbf B \bm P_t + \bm u_t 
        = \mathbf V_K \mathbf V_K' \bm X_t + \mathbf V_R \mathbf V_R' \bm X_t 
        = \mathbf V_K \mathbf V_K' \bm X_t + (\mathbf I_p - \mathbf V_K \mathbf V_K') \bm X_t 
        = \bm X_t.
    \]
    We also note that
    $\mathbf B' \mathbf B  = \bm \Lambda_K^{1/2} \mathbf V_K'\mathbf V_K \bm \Lambda_K^{1/2} = \bm \Lambda_K$.
    Furthermore, as 
    $\bm \Sigma = \mathbf V_K \bm \Lambda_K \mathbf V_K' + \mathbf V_R \bm \Lambda_R \mathbf V_R'$,
    and $\mathbf V_K' \mathbf V_R = \mathbf 0_{K \times (p-K)}$, we immediately obtain
    \begin{align*}
	& \mathbb E( \bm P_t \bm P_t' ) 
        = \bm \Lambda_K^{-1/2} \mathbf V_K' \bm \Sigma \mathbf V_K \bm \Lambda_K^{-1/2} 
        = \bm \Lambda_K^{-1/2} \mathbf V_K' ( \mathbf V_K \bm \Lambda_K \mathbf V_K' 
        + \mathbf V_R \bm \Lambda_R \mathbf V_R') \mathbf V_K \bm \Lambda_K^{-1/2} 
        = \mathbf I_K \\
	& \mathbb E( \bm u_t \bm u_t' ) 
        = \mathbf V_R \mathbf V_R' (\mathbf V_K \bm \Lambda_K \mathbf V_K' 
        + \mathbf V_R \bm \Lambda_R \mathbf V_R') \mathbf V_R \mathbf V_R'
        = \mathbf V_R \bm \Lambda_R \mathbf V_R'\\
	& \mathbb E( \bm P_t \bm u_t' ) 
        = \bm \Lambda_K^{-1/2} \mathbf V_K' (\mathbf V_K \bm \Lambda_K \mathbf V_K' 
        + \mathbf V_R \bm \Lambda_R \mathbf V_R') \mathbf V_R \mathbf V_R' = \mathbf 0_{K \times p} ~.
    \end{align*}
    $(ii)$ It follows from part $(i)$ that
    \[
        \bm X_t '\bm \theta^* 
        = \bm P_t' \mathbf B' \bm \theta^* + \bm u_t'\bm \theta^*.
    \]
    The claim then follows from noting that
    $\bm P_t' \mathbf B' \bm \theta^* = \bm P_t'\bm \Lambda^{1/2}_K \mathbf V_K' \bm \theta^* 
    = \bm P_t'\bm \vartheta^*$ and 
    \[
        \bm u_t' \bm \theta^* 
        = \bm X_t' \mathbf V_R \mathbf V_R' \bm \theta^* 
        = \bm X_t' \mathbf V_R \mathbf V_R' \mathbf V_R \mathbf V_R' \bm \theta^* 
        = \bm u_t' \bm \gamma^* ~.
    \]
    $(iii)$ First note that if $\bm\theta_1^*$ and $\bm\theta_2^*$ are two distinct vectors of coefficients 
    associated with the best linear predictor, we have
    \[
        \mathbb E[ (\bm X_t'(\bm\theta_1^*-\bm\theta_2^*))^2]
        =(\bm\theta_1^* - \bm\theta_2^*)' \bm\Sigma (\bm\theta_1^*-\bm\theta_2^*) = 0 ~,
    \]
    so that $\bm\theta_1^*-\bm\theta_2^*\in \ker(\bm\Sigma)$.
    %As a consequence, while the best linear predictor is unique, $\bm\theta^*$ is unique only if 
    %$\ker(\bm\Sigma) = \lbrace \bm 0 \rbrace$, that is, if $\lambda_p > 0$.
    %
    Since we assume $\lambda_K>0$,
    we have that $\operatorname{span}(\mathbf V_K)$ is orthogonal to $\ker(\bm\Sigma)$. 
    Hence, for any two minimisers $\mathbf V_K'(\bm\theta_1^* - \bm\theta_2^*) = \bm 0$,
    and thus
    \[
        \bm\vartheta_1^*
        =\bm\Lambda_K^{1/2}\mathbf V_K'\bm\theta_1^*
        =\bm\Lambda_K^{1/2}\mathbf V_K'\bm\theta_2^*
        =\bm\vartheta_2^* ~.
    \]
    Moreover, $\mathbf V_R \mathbf V_R'$ is an orthogonal projector onto 
    the complement space of $\operatorname{span}(\mathbf V_K)$,
    and thus $\mathbf V_R \mathbf V_R'(\bm \theta_1^* - \bm \theta_2^*) = \bm \theta_1^* - \bm \theta_2^*$.
    As a consequence
    \[
        \mathbb E[ (\bm u_t'\bm\gamma_1^*- \bm u_t'\bm\gamma_2^*)^2 ]
        = \mathbb E[ (\bm X_t' \mathbf V_R \mathbf V_R' (\bm\theta_1^*-\bm\theta_2^*))^2]
        = \mathbb E[ (\bm X_t' (\bm\theta_1^*-\bm\theta_2^*))^2 ] 
        = 0 ~,
    \]
    so the random variable $\bm u_t'\bm\gamma^*$ is almost surely unique.\\
    $(iv)$ Note that the normal equations after pre-multiplying with $(\bm \theta^*)'$ yield
    $\mathbb E[ (\bm \theta^*)' \bm X_t \bm X_t' \bm\theta^* ] = \mathbb E[ (\bm \theta^*)' \bm X_tY_t ]$.
    Then by Cauchy-Schwarz
    \[
        \mathbb E[ (\bm X_t'\bm\theta^*)^2]
        = \mathbb E[(\bm X_t'\bm\theta^*)Y_t]
        \le \mathbb E[ (\bm X_t'\bm\theta^*)^2 ]^{1/2} \mathbb E [Y_t^2]^{1/2},
    \]
    which in turn implies 
    $\mathbb E[ (\bm X_t'\bm\theta^*)^2 ]^{1/2} \le \mathbb E[Y_t^2]^{1/2} \le C$.
    But we also have 
    \[
        \mathbb E[(\bm X_t'\bm\theta^*)^2]
        = (\bm\theta^*)' \bm\Sigma \bm\theta^*
        = \|\bm\Sigma^{1/2} \bm\theta^*\|_2^2,
    \]
    which then yields $\|\bm\Sigma^{1/2} \bm\theta^* \|_2 \leq C$.
    Notice next that
    \begin{align*}
        \|\bm\Sigma^{1/2}\bm\theta^*\|_2^2
        &= (\bm \theta^*)' (\mathbf V_K\bm\Lambda_K\mathbf V_K' 
        + \mathbf V_R\bm\Lambda_R\mathbf V_R') \bm\theta^* \\
        &= \|\bm\Lambda_K^{1/2} \mathbf V_K'\bm\theta^*\|_2^2
        + \|\bm\Lambda_R^{1/2} \mathbf V_R'\bm\theta^*\|_2^2.
    \end{align*}
    Using the definitions $\bm\vartheta^* =  \bm\Lambda_K^{1/2} \mathbf V_K' \bm\theta^*$
    and $\bm \gamma^* = \mathbf V_R \mathbf V_R' \bm \theta^*$, and the fact that
    \[
        \| \bm \Lambda_R^{1/2} \mathbf V_R' \bm\theta^* \|_2^2 
        = (\bm \theta^*)' \mathbf V_R \bm\Lambda_R \mathbf V_R' \bm \theta^*
        = (\bm \theta^*)' \mathbf V_R \mathbf V_R' \mathbf V_R 
        \bm\Lambda_R \mathbf V_R' \mathbf V_R\mathbf V_R' \bm \theta^*
        = (\bm\gamma^*)'\mathbf V_R \bm\Lambda_R \mathbf V_R'\bm \gamma^*,
    \]
    we obtain
    \[
        \|\bm\Sigma^{1/2}\bm\theta^*\|_2^2
        = \|\bm\vartheta^*\|_2^2 + (\bm\gamma^*)'\mathbf V_R\bm\Lambda_R\mathbf V_R'\bm\gamma^*
        \leq C^2,
    \]
    which proves the claim. 
\end{proof}

\begin{proof}[Proof of Lemma \ref{lem:decomp}]
    Consider the excess risk decomposition given by
    \begin{align}
	    R( \hat{\bm \vartheta} ) - R^* 
	    &= \| Y_{T+1} - \widehat{\bm P}'_{T+1} \hat{\bm \vartheta} \|^2_{L_2, \mathcal D_T} 
        - \mathbb E[ (Y_{T+1} - \bm X_{T+1}' \bm \theta^*) ^2 ] \notag \\
	    &= \| Y_{T+1} - \widehat{\bm P}'_{T+1} \hat{\bm \vartheta} \|^2_{L_2, \mathcal D_T} 
        - \| Y_{T+1} - \bm X_{T+1}' \bm \theta^* \|^2_{L_2, \mathcal D_T} \notag \\
	    &= \| Y_{T+1} - \widehat{\bm P}'_{T+1} \hat{\bm \vartheta} \|^2_{L_2, \mathcal D_T} 
        - \| Y_{T+1} - {\bm P}'_{T+1} \bm H ' \hat{\bm \vartheta} \|^2_{L_2, \mathcal D_T} \notag \\
	    &+ \| Y_{T+1} - {\bm P}'_{T+1} \bm H' \hat{\bm \vartheta} \|^2_{L_2, \mathcal D_T} 
        - \| Y_{T+1} - {\bm P}'_{T+1} \tilde{\bm \vartheta} \|^2_{L_2, \mathcal D_T} \notag \\
	    &+ \| Y_{T+1} - {\bm P}'_{T+1} \tilde{\bm \vartheta} \|^2_{L_2, \mathcal D_T} 
        - \| Y_{T+1} - {\bm P}'_{T+1} \bm \vartheta^* \|^2_{L_2, \mathcal D_T} \notag \\
	    &+ \| Y_{T+1} - {\bm P}'_{T+1} \bm \vartheta^* \|^2_{L_2, \mathcal D_T} 
        - \| Y_{T+1} - \bm X_{T+1}' \bm \theta^* \|^2_{L_2, \mathcal D_T} \notag \\
        &\equiv S_1 + S_2 + S_3 + S_4 ~, \label{eq:risk-decomp}
    \end{align}
    where the first equality follows as $\mathbb E[ (Y_{T+1} - \bm X_{T+1}' \bm \theta^*) ^2 \mid \mathcal D_T ]
    = \mathbb E[ (Y_{T+1} - \bm X_{T+1}' \bm \theta^*) ^2 ]$.
    We remark that all quantities in this expression exist with high probability for sufficiently large $T$.
    We begin by applying the projection theorem to the last two terms to obtain
    \begin{align}
	    S_3 &= \| \bm P_{T+1}' \tilde{\bm \vartheta} - \bm P_{T+1}' \bm \vartheta^* \|^2_{L_2, \mathcal D_T} 
	    = \| \tilde{\bm \vartheta} - \bm \vartheta^* \|^2_2, \label{eq:s3} \\
	    S_4 &= \| \bm u_{T+1}' \bm \gamma^* \|^2_{L_2, \mathcal D_T} \label{eq:s4} ~.
    \end{align}
    Next we note that for the first term we have that 
    \begin{align*}
        S_1 
        &= \| Y_{T+1} - \hat{\bm P}_{T+1}' \hat{\bm \vartheta} \|^2_{L_2, \mathcal D_T} 
        - \| Y_{T+1} - \bm P_{T+1}' \mathbf H' \hat{\bm \vartheta} \|^2_{L_2, \mathcal D_T} \\
        &= \| Y_{T+1} - \bm X_{T+1}' \bm \theta^* 
        + \bm X_{T+1}' \bm \theta^* - \hat{\bm P}_{T+1}' \hat{\bm \vartheta} \|^2_{L_2, \mathcal D_T} 
        - \| Y_{T+1} - \bm X_{T+1}' \bm \theta^* 
        + \bm X_{T+1}' \bm \theta^* - \bm P_{T+1}' \mathbf H' \hat{\bm \vartheta} \|^2_{L_2, \mathcal D_T} \\
        &= \| \bm X_{T+1}' \bm \theta^* - \hat{\bm P}_{T+1}' \hat{\bm \vartheta} \|^2_{L_2, \mathcal D_T} 
        - \| \bm X_{T+1}' \bm \theta^* - \bm P_{T+1}' \mathbf H' \hat{\bm \vartheta} \|^2_{L_2, \mathcal D_T} ~,
    \end{align*}
    by adding and subtracting, expanding the squares and applying the projection theorem.
    Adding and subtracting again we obtain
    \begin{align}
        S_1
        &= \| \bm X_{T+1}' \bm \theta^* - \bm P_{T+1}' \mathbf H' \hat{\bm \vartheta} 
        - (\hat{\bm P}_{T+1}' - \bm P_{T+1}' \mathbf H') \hat{\bm \vartheta} \|^2_{L_2, \mathcal D_T} 
        - \| \bm X_{T+1}' \bm \theta^* - \bm P_{T+1}' \mathbf H' \hat{\bm \vartheta} \|^2_{L_2, \mathcal D_T} 
        \notag \\
        &= \| (\hat{\bm P}_{T+1}' - \bm P_{T+1}' \mathbf H') \hat{\bm \vartheta} \|^2_{L_2, \mathcal D_T} 
        - 2 \mathbb E[ (\bm X_{T+1}' \bm \theta^* - \bm P_{T+1}' \mathbf H' \hat{\bm \vartheta})
        (\hat{\bm P}_{T+1}' - \bm P_{T+1}' \mathbf H') \hat{\bm \vartheta} \mid \mathcal D_T] \notag \\
        &= \| (\hat{\bm P}_{T+1}' - \bm P_{T+1}' \mathbf H') \hat{\bm \vartheta} \|^2_{L_2, \mathcal D_T} 
        - 2 \mathbb E[ \bm X_{T+1}' \bm \theta^* 
        (\hat{\bm P}_{T+1}' - \bm P_{T+1}' \mathbf H') \hat{\bm \vartheta} \mid \mathcal D_T] \notag \\
        &+ 2 \mathbb E[ \bm P_{T+1}' \mathbf H' \hat{\bm \vartheta}
        (\hat{\bm P}_{T+1}' - \bm P_{T+1}' \mathbf H') \hat{\bm \vartheta} \mid \mathcal D_T]. 
        \label{eq:s1-terms}
    \end{align}
    To deal with the cross terms, we will use the fact that
    \begin{equation} \label{eq:epp-h}
        \mathbb E [\widehat{\bm P}_{T+1} \bm P_{T+1}' \mid \mathcal D_T] 
        = \widehat{\bm \Lambda}_K^{-1/2} \widehat{\mathbf V}_K' \bm \Sigma \mathbf V_K \bm \Lambda_K^{-1/2} 
        = \widehat{\bm \Lambda}_K^{-1/2} \widehat{\mathbf V}_K' \mathbf V_K \bm \Lambda_K^{1/2}
        = \mathbf H,
    \end{equation}
    which follows by the definitions of $\bm P_{T+1}, \widehat{\bm P}_{T+1}$, 
    the structure of $\bm \Sigma = \mathbb E[ \bm X_{T+1} \bm X_{T+1}' \mid \mathcal D_T]$ 
    and the definition of $\mathbf H$.
    Now use this fact to see that the second cross term is zero
    \begin{align*}
        2 \mathbb E[ \bm P_{T+1}' \mathbf H' \hat{\bm \vartheta}
        (\hat{\bm P}_{T+1}' - \bm P_{T+1}' \mathbf H') \hat{\bm \vartheta} \mid \mathcal D_T]
        &= 2 \hat{\bm \vartheta}' \mathbf H \mathbb E[ \bm P_{T+1} \hat{\bm P}_{T+1}' \mid \mathcal D_T] 
        \hat{\bm \vartheta}
        -2 \hat{\bm \vartheta}' \mathbf H \mathbb E[ \bm P_{T+1} \bm P_{T+1}' \mid \mathcal D_T]
        \mathbf H' \hat{\bm \vartheta} \\
        &= 2 \hat{\bm \vartheta}' \mathbf H \mathbf H' \hat{\bm \vartheta}
        -2 \hat{\bm \vartheta}' \mathbf H \mathbf H' \hat{\bm \vartheta} = 0 ~.
    \end{align*}
    Going back to \eqref{eq:s1-terms} and using the fact that Lemma \ref{lem:pcr} implies 
    $\bm X_{T+1}' \bm \theta^* = \bm P_{T+1}' \bm \vartheta^* + \bm u_{T+1}' \bm \gamma^*$,
    we thus obtain
    \begin{align*}
        S_1
        &= \| (\hat{\bm P}_{T+1}' - \bm P_{T+1}' \mathbf H') \hat{\bm \vartheta} \|^2_{L_2, \mathcal D_T} 
        - 2 \mathbb E[ \bm X_{T+1}' \bm \theta^* 
        (\hat{\bm P}_{T+1}' - \bm P_{T+1}' \mathbf H') \hat{\bm \vartheta} \mid \mathcal D_T] \\
        &= \| (\hat{\bm P}_{T+1}' - \bm P_{T+1}' \mathbf H') \hat{\bm \vartheta} \|^2_{L_2, \mathcal D_T} 
        - 2 \mathbb E[ \bm P_{T+1}' \bm \vartheta^* 
        (\hat{\bm P}_{T+1}' - \bm P_{T+1}' \mathbf H') \hat{\bm \vartheta} \mid \mathcal D_T] \\
        &- 2 \mathbb E[ \bm u_{T+1}' \bm \gamma^* 
        (\hat{\bm P}_{T+1}' - \bm P_{T+1}' \mathbf H') \hat{\bm \vartheta} \mid \mathcal D_T].
    \end{align*}
    We use \eqref{eq:epp-h} again to show that the second cross term is zero
    \[
        - 2 \mathbb E[ \bm P_{T+1}' \bm \vartheta^* 
        (\hat{\bm P}_{T+1}' - \bm P_{T+1}' \mathbf H') \hat{\bm \vartheta} \mid \mathcal D_T]
        = -2 (\bm \vartheta^*)'  \mathbb E[ \bm P_{T+1} \hat{\bm P}_{T+1}' \mid \mathcal D_T] 
        \hat{\bm \vartheta}
        + 2 (\bm \vartheta^*)'  \mathbb E[ \bm P_{T+1} \bm P_{T+1}' \mid \mathcal D_T] 
        \mathbf H' \hat{\bm \vartheta}
        = 0,
    \]
    and therefore have obtained
    \[
        S_1
        = \| (\hat{\bm P}_{T+1}' - \bm P_{T+1}' \mathbf H') \hat{\bm \vartheta} \|^2_{L_2, \mathcal D_T} 
        - 2 \mathbb E[ \bm u_{T+1}' \bm \gamma^* 
        (\hat{\bm P}_{T+1}' - \bm P_{T+1}' \mathbf H') \hat{\bm \vartheta} \mid \mathcal D_T].
    \]
    Now use \eqref{eq:s4} and consider the sum of the first and fourth terms
    \begin{align*}
        S_1 + S_4
        &= \| (\hat{\bm P}_{T+1}' - \bm P_{T+1}' \mathbf H') \hat{\bm \vartheta} \|^2_{L_2, \mathcal D_T} 
        - 2 \mathbb E[ \bm u_{T+1}' \bm \gamma^* 
        (\hat{\bm P}_{T+1}' - \bm P_{T+1}' \mathbf H') \hat{\bm \vartheta} \mid \mathcal D_T]
        + \| \bm u_{T+1}' \bm \gamma^* \|^2_{L_2, \mathcal D_T} \\
        &= \| (\hat{\bm P}_{T+1}' - \bm P_{T+1}' \mathbf H') \hat{\bm \vartheta} 
        - \bm u_{T+1}' \bm \gamma^* \|^2_{L_2, \mathcal D_T} \\
        &\le 2 \| (\hat{\bm P}_{T+1}' - \bm P_{T+1}' \mathbf H') \hat{\bm \vartheta} \|^2_{L_2, \mathcal D_T}
        + 2 \| \bm u_{T+1}' \bm \gamma^* \|^2_{L_2, \mathcal D_T},
    \end{align*}
    where we used the fact that $(a-b)^2 \le 2a^2 + 2b^2$.
    Now note that $\| \bm u_{T+1}' \bm \gamma^* \|^2_{L_2, \mathcal D_T}
    = (\bm \gamma^*)' \bm V_R \bm \Lambda_R \bm V_R' \bm \gamma^*$.
    Moreover, by Cauchy-Schwarz
    \[
        \| (\hat{\bm P}_{T+1}' - \bm P_{T+1}' \mathbf H') \hat{\bm \vartheta} \|^2_{L_2, \mathcal D_T}
        = \mathbb E[ ((\hat{\bm P}_{T+1} - \mathbf H \bm P_{T+1})'\hat{\bm \vartheta})^2 \mid \mathcal D_T]
        \le \| \hat{\bm \vartheta} \|_2^2 
        \mathbb E[ \| \hat{\bm P}_{T+1} - \mathbf H \bm P_{T+1} \|_2^2 \mid \mathcal D_T].
    \]
    We therefore have
    \begin{equation}
        S_1 + S_4
        \le 2  \| \hat{\bm \vartheta} \|_2^2 
        \mathbb E[ \| \hat{\bm P}_{T+1} - \mathbf H \bm P_{T+1} \|_2^2 \mid \mathcal D_T]
        + 2 (\bm \gamma^*)' \bm V_R \bm \Lambda_R \bm V_R' \bm \gamma^*. \label{eq:s1s4}
    \end{equation}

    For the second term, the following identity will be useful.
    For any $\bm a \in \mathbb R^K$, we have 
    $Y_{T+1} - \bm P_{T+1}' \bm a = Y_{T+1} - \bm P_{T+1}' \bm \vartheta^* - \bm P_{T+1}' (\bm a - \bm \vartheta^*)$.
    By the projection theorem, the two right hand side terms are orthogonal so the Pythagorean identity gives
    \begin{equation} \label{eq:pyth}
        \| Y_{T+1} - \bm P_{T+1}' \bm a \|_{L_2, \mathcal D_T}^2 
        = \| Y_{T+1} - \bm P_{T+1}' \bm \vartheta^* \|_{L_2, \mathcal D_T}^2
        + \| \bm P_{T+1}' (\bm a - \bm \vartheta^*) \|_{L_2, \mathcal D_T}^2.
    \end{equation}
    Applying \eqref{eq:pyth} first with $\bm a = \mathbf H' \hat{\bm \vartheta}$
    and then $\bm a = \tilde{\bm \vartheta}$ gives
    \begin{align*}
	    S_2 
        &= \| Y_{T+1} - \bm P'_{T+1} \mathbf H ' \hat{\bm \vartheta} \|^2_{L_2, \mathcal D_T} 
        - \| Y_{T+1} - \bm P'_{T+1} \tilde{\bm \vartheta} \|^2_{L_2, \mathcal D_T} \\
        &= \| \bm P_{T+1}' (\mathbf H' \hat{\bm \vartheta} - \bm \vartheta^*) \|_{L_2, \mathcal D_T}^2
        -\| \bm P_{T+1}' (\tilde{\bm \vartheta} - \bm \vartheta^*) \|_{L_2, \mathcal D_T}^2 \\
        &= \| \mathbf H' \hat{\bm \vartheta} - \bm \vartheta^* \|_2^2
        -\| \tilde{\bm \vartheta} - \bm \vartheta^* \|_2^2,
    \end{align*}
    where the final step follows as $\mathbb E[\bm P_{T+1} \bm P_{T+1}' \mid \mathcal D_T] = \mathbf I_K$.
    Adding and subtracting yields
    \begin{align}
        S_2 &= \| \mathbf H' \hat{\bm \vartheta} - \tilde{\bm \vartheta} 
        + \tilde{\bm \vartheta} - \bm \vartheta^* \|_2^2
        - \| \tilde{\bm \vartheta} - \bm \vartheta^* \|_2^2 \notag \\
        &= \| \mathbf H' \hat{\bm \vartheta} - \tilde{\bm \vartheta} \|_2^2
        + 2(\tilde{\bm \vartheta} - \bm \vartheta^*)'
        (\mathbf H' \hat{\bm \vartheta} - \tilde{\bm \vartheta}) \notag \\
	    &\le \| \tilde{\bm \vartheta} - \mathbf H ' \hat{\bm \vartheta} \|^2_2
	    + 2 \| \bm \vartheta^* - \tilde{\bm \vartheta} \|_2 
        \| \tilde{\bm \vartheta} - \mathbf H' \hat{\bm \vartheta} \|_2 \notag \\
        &\le 2 \| \tilde{\bm \vartheta} - \mathbf H' \hat{\bm \vartheta} \|^2_2 
        +  \| \bm \vartheta^* - \tilde{\bm \vartheta} \|^2_2,  \label{eq:s2}
    \end{align}
    where the inequalities are Cauchy-Schwarz followed by $2 a b \leq a^2 + b^2$.

    The claim then follows by inserting \eqref{eq:s3}, \eqref{eq:s1s4} and \eqref{eq:s2} 
    into \eqref{eq:risk-decomp}.
\end{proof}

\subsection{Proof of Theorem \ref{thm:erm-pcr}}

\subsubsection{Proof of Proposition \ref{prop:p-est-oos}}

\begin{proof}[Proof of Proposition \ref{prop:p-est-oos}]
    Begin by noting that as
    \[
        \mathbf H \bm P_{T+1}
        = \widehat{\bm \Lambda}_K^{-1/2} \widehat{\mathbf V}_K' \mathbf V_K 
        \bm \Lambda_K^{1/2} \bm \Lambda_K^{-1/2} \mathbf V_K' \bm X_{T+1}
        = \widehat{\bm \Lambda}_K^{-1/2} \widehat{\mathbf V}_K' \mathbf V_K \mathbf V_K' \bm X_{T+1},
    \]
    we have
    \[
        \hat{\bm P}_{T+1} - \mathbf H \bm P_{T+1}
        = \widehat{\bm \Lambda}_K^{-1/2} \widehat{\mathbf V}_K' (\mathbf I_p - \mathbf V_K\mathbf V_K') \bm X_{T+1}
        = \widehat{\bm \Lambda}_K^{-1/2} \widehat{\mathbf V}_K' \mathbf V_R \mathbf V_R' \bm X_{T+1}
        = \widehat{\bm \Lambda}_K^{-1/2} \widehat{\mathbf V}_K' \bm u_{T+1},
    \]
    where the last step follows as $\bm u_{T+1} = \mathbf V_R \mathbf V_R' \bm X_{T+1}$.
    Thus
    \[
        \| \widehat{\bm P}_{T+1} - \mathbf H \bm P_{T+1}\|_2^2
        = \bm u_{T+1}' \widehat{\mathbf V}_K \widehat{\bm \Lambda}_K^{-1} \widehat{\mathbf V}_K' \bm u_{T+1}.
    \]
    We may rewrite this as 
    \[
        \mathbb E( \| \hat{\bm P}_{T+1} - \mathbf H \bm P_{T+1}\|_2^2 \mid \mathcal D_T)
        = \operatorname{tr} ( \widehat{\bm \Lambda}_K^{-1} \widehat{\mathbf V}_K' 
        \mathbb E(\bm u_{T+1} \bm u_{T+1}') \widehat{\mathbf V}_K )
        = \operatorname{tr}( \widehat{\bm \Lambda}_K^{-1} \widehat{\mathbf V}_K' \mathbf V_R
        \bm \Lambda_R \mathbf V_R' \widehat{\mathbf V}_K),
    \]
    as $\mathbb E(\bm u_{T+1} \bm u_{T+1}') = \mathbf V_R \bm{\Lambda}_R \mathbf V_R'$. 
    Using the definition $\mathbf S = \mathbf V_R' \widehat{\mathbf V}_K$, we may upper bound with
    \[
        \operatorname{tr}( \widehat{\bm \Lambda}_K^{-1} \mathbf S' \bm \Lambda_R \mathbf S)
        \le \frac{1}{\hat\lambda_K} \operatorname{tr}(\bm \Lambda_R \mathbf S \mathbf S')
        \le \frac{\lambda_{K+1}}{\hat\lambda_K} \operatorname{tr}(\mathbf S' \mathbf S),
    \]
    where the first inequality follows as
    $\operatorname{tr}(\mathbf{AB}) \le \| \mathbf A \|_2 \operatorname{tr}(\mathbf B)$
    for two symmetric non-negative definite conformable matrices $\mathbf A, \mathbf B$,
    and the second because the diagonal of $\mathbf S \mathbf S'$ is a sum of squares and thus non-negative.
    As a consequence, we obtain
    \[
        \mathbb E( \| \hat{\bm P}_{T+1} - \mathbf H \bm P_{T+1}\|_2^2 \mid \mathcal D_T)
        \le \frac{\lambda_{K+1}}{\hat\lambda_K} \operatorname{tr}(\mathbf S' \mathbf S)
        = \frac{\lambda_{K+1}}{\hat\lambda_K} \| \mathbf S \|_F^2
        = \frac{\lambda_{K+1}}{\hat\lambda_K} \| \mathbf V_R' \widehat{\mathbf V}_K \|_F^2.
    \]
    Let $\eta' = \eta + \delta$ with $\delta > 0$.
    Proposition \ref{prop:cov-est} implies that for all sufficiently large $T$, 
    we have $\hat \lambda_K \ge c_K p^\alpha / 2$ with probability at least $1 - T^{-\eta'}$.
    This together with Assumption \ref{asm:eigen} implies that there exists a constant $C_1 > 0$ such that
    \[
        \mathbb E( \| \hat{\bm P}_{T+1} - \mathbf H \bm P_{T+1}\|_2^2 \mid \mathcal D_T)
        \le C_1 \frac{1}{p^\alpha} \| \mathbf V_R' \widehat{\mathbf V}_K \|_F^2,
    \]
    with probability at least $1 - T^{-\eta'}$.
    Proposition \ref{prop:leak} then implies that there exists a constant $C > 0$ such that, 
    for sufficiently large $T$
    \[
        \mathbb E( \| \hat{\bm P}_{T+1} - \mathbf H \bm P_{T+1}\|_2^2 \mid \mathcal D_T)
        \le C \frac{p^{1 - \alpha}}{p^{\alpha}} \frac{K \log(T)}{T}
        = C \frac{1}{p^{2 \alpha - 1}} \frac{K \log(T)}{T},
    \]
    holds with probability at least $1 - 2 T^{-\eta'} \ge 1 - T^{-\eta}$, which proves the claim.
\end{proof}

\begin{prop} \label{prop:cov-est} 
    Suppose Assumptions \ref{asm:dist}--\ref{asm:npk} are satisfied. 
    Then for every $\eta>0$, for all $T$ sufficiently large
    \begin{align*} 
        &(i) \quad \hat \lambda_K \ge \frac{c_K}{2}  p^\alpha, \\
        &(ii) \quad \| \mathbf V_R' (\widehat{\bm \Sigma} - \bm \Sigma) \mathbf V_R \|_2
        \le \frac{\hat \lambda_K - \lambda_{K+1}}{2}, \\
        &(iii) \quad \| \widehat{\mathbf V}_K \widehat{\mathbf V}_K' - \mathbf V_K \mathbf V_K' \|_2
        \le \sqrt{5/6}, \\
        &(iv) \quad \lambda_{\min}( \widehat{\bm \Lambda}_K^{-1/2} \widehat{\mathbf V}_K' \bm\Sigma
        \widehat{\mathbf V}_K \widehat{\bm\Lambda}_K^{-1/2} ) \ge 1/2,
    \end{align*}
    hold with probability at least $1 - T^{-\eta}$.
\end{prop}
\begin{proof}
    $(i)$ Using the representation of Lemma \ref{lem:pcr}  we get
    \begin{align}
    	\widehat{\bm \Sigma}  - \bm \Sigma 
        &= \frac1T \sum_{t=1}^T (\mathbf B \bm P_t + \bm u_t )(\mathbf B \bm P_t+\bm u_t )'
        - \mathbb E(\bm X_t \bm X_t') \notag \\
    	&= {1 \over T } \mathbf B  \sum_{t=1}^T (\bm P_t \bm P_t' - \bm I_K) \mathbf B'
    	+ {1 \over T} \sum_{t=1}^T(\bm u_t \bm u_t' - \mathbb E(\bm u_t \bm u_t'))
    	+ {1 \over T } \sum_{t=1}^T (\mathbf B \bm P_t \bm u_t')
    	+ {1 \over T }  \sum_{t=1}^T ( \bm u_t \bm P_t' \mathbf B') \notag \\ 
    	&= \mathbf D_1 + \mathbf D_2 + \mathbf D_3 + \mathbf D_4 \label{eq:sigma-decomp}.
    \end{align}
    This, in turn, implies that $\mathbb P(\| \widehat{\bm \Sigma} - \bm \Sigma \|_2 > \varepsilon)
    \leq \sum_{i = 1}^4 \mathbb P( \| \mathbf D_i \|_2 > \varepsilon_i)$, or  
    \begin{equation} \label{eq:cov-est-terms}
    	\mathbb P\left( \| \widehat{\bm \Sigma} - \bm \Sigma \|_2 \le \varepsilon \right) 
    	\ge 1 - \sum_{i = 1}^4 \mathbb P\left( \| \mathbf D_i \|_2 > \varepsilon_i \right), 
    \end{equation}
    for some $\varepsilon_1, \varepsilon_2, \varepsilon_3, \varepsilon_4 > 0$ 
    such that $\varepsilon = \varepsilon_1 + \varepsilon_2 + \varepsilon_3 + \varepsilon_4$.
    Let $\eta' = \eta + \delta > 0$ with $\delta > 0$.
    Proposition \ref{prop:cov-est-p} and Assumption \ref{asm:eigen} imply that there is a constant $C_1 > 0$ 
    such that 
    \begin{equation}\label{eq:d1}
    	\| \mathbf D_1 \|_2 
    	\le \left\| \frac1T \sum_{t = 1}^T (\bm P_t \bm P_t' - \mathbf I_K) \right\|_2 \| \mathbf{BB}' \|_2   
    	\le p^{\alpha} C_1 \sqrt{ { (K + \log(T) )^{ {r_\beta +1 \over \, r_\beta}  } \over T } }
        = \varepsilon_1,
    \end{equation} 
    holds with probability at least $1 - T^{-\eta'}$.
    By Proposition \ref{prop:cov-est-u}, there exists a constant $C_2 > 0$ such that 
    \begin{equation} \label{eq:d2}
    	\| \mathbf D_2 \|_2 
        \le C_2 \left({ ( p + \log(T) )^{ {r_\beta +1 \over \, r_\beta}  } \over T }
        + \sqrt{ { ( p + \log(T) )^{ {r_\beta +1 \over \, r_\beta}  } \over T }} \right)
        = \varepsilon_2,
     \end{equation}
    holds with probability at least $1 - T^{-\eta'}$.
    Proposition \ref{prop:cov-est-pu}, Assumption \ref{asm:eigen} then imply that there is a constant 
    $C_3 > 0$ such that
    \begin{equation}\label{eq:d3}
    	\| \mathbf D_3 \|_2 
    	\le \left\| \frac1T \sum_{t = 1}^T \bm P_t \bm u_t' \right\|_2 \| \mathbf B \|_2 
    	\le p^{\alpha/2} C_3 \sqrt{ \frac{p K \log T}{T}}  = \varepsilon_3,
    \end{equation} 
    holds with probability at least $1 - T^{-\eta'}$.
    Last, note that $\| \mathbf D_3 \|_2 = \| \mathbf D_4 \|_2$. 
    Combining the inequalities in \eqref{eq:cov-est-terms}, \eqref{eq:d1}, \eqref{eq:d2}, \eqref{eq:d3} 
    we get 
    \begin{equation*}
    	\mathbb P\left( \| \widehat{\bm \Sigma} - \bm \Sigma \|_2 \le \varepsilon \right) 
    	\ge 1 - 4 T^{-\eta'} \ge 1 - T^{-\eta},
    \end{equation*}
    where $\varepsilon$ is defined as 
    \begin{align*}
    	\varepsilon &= p^{\alpha} C_1 \sqrt{ { (K + \log(T) )^{ {r_\beta +1 \over \, r_\beta}  } \over T } }
        + C_2 \left( \frac{ (p + \log(T))^{ \frac{r_\beta + 1}{r_\beta} }}{T}
        + \sqrt{ \frac{ (p + \log(T))^{ \frac{r_\beta + 1}{r_\beta} }}{T} } \right) \\
    	&+ 2 p^{\alpha/2} C_3 \sqrt{ \frac{p K \log T}{T}}.
    \end{align*}
    Claim $(i)$ follows from noting that
    \begin{align*}
    	\varepsilon &= p^{\alpha} C_1 \sqrt{ { (K + \log(T) )^{ {r_\beta +1 \over \, r_\beta}  } \over T } }
        + p^\alpha C_2 \left(\frac{1}{p^\alpha} \frac{ (p + \log(T))^{ \frac{r_\beta + 1}{r_\beta} }}{T}
        +  \frac{1}{p^\alpha} \sqrt{ \frac{ (p + \log(T))^{ \frac{r_\beta + 1}{r_\beta} }}{T} } \right) \\
    	&+ 2 p^\alpha C_3 \sqrt{ \frac{p^{1 - \alpha} K \log T}{T}},
    \end{align*}
    so that by Assumption \ref{asm:npk}, for all $T$ sufficiently large, 
    $\varepsilon \leq c_K p^\alpha / 2$.
    Then Assumption \ref{asm:eigen} implies 
    \[ 
        \hat{\lambda}_K 
        \ge \lambda_K - | \lambda_K - \hat{\lambda}_K| 
        \ge c_K p^\alpha - \| \widehat{\bm \Sigma}  - \bm \Sigma \|_2
        \ge c_K p^\alpha - \varepsilon
        = \frac{c_K}{2} p^\alpha,
    \]
    where the second inequality follows from Weyl's inequality. \\
    $(ii)$ Note that $\hat \lambda_K - \lambda_{K+1} \ge C' p^\alpha$ for some constant $C' > 0$ 
    by Assumption \ref{asm:eigen}.
    The claim then follows from Assumption \ref{asm:npk} as 
    $\|\mathbf V_R'(\widehat{\bm \Sigma} - \bm \Sigma) \mathbf V_R\|_2
    \le \|\widehat{\bm \Sigma} - \bm \Sigma\|_2 \le \varepsilon$.\\
    $(iii)$ follows by the Davis-Kahan theorem of \citet[Theorem 2]{yu-wang-samworth-2015},
    which implies
    \[
        \| \widehat{\mathbf V}_K \widehat{\mathbf V}_K' - \mathbf V_K \mathbf V_K' \|_2
        \le \frac{2\|\hat{\bm\Sigma}-\bm\Sigma\|_2}{c_K p^\alpha - c_{K+1}}
        \le \frac{2\varepsilon}{c_K p^\alpha - c_{K+1}}
        \le \sqrt{5/6},
    \]
    where the final step follows for large enough $T$. \\
    $(iv)$ Consider the matrix
    \[
        \widehat{\bm \Lambda}_K^{-1/2} \widehat{\mathbf V}_K' \bm\Sigma
        \widehat{\mathbf V}_K \widehat{\bm\Lambda}_K^{-1/2} - \mathbf I_K
        = \widehat{\bm \Lambda}_K^{-1/2} \widehat{\mathbf V}_K'(\bm\Sigma - \widehat{\bm\Sigma})
        \widehat{\mathbf V}_K \widehat{\bm\Lambda}_K^{-1/2},
    \]
    which follows as $\widehat{\mathbf V}_K'\widehat{\bm\Sigma} \widehat{\mathbf V}_K = \widehat{\bm\Lambda}_K$.
    Then
    \[
        \| \widehat{\bm \Lambda}_K^{-1/2} \widehat{\mathbf V}_K' \bm\Sigma
        \widehat{\mathbf V}_K \widehat{\bm\Lambda}_K^{-1/2} - \mathbf I_K \|_2
        \le \| \widehat{\bm\Lambda}_K^{-1/2} \|_2^2 \|\widehat{\bm\Sigma} - \bm\Sigma\|_2
        = \frac{ \|\widehat{\bm\Sigma} - \bm\Sigma\|_2}{\hat{\lambda}_K} 
        \le \frac{2 \|\widehat{\bm\Sigma} - \bm\Sigma\|_2}{c_K p^\alpha}.
    \]
    It then follows that
    \[
        \lambda_{\min}(\widehat{\bm \Lambda}_K^{-1/2} \widehat{\mathbf V}_K' \bm\Sigma
        \widehat{\mathbf V}_K \widehat{\bm\Lambda}_K^{-1/2})
        \ge 1 - \| \widehat{\bm \Lambda}_K^{-1/2} \widehat{\mathbf V}_K' \bm\Sigma
        \widehat{\mathbf V}_K \widehat{\bm\Lambda}_K^{-1/2} - \mathbf I_K \|_2
        \ge 1 - \frac{2 \|\widehat{\bm\Sigma} - \bm\Sigma\|_2}{c_K p^\alpha}
        \ge 1/2,
    \]
    by Weyl's inequality, for large enough $T$.
\end{proof}

\begin{prop} \label{prop:cov-est-u} 
    Suppose Assumptions \ref{asm:dist}--\ref{asm:npk} are satisfied. 
    Then for every $\eta>0$ there exists a constant $C > 0$ such that, for all $T$ sufficiently large
    \[
        \left \| {1 \over T } \sum_{t = 1}^T \bm u_t \bm u_t' - \mathbb E( \bm u_t \bm u_t' ) \right \|_2
        \le C \left({ ( p + \log(T) )^{ {r_\beta +1 \over \, r_\beta}  } \over T }
        + \sqrt{ { ( p + \log(T) )^{ {r_\beta +1 \over \, r_\beta}  } \over T }} \right),
    \]
    holds with probability at least $1 - T^{-\eta}$.
\end{prop}
\begin{proof}
    When $c_{K+1} = 0$ the claim is trivial as $\bm u_t = \bm 0$ almost surely, so we may assume otherwise.
    We check the conditions of Lemma \ref{lem:conc-bosq} for $\bm u_t$. 
    Let $\bm \Sigma_u = \mathbb E( \bm u_t \bm u_t' ) = \bm V_R \bm V_R' \bm \Sigma \bm V_R  \bm V_R'$.
    Lemma \ref{lem:sub-g} establishes that $\bm \Sigma_u^{-1/2} \bm u_t$ is sub-Gaussian.
    Additionally, standard properties of strong mixing processes imply that $\bm u_t$ inherits 
    the mixing properties of the sequence $\{(Y_t,\bm X_t)'\}_{t=1}^T$ from Assumption \ref{asm:mix}.
    Finally, Assumption \ref{asm:npk} implies that $p = \lfloor C_p T^{r_p}\rfloor$ where 
    $r_p < r_\beta$.
    We may then apply Lemma \ref{lem:conc-bosq} to obtain, for all $T$ sufficiently large
    \[
        \mathbb P \left( \left \| {1 \over T } \sum_{t=1}^T \bm u_t \bm u_t' - \bm \Sigma_u \right \|_2  
        > \| \bm \Sigma_u \|_2 C_1 \left( \frac{ ( p + \log(T) )^{ {r_\beta +1 \over r_\beta}  } }{T} 
        + \sqrt{ { ( p + \log(T) )^{ {r_\beta +1 \over \, r_\beta}  } \over T } } \right) \right) 
        \le T^{-\eta},
    \]
    for some constant $C_1 > 0$.
    The claim then follows by noting that Assumption \ref{asm:eigen} implies that 
    $\| \bm \Sigma_u \|_2 = c_{K+1}$ and setting $C = C_1 c_{K+1}$. 
\end{proof}

\begin{prop} \label{prop:cov-est-p}
    Suppose Assumptions \ref{asm:dist}, \ref{asm:mix} and \ref{asm:npk} are satisfied. 
    Then for every $\eta > 0$ there exists a constant $C > 0$ such that, for all $T$ sufficiently large
    \begin{align*}
        &(i) \quad \left\| \frac1T \sum_{t=1}^T \bm P_t \bm P_t' - \mathbf I_K \right \|_2
        \le C_2 \sqrt{ { (K + \log(T) )^{ {r_\beta +1 \over \, r_\beta}  } \over T } }, \\
	    &(ii) \quad \lambda_{\min}\left( \frac1T \sum_{t=1}^T \bm P_t \bm P_t' \right) \ge \frac12, 
    \end{align*}
    hold with probability at least $1 - T^{-\eta}$.
\end{prop}
\begin{proof}
    $(i)$ Following similar arguments as in the proof of Proposition \ref{prop:cov-est-u},
    Lemma \ref{lem:sub-g} establishes that $\bm P_t$ is sub-Gaussian.
    Standard properties of strong mixing processes imply that $\bm P_t$ 
    inherits the mixing properties from Assumption \ref{asm:mix}.
    Finally, Assumption \ref{asm:npk} implies that $K = \lfloor C_K T^{r_K}\rfloor$ where 
    $r_K < r_\beta$.
    We may then apply Lemma \ref{lem:conc-bosq} and Assumption \ref{asm:npk} 
    to obtain for some constant $C_1 > 0$, for all $T$ sufficiently large
    \begin{align}
        \left \| {1 \over T } \sum_{t=1}^T \bm P_t \bm P_t' - \mathbf I_K \right \|_2  
        &\le C_1 \left( \frac{ ( K + \log(T) )^{ {r_\beta +1 \over r_\beta}  } }{T} 
        + \sqrt{ { ( K + \log(T) )^{ {r_\beta +1 \over \, r_\beta}  } \over T } } \right) \notag \\
        &\le C_2 \sqrt{ { (K + \log(T) )^{ {r_\beta +1 \over \, r_\beta}  } \over T } }, \label{eq:p-cov}
    \end{align}
    with probability at least $1 - T^{-\eta}$.\\
    $(ii)$ By Weyl's inequality
    \[
        \lambda_{\min}\left( \frac1T \sum_{t=1}^T \bm P_t \bm P_t'\right) 
        \geq 1 - \left \| {1 \over T } \sum_{t=1}^T \bm P_t \bm P_t' - \bm I_K \right \|_2. 
    \]
    The claim then follows from \eqref{eq:p-cov} and Assumption \ref{asm:npk}, since for every $\eta>0$, 
    for all $T$ sufficiently large, $\| (1/T) \sum_{t=1}^T \bm P_t \bm P_t' - \bm I_K \|_2 \le 1/2$
    with probability at least $1-T^{-\eta}$. 
\end{proof}

\begin{prop} \label{prop:cov-est-pu}
    Suppose Assumptions \ref{asm:dist}--\ref{asm:npk} are satisfied. 
    Then for every $\eta>0$ there exists a constant $C > 0$ such that, for all $T$ sufficiently large 
    \begin{equation*}
        \left\| {1 \over T} \sum_{t = 1}^T \bm P_t \bm u_t' \right\|_F \le C \sqrt{p K \log(T) \over T},
    \end{equation*}
    holds with probability at least $1 - T^{-\eta}$.
\end{prop}
\begin{proof}
    We assume that $c_{K+1}$ is positive (when $c_{K+1}$ is zero the claim is trivial).
    Let $V_{ij,t}$ denote $P_{it} u_{jt}$ and $\bm V_{j,t}$ denote the $j$-th column of 
    $\bm P_t \bm u_t'$ where $1 \leq  i \leq K$ and $1 \leq j \leq p$.  
    We begin by showing that, for each $j$, the sequence of $K$-dimensional random vectors 
    $\{\bm V_{j,t}\}_{t=1}^T$ satisfies the conditions required by Lemma \ref{lem:conc-lieb}.
    Lemma \ref{lem:sub-g} establishes that $\bm P_t$ and $\bm u_t$ are sub-Gaussian random vectors.
    Standard results on sub-Gaussian random variables then imply that for some $C_m' > 0$
    \[
        \sup_{\bm v : \| \bm v \|_2 = 1 } \mathbb P( | \bm V_{j\,t}' \bm v | > \varepsilon) 
        \leq \sup_{\bm v : \| \bm v \|_2 = 1 } \mathbb P( |\bm P_t' \bm v | | u_{jt} | > \varepsilon) 
        \leq \exp(-C_m' \varepsilon) ~,
    \]
    and Lemma \ref{lem:pcr} implies that $\mathbb E( \bm V_{j,t}) =  \mathbb E(\bm P_{t}u_{jt}) = \bm 0$, 
    so $\bm V_{j,t}$ is a zero-mean sub-exponential vector.
    Moreover, standard properties of strong mixing processes imply that $\{ \bm V_{j,t}\}_{t=1}^T$ inherits 
    the mixing properties of the sequence $\{(Y_t,\bm X_t)'\}_{t=1}^T$ from Assumption \ref{asm:mix}.
    Lastly, Assumption \ref{asm:npk} implies that $K = \lfloor C_K T^{r_K}\rfloor$ where $r_K < 1$.
    We may then apply Lemma \ref{lem:conc-lieb} to the sequence $\{\bm V_{j,t}\}_{t=1}^T$.
    Choosing $\eta' = \eta + r_p + \delta$ with $\delta > 0$, 
    there exists a $C_{\eta'} > 0$ such that, for all $T$ sufficiently large
    \[
	    \mathbb P\left( \left \| {1 \over T} \sum_{t=1}^T \bm V_{j,t} \right \|_2 
        > C_{\eta'} \sqrt{ K \log(T) \over T} \right) \leq T^{-\eta'} ~.
    \]
    The claim then follows as 
    \[
        \left\| {1 \over T} \sum_{t = 1}^T \bm P_t \bm u_t' \right\|_F^2
	    = \sum_{j = 1}^p \left \| {1 \over T} \sum_{t=1}^T \bm V_{j,t} \right \|_2^2 
	    \le p \max_{1 \leq j \leq p} \left \| {1 \over T} \sum_{t=1}^T \bm V_{j,t} \right \|_2^2 ~,
    \]
    which implies by Assumption \ref{asm:npk}
    \begin{align*}
        \mathbb P\left( \left\| {1 \over T} \sum_{t = 1}^T \bm P_t \bm u_t' \right\|_F^2 
        > C_{\eta'}^2 {p K \log(T) \over T} \right)
        &\le \mathbb P \left(p \max_{1 \le j \le p} \left\| \frac1T \sum_{t=1}^T \bm V_{j,t} \right\|_2^2
        > C_{\eta'}^2 {p K \log(T) \over T} \right) \\
    	&\le p \max_{1 \leq j \leq p} \mathbb P \left( \left\| {1 \over T} \sum_{t=1}^T \bm V_{j,t} \right\|_2 
        > C_{\eta'} \sqrt{ K \log(T) \over T}  \right) \\
    	&\le {p \over T^{\eta'} } = C_p { T^{r_p} \over T^{\eta + r_p + \delta} } 
        = C_p T^{-\eta - \delta} \leq T^{-\eta} ~.
    \end{align*}
\end{proof}

\begin{prop} \label{prop:cov-est-leak}
    Suppose Assumptions \ref{asm:dist}--\ref{asm:npk} are satisfied. 
    Then for every $\eta>0$ there exists a constant $C > 0$ such that, for all $T$ sufficiently large
    \[
        \|\mathbf V_R' (\widehat{\bm \Sigma} - \bm \Sigma) \mathbf V_K \|_F^2
        \le C p^{1 + \alpha} \frac{K \log(T)}{T},
    \]
    holds with probability at least $1 - T^{-\eta}$.
\end{prop}
\begin{proof}
    Consider the decomposition 
    $\widehat{\bm \Sigma} - \bm \Sigma = \mathbf D_1 + \mathbf D_2 + \mathbf D_3 + \mathbf D_4$ 
    from \eqref{eq:sigma-decomp}.
    Projecting it into the relevant block, we obtain
    \[
        \mathbf V_R' (\widehat{\bm \Sigma} - \bm \Sigma) \mathbf V_K 
        = \sum_{j=1}^4 \mathbf V_R' \mathbf D_j \mathbf V_K
        = \mathbf V_R' \mathbf D_4 \mathbf V_K,
    \]
    because $\mathbf V_R' \mathbf D_1 = \mathbf 0$ and $\mathbf V_R' \mathbf D_3 = \mathbf 0$
    as $\mathbf B = \mathbf V_K \bm \Lambda_K^{1/2}$,
    and $\mathbf D_2 \mathbf V_K = \mathbf 0$ as $\bm u_t = \mathbf V_R \mathbf V_R' \bm X_t$
    and $\mathbb E[\bm u_t \bm u_t'] = \mathbf V_R \bm \Lambda_R \mathbf V_R'$,    
    all of which follows from Lemma \ref{lem:pcr}.
    Using $\mathbf B'\mathbf V_K = \bm{\Lambda}_K^{1/2}$ again yields
    \[
        \mathbf V_R' (\widehat{\bm \Sigma} - \bm \Sigma) \mathbf V_K 
        = \mathbf V_R' \left( \frac1T \sum_{t=1}^T \bm u_t \bm P_t' \mathbf B' \right) \mathbf V_K
        = \mathbf V_R' \left( \frac1T \sum_{t=1}^T \bm u_t \bm P_t' \right) \bm{\Lambda}_K^{1/2}.
    \]
    Taking squared Frobenius norms and upper bounding gives
    \[
        \| \mathbf V_R' (\widehat{\bm \Sigma} - \bm \Sigma) \mathbf V_K \|_F^2
        = \left\| \mathbf V_R' \left( \frac1T \sum_{t=1}^T \bm P_t \bm u_t' \right) 
        \bm{\Lambda}_K^{1/2} \right\|_F^2 
        \leq c_1 p^\alpha \left\| \frac1T \sum_{t=1}^T \bm P_t \bm u_t' \right\|_F^2,
    \]
    as $\lambda_1 = c_1 p^\alpha$ by Assumption \ref{asm:eigen}.
    Then by Proposition \ref{prop:cov-est-pu}, for every $\eta>0$ there exists a constant $C > 0$ such that, 
    for all $T$ sufficiently large 
    \begin{equation*}
        \| \mathbf V_R' (\widehat{\bm \Sigma} - \bm \Sigma) \mathbf V_K \|_F^2
        \le C p^{1 + \alpha} { K \log(T) \over T},
    \end{equation*}
    holds with probability at least $1 - T^{-\eta}$ which proves the claim.
\end{proof}

\begin{prop} \label{prop:leak}
    Suppose Assumptions \ref{asm:dist}--\ref{asm:npk} are satisfied. 
    Then for every $\eta>0$ there exists a constant $C > 0$ such that, for all $T$ sufficiently large
    \[
        \| \mathbf V_R' \widehat{\mathbf V}_K \|_F^2
        \le C p^{1 - \alpha} \frac{K \log(T)}{T},
    \]
    holds with probability at least $1 - T^{-\eta}$.
\end{prop}
\begin{proof}
    Let $\mathbf S \equiv \mathbf V_R' \widehat{\mathbf V}_K$ denote the matrix of interest.
    Our argument is inspired by the beginning of the proof of Theorem 2 in \citet{yu-wang-samworth-2015},
    although we apply it to different matrices.
    We also note that our object of interest, $\mathbf S$, appears in their equation A4.
    Start from
    \[
        \widehat{\mathbf V}_K \widehat{\bm \Lambda}_K
        = \widehat{\bm \Sigma} \widehat{\mathbf V}_K 
        = (\bm \Sigma + (\widehat{\bm \Sigma} - \bm \Sigma)) \widehat{\mathbf V}_K.
    \]
    Multiplying by $\mathbf V_R'$, we obtain
    \[
        \mathbf V_R' \widehat{\mathbf V}_K \widehat{\bm \Lambda}_K
        = \mathbf V_R' \bm \Sigma \widehat{\mathbf V}_K
        + \mathbf V_R' (\widehat{\bm \Sigma} - \bm \Sigma) \widehat{\mathbf V}_K
        = \bm \Lambda_R \mathbf V_R' \widehat{\mathbf V}_K
        + \mathbf V_R' (\widehat{\bm \Sigma} - \bm \Sigma) \widehat{\mathbf V}_K,
    \]
    or
    \[
        \mathbf S \widehat{\bm \Lambda}_K - \bm \Lambda_R \mathbf S
        = \mathbf V_R' (\widehat{\bm \Sigma} - \bm \Sigma) \widehat{\mathbf V}_K.
    \]
    As $\mathbf V_K \mathbf V_K' + \mathbf V_R \mathbf V_R' = \mathbf I_p$, we may continue with
    \begin{align*}
        \mathbf S \widehat{\bm \Lambda}_K - \bm \Lambda_R \mathbf S
        &= \mathbf V_R' (\widehat{\bm \Sigma} - \bm \Sigma)
(\mathbf V_K \mathbf V_K' + \mathbf V_R \mathbf V_R') \widehat{\mathbf V}_K \\
        &= \mathbf V_R' (\widehat{\bm \Sigma} - \bm \Sigma) \mathbf V_K \mathbf V_K' \widehat{\mathbf V}_K
        + \mathbf V_R' (\widehat{\bm \Sigma} - \bm \Sigma) \mathbf V_R \mathbf S.
    \end{align*}
    Taking the Frobenius norm on both sides, using the triangle inequality and the fact that 
    $\| \mathbf V_K' \widehat{\mathbf V}_K \|_2 \leq 1$ yields
    \begin{equation} \label{eq:sylv-fn}
        \| \mathbf S \widehat{\bm \Lambda}_K - \bm \Lambda_R \mathbf S \|_F
        \le \| \mathbf V_R' (\widehat{\bm \Sigma} - \bm \Sigma) \mathbf V_K \|_F
        + \| \mathbf V_R' (\widehat{\bm \Sigma} - \bm \Sigma) \mathbf V_R \|_2 \| \mathbf S \|_F,
    \end{equation}
    where we also used the fact that $\| \mathbf A \mathbf B \|_F \leq \| \mathbf A \|_2 \| \mathbf B \|_F$
    for two conformable matrices $\mathbf A$ and $\mathbf B$.
    This Sylvester-type equation is an analogue to equation A2 in \citet{yu-wang-samworth-2015}.
    Now we lower bound the left hand side.
    Notice that the $(i,j)$-th entry of the $(p-K) \times K$ matrix 
    $\mathbf S \widehat{\bm \Lambda}_K - \bm \Lambda_R \mathbf S$ is $S_{ij}(\hat\lambda_j - \lambda_{K+i})$, 
    where $S_{ij}$ is the $(i,j)$-entry of $\mathbf S$.
    Then we see that the square of the left hand side quantity from \eqref{eq:sylv-fn} is
    \[
        \| \mathbf S \widehat{\bm \Lambda}_K - \bm \Lambda_R \mathbf S \|_F^2
        = \sum_{i=1}^{p-K} \sum_{j=1}^K S_{ij}^2 (\hat\lambda_j-\lambda_{K+i})^2
        \geq (\hat \lambda_K - \lambda_{K+1})^2 \| \mathbf S \|_F^2,
    \]
    as $\hat\lambda_j - \lambda_{K+i} \ge \hat\lambda_K - \lambda_{K+1}$.
    Combining this with \eqref{eq:sylv-fn}, we obtain
    \[
        (\hat \lambda_K - \lambda_{K+1}) \| \mathbf S \|_F
        \leq \| \mathbf V_R' (\widehat{\bm \Sigma} - \bm \Sigma) \mathbf V_K \|_F
        + \| \mathbf V_R' (\widehat{\bm \Sigma} - \bm \Sigma) \mathbf V_R \|_2 \| \mathbf S \|_F,
    \]
    which implies
    \begin{equation} \label{eq:leak}
        \| \mathbf S \|_F
        \leq \frac{ \| \mathbf V_R' (\widehat{\bm \Sigma} - \bm \Sigma) \mathbf V_K \|_F }
        { (\hat \lambda_K - \lambda_{K+1}) 
        - \| \mathbf V_R' (\widehat{\bm \Sigma} - \bm \Sigma) \mathbf V_R \|_2 }.
    \end{equation}
    Let $\eta' = \eta + \delta$ with $\delta > 0$.
    Proposition \ref{prop:cov-est} implies that for all sufficiently large $T$, 
    we have $\hat \lambda_K \ge c_K p^\alpha / 2$ and
    $\|\mathbf V_R'(\widehat{\bm \Sigma} - \bm \Sigma) \mathbf V_R\|_2 \le (\hat \lambda_K - \lambda_{K+1})/2$
    with probability at least $1 - T^{-\eta'}$.
    In particular, this implies
    \[
        (\hat \lambda_K - \lambda_{K+1}) - \|\mathbf V_R'(\widehat{\bm \Sigma} - \bm \Sigma) \mathbf V_R\|_2
        \ge \frac{\hat \lambda_K - \lambda_{K+1}}{2}.
    \]
    Therefore, there exists a constant $C_1 > 0$ such that, for $T$ sufficiently large, 
    we obtain from \eqref{eq:leak}
    \[
        \| \mathbf S \|_F^2
        \le \frac{4 C_1}{p^{2\alpha}} \| \mathbf V_R' (\widehat{\bm \Sigma} - \bm \Sigma) \mathbf V_K \|_F^2,
    \]
    with probability at least $1 - T^{-\eta'}$.
    Finally, by Proposition \ref{prop:cov-est-leak}, there exists a constant $C > 0$ such that, 
    for all $T$ sufficiently large
    \[
        \| \mathbf S \|_F^2
        = \| \mathbf V_R' \widehat{\mathbf V}_K \|_F^2
        \le C \frac{p^{1 + \alpha}}{p^{2 \alpha}} \frac{K \log(T)}{T}
        = C \frac{1}{p^{\alpha - 1}} \frac{K \log(T)}{T},
    \]
    holds with probability at least $1 - 2 T^{-\eta'} \ge 1 - T^{-\eta}$.
\end{proof}

\begin{prop} \label{prop:norm-y}
    Suppose Assumptions \ref{asm:dist} and \ref{asm:mix} are satisfied. 
    Then for every $\eta>0$ there exists a constant $C > 0$ such that, for all $T$ sufficiently large
    \[
        \| \hat{\bm \vartheta} \|_2^2
        \le \frac{1}{T} \| \bm Y \|_2^2 
        \le C,
    \]
    holds with probability at least $1 - T^{-\eta}$.
\end{prop}
\begin{proof}
    First note that we have 
    \[
        \| \hat{\bm \vartheta} \|_2^2 
        = \left \| {1\over T} \widehat{\bm P}' \bm Y  \right \|_2^2
        \le \left\| \sqrt{ {1 \over T}} \widehat{\bm P}' \right \|_2^2
        \left\| \sqrt{ {1\over T}} \bm Y  \right \|_2^2
        = \frac{1}{T} \| \bm Y \|_2^2.
    \]
    The claim then follows from Assumption \ref{asm:dist} and Lemma \ref{lem:conc-lieb},
    which implies that for every $\eta > 0$, there exists a constant $C > 0$ such that,
    for all $T$ sufficiently large
    \begin{equation*}
        \frac{1}{T} \| \bm Y \|_2^2 
        = \mathbb E[ Y_t^2 ] + \frac{1}{T} \sum_{t = 1}^T (Y_t^2 - \mathbb E[ Y_t^2 ])
        \leq C,
    \end{equation*}
    with probability at least $1 - T^{-\eta}$.
\end{proof}

\subsubsection{Proof of Proposition \ref{prop:p-est-is}}

\begin{proof}[Proof of Proposition \ref{prop:p-est-is}]
    Let $\eta' = \eta + \delta$ with $\delta > 0$.
    By Proposition \ref{prop:cov-est-p}, for all $T$ sufficiently large
	\[
        \lambda_{\min}\left( \frac1T \sum_{t=1}^T \bm P_t \bm P_t' \right) \ge \frac12, 
    \]
    holds with probability at least $1 - T^{-\eta'}$.
    By Proposition \ref{prop:norm-y}, there exists a constant $C_1$ such that,
    for all $T$ sufficiently large, $\| \bm Y \|_2^2 / T \le C_1$
    holds with probability at least $1 - T^{-\eta'}$.
    Then by Proposition \ref{prop:decomp}, there exists a constant $C_2 > 0$ such that
    \[
        \| \tilde{\bm \vartheta} - \mathbf H' \hat{\bm\vartheta} \|_2^2
        \le 4 \lambda_{\min}\left(\frac{\mathbf P' \mathbf P}{T}\right)^{-1}
        \frac{\|\widehat{\mathbf P}-\mathbf P \mathbf H'\|_2^2}{T} \frac{\| \bm Y \|_2^2}{T}
        \le C_2 \frac{\|\widehat{\mathbf P}-\mathbf P \mathbf H'\|_2^2}{T},
    \]
    with probability at least $1 - 2T^{-\eta'}$.
    %    %    %    %    %    %    %    %
    Recall that $\mathbf X = (\bm X_1, \bm X_2, \ldots, \bm X_T )'$, and note that
    $\mathbf P = \mathbf X \mathbf V_K \bm \Lambda_K^{-1/2}$ and 
    $\widehat{\mathbf P} = \mathbf X \widehat{\mathbf V}_K \widehat{\bm \Lambda}_K^{-1/2}$.
    The proof is based on an argument very similar in nature to the one used in the 
    proof of Proposition \ref{prop:p-est-oos}.
    Note that
    \begin{align*}
        \widehat{\mathbf P} - \mathbf P \mathbf H'
        &= \mathbf X \widehat{\mathbf V}_K \widehat{\bm \Lambda}_K^{-1/2} 
        - \mathbf X \mathbf V_K \bm \Lambda_K ^{1/2} \bm \Lambda_K^{-1/2} 
        \mathbf V_K' \widehat{\mathbf V}_K \widehat{\bm \Lambda}_K ^{-1/2} \\
        &= \mathbf X (\mathbf I_p - \mathbf V_K \mathbf V_K') 
        \widehat{\mathbf V}_K \widehat{\bm \Lambda}_K^{-1/2} \\
        &= \mathbf X \mathbf V_R \mathbf V_R' \mathbf V_R \mathbf V_R' 
        \widehat{\mathbf V}_K \widehat{\bm \Lambda}_K^{-1/2} \\
        &= \mathbf U \mathbf V_R \mathbf V_R' \widehat{\mathbf V}_K \widehat{\bm \Lambda}_K^{-1/2},
    \end{align*}
    where $\mathbf U = \mathbf X \mathbf V_R \mathbf V_R'$.
    We then have
    \begin{equation} \label{eq:phat-ph-frob}
        \frac1T \| \widehat{\mathbf P} - \mathbf P \mathbf H' \|_F^2
        = \frac1T \|\mathbf U \mathbf V_R \mathbf V_R'\widehat{\mathbf V}_K 
        \widehat{\bm \Lambda}_K^{-1/2}\|_F^2
        \le \left\| \frac1{\sqrt T} \mathbf U \right\|_2^2
        \| \mathbf V_R' \widehat{\mathbf V}_K \|_F^2 \| \widehat{\bm \Lambda}_K^{-1/2} \|_2^2.
    \end{equation}
    We then bound each term.
    First, we have 
    $\| \mathbf U / \sqrt{T} \|_2^2 = \| \mathbf U' \mathbf U / T \|_2
    = \| \sum_{t = 1}^T \bm u_t \bm u_t' / T\|_2$ and
    \[
        \left\| \frac1T \sum_{t = 1}^T \bm u_t \bm u_t' \right\|_2
        \le \left\| \frac1T \sum_{t=1}^T \bm u_t \bm u_t' - \mathbb E(\bm u_t \bm u_t') \right\|_2 
        + \| \mathbb E(\bm u_t \bm u_t') \|_2 
        = \left\| \frac1T \sum_{t=1}^T \bm u_t \bm u_t' - \mathbb E \left( \bm u_t \bm u_t' \right) \right\|_2  
        + c_{K+1}.
    \]
    by Assumption \ref{asm:eigen}.
    Proposition \ref{prop:cov-est-u} implies that there exists a constant $C_3 > 0$ such that, 
    for all $T$ sufficiently large
    \[
        \left\| \frac1{\sqrt T} \mathbf U \right\|_2^2
        \le C_3 \left(1 + { ( p + \log(T) )^{ {r_\beta + 1 \over \, r_\beta}  } \over T }
        + \sqrt{ { ( p + \log(T) )^{ {r_\beta + 1 \over \, r_\beta}  } \over T }} \right),
    \]
    holds with probability at least $1 - T^{-\eta'}$.
    Proposition \ref{prop:cov-est} and Assumption \ref{asm:eigen} imply that there exists a constant 
    $C_4 > 0$ such that, for all $T$ sufficiently large
    \[
        \| \widehat{\bm \Lambda}_K^{-1/2} \|_2^2
        = \frac{1}{\hat \lambda_K}
        \le C_4 p^{-\alpha},
    \]
    with probability at least $1 - T^{-\eta'}$.
    Finally, Proposition \ref{prop:leak} implies that there exists a constant $C_5 > 0$ such that, 
    for all $T$ sufficiently large
    \[
        \| \mathbf V_R' \widehat{\mathbf V}_K \|_F^2
        \le C_5 p^{1 - \alpha} \frac{K \log(T)}{T},
    \]
    with probability at least $1 - T^{-\eta'}$.
    Combining these three bounds with \eqref{eq:phat-ph-frob} then yields
    \[
        \frac1T \| \widehat{\mathbf P} - \mathbf P \mathbf H' \|_F^2
        \le C_6 \left(1 + { ( p + \log(T) )^{ {r_\beta +1 \over \, r_\beta}  } \over T }
        + \sqrt{ { ( p + \log(T) )^{ {r_\beta +1 \over \, r_\beta}  } \over T }} \right)
        \frac{1}{p^{2\alpha - 1}} \frac{K \log(T)}{T},
    \]
    for some constant $C_6 > 0$, with probability at least $1 - 3 T^{-\eta'}$.
    This then implies that there exists a constant $C > 0$ such that the claim holds
    with probability at least $1 - 5T^{-\eta'} \ge 1 - T^{-\eta}$.
\end{proof}

\begin{prop} \label{prop:decomp}
    Assume that $\mathbf P$ has full column rank. Then
    \[
        \| \tilde{\bm \vartheta} - \mathbf H' \hat{\bm\vartheta} \|_2^2
        \le 4 \lambda_{\min}\left(\frac{\mathbf P' \mathbf P}{T}\right)^{-1}
        \frac{\|\widehat{\mathbf P}-\mathbf P \mathbf H'\|_2^2}{T} \frac{\| \bm Y \|_2^2}{T}.
    \]
\end{prop}
\begin{proof}
    Recall that 
    $\tilde{\bm \vartheta} = \operatorname{argmin}_{\bm \vartheta \in \mathbb R^K} 
    \|\bm Y - \mathbf P\bm\vartheta\|_2^2$
    and $\hat{\bm \vartheta} = \widehat{\mathbf P}'\bm Y/T$, because 
    $\widehat{\mathbf P}'\widehat{\mathbf P}/T = \mathbf I_K$. 
    We thus have $\tilde{\bm \vartheta} = (\mathbf P'\mathbf P/T)^{-1} \mathbf P'\bm Y/T$.
    Define the projectors 
    $\bm \Pi_{\widehat{\mathbf P}} 
    = \widehat{\mathbf P} (\widehat{\mathbf P}' \widehat{\mathbf P})^{-1} \widehat{\mathbf P}' 
    = \frac1T \widehat{\mathbf P} \widehat{\mathbf P}'$ and 
    $ \bm \Pi_{\mathbf P} = \mathbf P (\mathbf P' \mathbf P)^{-1} \mathbf P'$.
    Then consider the fitted values for $\hat{\bm \vartheta}$ and $\widetilde{\bm\vartheta}$
    \begin{align*}
        \widehat{\mathbf P} \hat{\bm \vartheta}
        &= \frac1T \widehat{\mathbf P} \widehat{\mathbf P}' \bm Y
        = \bm \Pi_{\widehat{\mathbf P}} \bm Y, \\
        \mathbf P \tilde{\bm \vartheta} 
        &= \mathbf P \left(\frac1T \mathbf P'\mathbf P \right)^{-1} \frac1T \mathbf P'\bm Y
        = \bm \Pi_{\mathbf P} \bm Y.
    \end{align*}
    Then we have by addition and subtraction
    \begin{align*}
        \mathbf P (\tilde{\bm \vartheta} - \mathbf H ' \hat{\bm \vartheta})
        &= \bm \Pi_{\mathbf P} \bm Y - \mathbf P \mathbf H' \hat{\bm \vartheta} \\
        &= \bm\Pi_{\mathbf P}\bm Y - \widehat{\mathbf P} \hat{\bm\vartheta}
        + \widehat{\mathbf P} \hat{\bm\vartheta} 
        - \mathbf P \mathbf H' \hat{\bm \vartheta} \\
        &= (\bm\Pi_{\mathbf P} - \bm\Pi_{\widehat{\mathbf P}}) \bm Y
        + (\widehat{\mathbf P} - \mathbf P \mathbf H') \hat{\bm\vartheta}.
    \end{align*}
    As a consequence we can upper bound its norm as
    \begin{equation} \label{eq:p-vt-diff}
        \| \mathbf P (\tilde{\bm \vartheta} - \mathbf H ' \hat{\bm \vartheta}) \|_2
        \le \| \bm\Pi_{\mathbf P} - \bm\Pi_{\widehat{\mathbf P}} \|_2 \| \bm Y \|_2
        + \| \widehat{\mathbf P} - \mathbf P \mathbf H' \|_2
        \frac{\| \bm Y \|_2}{\sqrt{T}},
    \end{equation}
    where we used Proposition \ref{prop:norm-y}.
    We now apply Theorem 2.5.1 from \citet{golub-van_loan-2013}.
    In their notation, let $S_1 = \operatorname{span}(\mathbf P)$ and 
    $S_2 = \operatorname{span}(\widehat{\mathbf P})$.
    We may then choose $\mathbf Z_1 = \widehat{\mathbf P} / \sqrt{T}$ 
    so that $\mathbf Z_1' \mathbf Z_1 = \mathbf I_K$
    and $\mathbf W_1 = \mathbf P (\mathbf P'\mathbf P)^{-1/2}$ 
    so that $\mathbf W_1' \mathbf W_1 = \mathbf I_K$.
    We then have 
    \[
        \mathbf W_2 \mathbf W_2' 
        = \mathbf I_T - \mathbf W_1 \mathbf W_1'
        = \mathbf I_T - \mathbf P (\mathbf P'\mathbf P)^{-1/2} (\mathbf P'\mathbf P)^{-1/2} \mathbf P'
        = \mathbf I_T - \bm \Pi_{\mathbf P}.
    \]
    Their Theorem 2.5.1 along with equation 2.5.1 then yields
    \[
        \| \bm \Pi_{\mathbf P} - \bm \Pi_{\widehat{\mathbf P}} \|_2
        = \| \mathbf W_2' \mathbf Z_1 \|_2
        = \| \mathbf W_2 \mathbf W_2' \mathbf Z_1 \|_2
        = \left\| (\mathbf I_T - \bm \Pi_{\mathbf P}) \frac{1}{\sqrt{T}} \widehat{\mathbf P} \right\|_2.
    \]
    Since $\bm \Pi_{\mathbf P}$ is the orthogonal projector onto the column space of $\mathbf P$,
    $\mathbf I_T - \bm\Pi_{\mathbf P}$ is the orthogonal projector onto its orthogonal complement.
    Moreover, $\operatorname{span}(\mathbf P \mathbf H') \subseteq \operatorname{span}(\mathbf P)$,
    so it follows that $(\mathbf I_T - \bm\Pi_{\mathbf P})\mathbf P \mathbf H' = \mathbf 0$.
    This means that
    \[
        \left\| (\mathbf I_T - \bm \Pi_{\mathbf P}) \frac{1}{\sqrt{T}} \widehat{\mathbf P} \right\|_2
        = \left\| (\mathbf I_T - \bm\Pi_{\mathbf P}) 
        \frac{\widehat{\mathbf P} - \mathbf P \mathbf H'}{\sqrt T} \right\|_2
        \le \frac{\|\widehat{\mathbf P}-\mathbf P \mathbf H'\|_2}{\sqrt T}.
    \]
    We can then continue from \eqref{eq:p-vt-diff}
    \begin{align*}
        \| \mathbf P (\tilde{\bm \vartheta} - \mathbf H ' \hat{\bm \vartheta}) \|_2
        &\le \| \bm\Pi_{\mathbf P} - \bm\Pi_{\widehat{\mathbf P}} \|_2 \| \bm Y \|_2
        + \| \widehat{\mathbf P} - \mathbf P \mathbf H' \|_2
        \frac{\| \bm Y \|_2}{\sqrt{T}} \le 2 \|\widehat{\mathbf P}-\mathbf P \mathbf H'\|_2 \frac{\| \bm Y \|_2}{\sqrt{T}}.
    \end{align*}
    Finally, we translate from the fitted values to the coefficients.
    For every $\bm a\in\mathbb R^K$
    \[
        \frac1T \|\mathbf P \bm a\|_2^2
        = \bm a'\left( \frac{\mathbf P'\mathbf P}{T} \right)\bm a
        \ge \lambda_{\min} \left(\frac{\mathbf P'\mathbf P}{T}\right)\|\bm a\|_2^2.
    \]
    Applying this with $\bm a = \tilde{\bm \vartheta} - \mathbf H' \hat{\bm\vartheta}$ gives
    \[
        \lambda_{\min}\left(\frac{\mathbf P' \mathbf P}{T}\right)^{1/2}
        \| \tilde{\bm \vartheta} - \mathbf H' \hat{\bm\vartheta} \|_2
        \le \frac{1}{\sqrt T} \|\mathbf P (\tilde{\bm\vartheta} - \mathbf H' \hat{\bm\vartheta})\|_2
        \le 2 \frac{\|\widehat{\mathbf P}-\mathbf P \mathbf H'\|_2}{\sqrt{T}} \frac{\| \bm Y \|_2}{\sqrt{T}}.
    \]
\end{proof}

\subsubsection{Proof of Proposition \ref{prop:linreg}}

\begin{proof}[Proof of Proposition \ref{prop:linreg}]
    The normal equations for $\tilde{\bm \vartheta}$ imply
    \begin{equation} \label{eq:norm-eq}
        \left( \frac1T \sum_{t = 1}^T \bm P_t \bm P_t' \right) ( \tilde{\bm \vartheta} - \bm \vartheta^*)
        = \frac1T \sum_{t = 1}^T \bm P_t (Y_t - \bm P_t' \bm \vartheta^*). 
    \end{equation}
    Let $\eta' = \eta + \delta$ with $\delta > 0$.
    Proposition \ref{prop:cov-est-p} guarantees that 
    $\lambda_{\min}\left( \sum_{t=1}^T \bm P_t \bm P_t' / T \right) \ge 1 / 2$ 
    with probability at least $1 - T^{-\eta'}$.
    As a consequence, \eqref{eq:norm-eq} and Proposition \ref{prop:linreg-conc} imply that
    \[
        \| \tilde{\bm \vartheta} - \bm \vartheta^* \|_2
        \le 2 \left\| \frac1T \sum_{t = 1}^T \bm P_t (Y_t - \bm P_t' \bm \vartheta^*) \right\|_2
        \le 2 C \sqrt{ {K \log (T) \over T} },
    \]
    with probability at least $1 - 2T^{-\eta'} \geq 1 - T^{-\eta}$.
    The claim then follows after squaring both sides.
\end{proof}

%\begin{prop} \label{prop:linreg-1}
%    Suppose Assumptions \ref{asm:dist}--\ref{asm:npk} are satisfied.
%    Then for every $\eta>0$ and for all $T$ sufficiently large
%    \[
%        \frac1T \sum_{t = 1}^T ( \bm P_t' \bm \vartheta^* - \bm P_t' \bm \vartheta)^2 
%        \ge \frac12 \| \bm \vartheta^* - \bm \vartheta \|_2^2 ~,
%    \]
%    holds simultaneously for all $\bm \vartheta \in \mathbb R^K$ with probability at least $1 - T^{-\eta}$.
%\end{prop}
%\begin{proof}
%    By part $(ii)$ of Proposition \ref{prop:cov-est-p} we have with probability at least $1 - T^{-\eta}$
%    \[
%        \lambda_{\min}\left( \frac1T \sum_{t=1}^T \bm P_t \bm P_t' \right) \ge \frac12.
%    \]
%    Hence, simultaneously for all $\bm \vartheta \in \mathbb R^K$, we have
%    \[
%	    \frac1T \sum_{t=1}^T (\bm P_t' (\bm \vartheta - \bm \vartheta^*) )^2 
%        = (\bm \vartheta - \bm \vartheta^*)' \left( \frac1T \sum_{t = 1}^T \bm P_t \bm P_t' \right) 
%        (\bm \vartheta - \bm \vartheta^*)
%        \ge \frac12 \| \bm \vartheta - \bm \vartheta^* \|_2^2.
%    \]
%\end{proof}

%\begin{prop} \label{lem:1}
%    Suppose Assumptions \ref{asm:dist}--\ref{asm:smallball} are satisfied.
%    Then for every $\eta>0$ and for all $T$ sufficiently large
%    \[
%        \frac1T \sum_{t = 1}^T ( \bm P_t' \bm \vartheta^* - \bm P_t' \bm \vartheta)^2 
%        \ge { \kappa_1^2 \kappa_2 \over 2} 
%        \| \bm P_t' \bm \vartheta^* - \bm P_t' \bm \vartheta \|_{L_2, \mathcal D_T}^2 ~,
%    \]
%    holds simultaneously for all $\bm \vartheta \in \mathbb R^K$ with probability at least $1 - T^{-\eta}$.
%\end{prop}

\begin{prop} \label{prop:linreg-conc}
    Suppose Assumptions \ref{asm:dist}--\ref{asm:npk} are satisfied.
    Then for every $\eta > 0$ there exists a constant $C > 0$ such that, for all $T$ sufficiently large 
    \[
        \left\| \frac1T \sum^T_{t = 1} \bm P_t (Y_t - \bm P_t' \bm \vartheta^*) \right\|_2
        \le C \sqrt{ {K \log (T) \over T} },
    \]
    holds with probability at least $1 - T^{-\eta}$.
\end{prop}
\begin{proof}
    Define $\bm W_t = (Y_t - \bm P_t' \bm \vartheta^*) \bm P_t$.
    We first verify that $\bm W_t$ satisfies the conditions of Lemma \ref{lem:conc-lieb}.
    First, $\mathbb E[ \bm W_t] = \mathbb E[ (Y_t - \bm P_t' \bm \vartheta^*) \bm P_t] = \bm 0$,
    since $\bm \vartheta^*$ is the population least squares coefficient associated 
    with the regression of $Y_t$ on $\bm P_t$.
    Assumption \ref{asm:dist} and Lemma \ref{lem:sub-g} imply that $\bm W_t$ is sub-exponential. 
    Moreover, 
    $\{ \bm W_t\}$ inherits the mixing properties of $\{(Y_t, \bm X_t')'\}$ from Assumption \ref{asm:mix}. 
    Assumption \ref{asm:npk} ensures that the remaining conditions of Lemma \ref{lem:conc-lieb} are satisfied. 
    The lemma then implies that for every $\eta>0$ there exists a constant $C>0$ such that, 
    for all $T$ sufficiently large,
    \[
        \left\| \frac1T \sum_{t=1}^T \bm W_t \right\|_2
        \le C \sqrt{\frac{K\log T}{T}},
    \]
    with probability at least $1-T^{-\eta}$ which establishes the claim.
\end{proof}

\subsection{Proof of Theorem \ref{thm:lower-bound}}

\begin{proof}[Proof of Theorem \ref{thm:lower-bound}]
    The proof proceeds in four steps.
    First we decompose the excess risk into an estimation term and a bias term.
    Second, we lower bound the bias term.
    Third, we lower bound the estimation term conditional on the training regressors with high probability.
    Finally, we remove the conditioning and complete the argument.

    \textbf{Step 1.} 
    We begin by introducing the oracle PCR estimator on the sample eigenvectors $\widehat{\mathbf V}_K$
    \[
        \tilde{\bm\theta}
        = \operatorname*{argmin}_{\bm\theta \in \operatorname{span}(\widehat{\mathbf V}_K)} R(\bm\theta).
    \]
    Define $R(\bm \theta) = \mathbb E \left[ ( Y_{T+1} - \bm X_{T+1}' \bm \theta )^2 \mid \mathcal D_T \right]$ 
    for an arbitrary $\bm \theta \in \mathbb R^p$ and note that 
    \[
        R(\bm \theta^*) 
        = \min_{\bm \theta \in \mathbb R^p} \mathbb E \left[ ( Y_{T+1} - \bm X_{T+1}' \bm \theta )^2
        \mid \mathcal D_T \right]
        = R^*.
    \]
    We may then decompose the excess risk as
    \[
        R(\hat{\bm\theta}) - R^*
        = ( R(\hat{\bm\theta}) - R(\tilde{\bm\theta}))
        + (R(\tilde{\bm\theta}) - R^*),
    \]
    where the first term is the estimation error within the class of PCR estimators 
    using $\widehat{\mathbf V}_K$ (i.e. relative to the PCR oracle on the sample eigenvectors),
    and the second term is the bias relative to the oracle.

    \textbf{Step 2.} We next lower bound the bias term.
    For any $\bm\theta\in\operatorname{span}(\widehat{\mathbf V}_K)$, we have
    \begin{align*}
    R(\bm\theta) - R^*
    &= \mathbb E[ (Y_{T+1} - \bm X_{T+1}' \bm\theta)^2 \mid \mathcal D_T]
    - \mathbb E[ (Y_{T+1} - \bm X_{t+1}' \bm\theta^*)^2 \mid \mathcal D_T] \\
    &= \mathbb E[ (Y_{T+1} - \bm X_{T+1}' \bm\theta^* 
    + \bm X_{T+1}' \bm\theta^* - \bm X_{T+1}' \bm\theta)^2 \mid \mathcal D_T]
    - \mathbb E[ (Y_{T+1} - \bm X_{T+1}' \bm\theta^*)^2 \mid \mathcal D_T] \\
    &= \mathbb E [(\bm X_{T+1}' (\bm\theta - \bm\theta^*))^2 \mid \mathcal D_T] \\
    &= (\bm\theta - \bm\theta^*)' \bm\Sigma (\bm\theta - \bm\theta^*),
    \end{align*}
    which follows as $Y_{T+1} - \bm X_{T+1}' \bm\theta^* = \varepsilon_{T+1}$,
    $\mathbb E[ \varepsilon_{T+1} \bm X_{T+1} \mid \mathcal D_T] = \bm 0$
    and $\mathbb E[ \bm X_{T+1} \bm X_{T+1}' \mid \mathcal D_T] = \bm \Sigma$. 
    As a consequence
    \[
        R(\tilde{\bm\theta}) - R^*
        = \inf_{\bm\theta \in \operatorname{span}(\widehat{\mathbf V}_K)}
        (\bm\theta - \bm\theta^*)' \bm\Sigma (\bm\theta - \bm\theta^*)
        \ge \inf_{\bm\theta \in \operatorname{span}(\widehat{\mathbf V}_K)}
        \|\bm\theta - \bm\theta^*\|_2^2,
    \]
    because
    \[
        (\bm\theta - \bm\theta^*)' \bm\Sigma (\bm\theta - \bm\theta^*)
        = (\bm\theta - \bm\theta^*)' \mathbf V \bm \Lambda \mathbf V' (\bm\theta - \bm\theta^*)
        \ge \|\bm\theta - \bm\theta^*\|_2^2.
    \]

    Now define the projectors $\bm \Pi = \mathbf V_K \mathbf V_K'$ and 
    $\widehat{\bm \Pi} = \widehat{\mathbf V}_K \widehat{\mathbf V}_K'$.
    Then by the projection theorem
    \[
        \inf_{\bm\theta \in \operatorname{span}(\widehat{\mathbf V}_K)} \|\bm\theta - \bm\theta^* \|_2^2
        = \|(\mathbf I_p - \widehat{\bm \Pi}) \bm\theta^*\|_2^2
        = \|\bm\theta^*\|_2^2 - \|\widehat{\bm \Pi} \bm\theta^* \|_2^2,
    \]
    as this is the shortest distance from $\bm\theta^*$ to the subspace 
    $\operatorname{span}(\widehat{\mathbf V}_K)$,
    and the second step is by the Pythagorean theorem since $\widehat{\bm \Pi}$ is an orthogonal projector.
    In our setting, $\bm\theta^* = \mathbf V_R \mathbf V_R' \bm \theta^* = \bm\gamma^*$, 
    so $\bm \Pi \bm\theta^* = \bm 0$,
    so that $\widehat{\bm \Pi} \bm\theta^* = (\widehat{\bm \Pi} - \bm \Pi) \bm \theta^*$ and thus
    \[
        \|\widehat{\bm \Pi} \bm\theta^*\|_2
        \le \|\widehat{\bm \Pi} - \bm \Pi\|_2 \|\bm\theta^* \|_2.
    \]
    Putting things together, we obtain
    \begin{equation*} \label{eq:lb-bias}
        R(\tilde{\bm\theta}) - R^*
        \ge \inf_{\bm\theta \in \operatorname{span}(\widehat{\mathbf V}_K)}
        \|\bm\theta - \bm\theta^*\|_2^2
        \ge (1 - \|\widehat{\bm \Pi} - \bm \Pi\|_2^2) \|\bm\theta^* \|_2^2
        \ge (1 - \|\widehat{\bm \Pi} - \bm \Pi\|_2^2) \|\bm\gamma^* \|_2^2.
    \end{equation*}
    %where the last step follows as $\|\bm\theta^*\|_2^2 = \|\bm\gamma^*\|_2^2$,
    %because $\mathbf V_R$ has orthonormal columns.

    \textbf{Step 3.} We next turn to lower bounding the estimation term.
    We have
    \begin{align*}
        R(\hat{\bm\theta}) - R(\tilde{\bm\theta})
        &= \mathbb E[ (Y_{T+1} - \bm X_{T+1}' \hat{\bm\theta})^2 \mid \mathcal D_T]
        - \mathbb E[ (Y_{T+1} - \bm X_{T+1}' \tilde{\bm\theta})^2 \mid \mathcal D_T] \\
        &= \mathbb E[ (Y_{T+1} - \bm X_{T+1}' \tilde{\bm\theta}
        + \bm X_{T+1}' \tilde{\bm\theta} - \bm X_{T+1}' \hat{\bm\theta})^2 \mid \mathcal D_T]
        - \mathbb E[ (Y_{T+1} - \bm X_{T+1}' \tilde{\bm\theta})^2 \mid \mathcal D_T] \\
        &= \mathbb E[(\bm X_{T+1}' \tilde{\bm\theta} - \bm X_{T+1}' \hat{\bm\theta})^2 \mid \mathcal D_T] \\
        &= (\hat{\bm\theta} - \tilde{\bm\theta})' \bm\Sigma (\hat{\bm\theta} - \tilde{\bm\theta}),
    \end{align*}
    by the projection theorem.
    Now consider $\tilde{\bm \theta}$, the oracle PCR estimator using $\widehat{\mathbf V}_K$.
    Because every $\bm\theta \in \operatorname{span}(\widehat{\mathbf V}_K)$ has the form
    $\bm\theta = \widehat{\mathbf V}_K \hat{\bm \Lambda}_K^{-1/2} \bm\vartheta$,
    there exists a corresponding oracle coefficient $\tilde{\bm \vartheta} \in \mathbb R^K$ such that
    $\tilde{\bm\theta} = \widehat{\mathbf V}_K \hat{\bm \Lambda}_K^{-1/2} \tilde{\bm\vartheta}$.
    We can thus continue to obtain
    \begin{align}
        R(\hat{\bm\theta}) - R(\tilde{\bm\theta})
        &= (\hat{\bm\theta} - \tilde{\bm\theta})' \bm\Sigma 
        (\hat{\bm\theta} - \tilde{\bm\theta}) \notag \\
        &= (\hat{\bm \vartheta} - \tilde{\bm \vartheta})' \hat{\bm\Lambda}_K^{-1/2} \widehat{\mathbf V}_K'
        \bm\Sigma \widehat{\mathbf V}_K \hat{\bm\Lambda}_K^{-1/2}
        (\hat{\bm\vartheta} - \tilde{\bm\vartheta}) \notag \\
        &\equiv (\hat{\bm \vartheta} - \tilde{\bm \vartheta})' \mathbf G
        (\hat{\bm\vartheta} - \tilde{\bm\vartheta}), \label{eq:lb-est-risk}
    \end{align}
    as $\hat{\bm\theta} = \widehat{\mathbf V}_K \hat{\bm\Lambda}_K^{-1/2} \hat{\bm\vartheta}$,
    because $\hat{\bm\theta}_{PCR} \in \operatorname{span}(\widehat{\mathbf V}_K)$,
    and we define $\mathbf G = \hat{\bm\Lambda}_K^{-1/2}\widehat{\mathbf V}_K' \bm\Sigma 
    \widehat{\mathbf V}_K \hat{\bm\Lambda}_K^{-1/2}$. 

    To keep the proof self-contained, recall that 
    $\widehat{\bm P}_t = \hat{\bm\Lambda}_K^{-1/2}\widehat{\mathbf V}_K'\bm X_t$ and
    $\widehat{\mathbf P} = (\widehat{\bm P}_1, \ldots, \widehat{\bm P}_T)'$.
    We also have from the paper that $\widehat{\mathbf P}'\widehat{\mathbf P} / T = \mathbf I_K$
    and $\hat{\bm\vartheta} = \widehat{\mathbf P}' \bm Y / T$.
    We therefore obtain
    \begin{equation} \label{eq:lb-vartheta-diff}
        \hat{\bm\vartheta} - \tilde{\bm\vartheta}
        = \frac1T\widehat{\mathbf P}'\bm Y 
        - \frac1T\widehat{\mathbf P}'\widehat{\mathbf P} \tilde{\bm\vartheta}
        = \frac1T\widehat{\mathbf P}'(\bm Y-\widehat{\mathbf P}\tilde{\bm\vartheta}).
    \end{equation}
    Now rewrite the vector
    \[
        \bm Y - \widehat{\mathbf P} \tilde{\bm\vartheta}
        = \mathbf X \bm \theta^* + \bm \varepsilon - \widehat{\mathbf P}\tilde{\bm\vartheta}
        = \mathbf X (\bm \theta^* - \tilde{\bm\theta}) + \bm \varepsilon,
    \]
    because $\widehat{\mathbf P}\tilde{\bm\vartheta} = \mathbf X \widehat{\mathbf V}_K 
    \widehat{\bm \Lambda}_K^{-1/2} \tilde{\bm \vartheta} = \mathbf X \tilde{\bm \theta}$.
    Then we obtain from \eqref{eq:lb-vartheta-diff}
    \[
        \hat{\bm\vartheta} - \tilde{\bm\vartheta}
        = \frac1T\widehat{\mathbf P}'(\bm Y-\widehat{\mathbf P}\tilde{\bm\vartheta})
        = \frac1T\widehat{\mathbf P}' \mathbf X (\bm \theta^* - \tilde{\bm\theta})
        + \frac1T\widehat{\mathbf P}'\bm \varepsilon
        \equiv \bm a + \bm Z,
    \]
    where we defined $\bm a = \widehat{\mathbf P}' \mathbf X (\bm \theta^* - \tilde{\bm\theta}) / T$
    and $\bm Z = \widehat{\mathbf P}'\bm \varepsilon / T$.
    Combining this with \eqref{eq:lb-est-risk}, we obtain
    \[
        R(\hat{\bm\theta}) - R(\tilde{\bm\theta})
        = (\hat{\bm \vartheta} - \tilde{\bm \vartheta})' \mathbf G (\hat{\bm\vartheta} - \tilde{\bm\vartheta})
        = (\bm a + \bm Z)' \mathbf G (\bm a + \bm Z).
    \]

    Let $\mathcal D_{T,X} = \{ \bm X_s \}_{s=1}^T$ be the regressors from the training sample.
    Conditioning on $\mathcal D_{T,X}$, the vector $\bm a$ is deterministic,
    while $\bm Z$ satisfies
    \begin{equation} \label{eq:lb-z}
        \bm Z \sim \mathcal N \left(\bm 0, \frac{\sigma^2}{T} \mathbf I_K \right),
    \end{equation}
    because $\operatorname{Var}(\bm Z \mid \mathcal D_{T,X}) 
    = \operatorname{Var}( \widehat{\mathbf P}'\bm\varepsilon / T \mid \mathcal D_{T,X})
    = \sigma^2 \widehat{\mathbf P}'\widehat{\mathbf P} / T^2$.
    In other words, lower bounding the excess risk boils down to lower bounding
    $(\bm a + \bm Z)' \mathbf G (\bm a + \bm Z)$, which is a quadratic form in a shifted Gaussian vector
    if we condition on $\mathcal D_{T,X}$.
    Because $\mathbf G$ is non-negative definite, 
    we use Anderson's shifted ball inequality \citep{anderson-1955} to obtain for any $r > 0$
    \[
        \mathbb P\left( R(\hat{\bm\theta}) - R(\tilde{\bm\theta}) \ge r \,\Big\vert\, \mathcal D_{T,X} \right)
        = \mathbb P \left( (\bm a + \bm Z)' \mathbf G (\bm a + \bm Z ) \ge r \mid \mathcal D_{T,X} \right)
        \ge \mathbb P \left( \bm Z' \mathbf G \bm Z \ge r \mid \mathcal D_{T,X} \right).
    \]

    Define $\lambda_{min}(\mathbf G) \geq 0$ as the smallest eigenvalue of $\mathbf G$,
    so that $\bm Z' \mathbf G \bm Z \ge \lambda_{min}(\mathbf G) \| \bm Z \|_2^2$.
    By \eqref{eq:lb-z} we have
    \[
        \| \bm Z \|_2^2
        = \frac{\sigma^2}{T} \bigg\| \bm Z \sqrt{\frac{T}{\sigma^2}} \bigg\|_2^2
        \sim \frac{\sigma^2}{T} \chi_K^2,
    \]
    which implies 
    \[
        \mathbb P \left( R(\hat{\bm\theta}) - R(\tilde{\bm\theta}) \ge r \,\Big\vert\, \mathcal D_T \right)
        \ge \mathbb P \left( \lambda_{min}(\mathbf G) \frac{\sigma^2}{T}\chi_K^2
        \ge r \,\Big\vert\, \mathcal D_{T,X} \right).
    \]
    We may then apply the Laurent-Massart lower-tail bound for chi-square variables \citep{laurent-massart-2000}
    for $0\le x < K/4$
    \[
        \mathbb P\left(\chi_K^2 \le K - 2\sqrt{Kx} \right) \le e^{-x}.
    \]
    Take $x = K / 9$ and $r = \lambda_{min}(\mathbf G) \sigma^2 K / (3T)$ to obtain
    \[
        \mathbb P \left( R(\hat{\bm\theta}) - R(\tilde{\bm\theta}) 
        \ge \lambda_{min}(\mathbf G) \frac{\sigma^2}{3} \frac{K}{T} \,\Bigg\vert\, \mathcal D_{T,X} \right)
        \ge \mathbb P \left( \chi_K^2 \geq \frac{K}{3} \right)
        \geq 1 - e^{-K/9}.
    \]
    Finally, add back the bias term to obtain
    \[
        \mathbb P \left( R(\hat{\bm\theta}) - R^*
        \ge  (1 - \|\widehat{\bm \Pi} - \bm \Pi\|_2^2) \|\bm\gamma^* \|_2^2
        + \lambda_{min}(\mathbf G) \frac{\sigma^2}{3} \frac{K}{T} \,\Bigg\vert\, \mathcal D_{T,X} \right)
        \geq 1 - e^{-K/9}.
    \]

    \textbf{Step 4.} Proposition \ref{prop:cov-est} implies that for every $\eta$, 
    and for sufficiently large $T$, we have $\| \widehat{\bm \Pi} - \bm \Pi \|_2 \le \sqrt{5/6}$ and
    $\lambda_{min}( \mathbf G ) \ge 1/2$ with probability at least $1 - T^{-\eta}$.
    Let $\mathcal E_T$ denote this event.
    On $\mathcal E_T$, we thus obtain
    \[
        \mathbb P \left( R(\hat{\bm\theta}) - R^*
        \ge \frac16 \left(\|\bm\gamma^* \|_2^2 + \sigma^2 \frac{K}{T}\right) 
        \,\Bigg\vert\, \mathcal D_{T,X} \right)
        \geq 1 - e^{-K/9}.
    \]
    Next define the event
    \[
        \mathcal A_T = \left\lbrace R(\hat{\bm\theta}) - R^* 
        \ge \frac16 \left( \|\bm\gamma^* \|_2^2 + \sigma^2 \frac{K}{T} \right) \right\rbrace.
    \]
    We have shown $\mathbb P(\mathcal A_T \mid \mathcal D_{T,X}) \geq 1 - e^{-K/9}$ on the event $\mathcal E_T$,
    or equivalently that
    \begin{equation} \label{eq:lb-prob}
        \mathbf 1_{\mathcal E_T} \mathbb P( \mathcal A_T \mid \mathcal D_{T,X}) 
        \ge \mathbf 1_{\mathcal E_T} (1 - e^{-K/9}),
    \end{equation}
    almost surely.
    We may write
    \[
        \mathbf 1_{\mathcal E_T} \mathbb P( \mathcal A_T \mid \mathcal D_{T,X}) 
        = \mathbf 1_{\mathcal E_T} \mathbb E[ \mathbf 1_{\mathcal A_T} \mid \mathcal D_{T,X}]
        = \mathbb E[ \mathbf 1_{\mathcal E_T} \mathbf 1_{\mathcal A_T} \mid \mathcal D_{T,X}], 
    \]
    where the second step follows because $\mathbf 1_{\mathcal E_T}$ 
    is measurable with respect to the sigma algebra generated by $\mathcal D_{T,X}$.
    Plugging this into the left hand side of \eqref{eq:lb-prob} and taking expectations gives
    \[
        \mathbb E[ \mathbb E[ \mathbf 1_{\mathcal E_T} \mathbf 1_{\mathcal A_T} \mid \mathcal D_{T,X}] ]
        = \mathbb E[\mathbf 1_{\mathcal E_T} \mathbf 1_{\mathcal A_T} ]
        = \mathbb P( \mathcal A_T \cap \mathcal E_T ),
    \]
    by the law of iterated expectations.
    It follows that
    \[
        \mathbb P( \mathcal A_T \cap \mathcal E_T )
        \ge \mathbb E[\mathbf 1_{\mathcal E_T}] (1 - e^{-K/9})
        = \mathbb P( \mathcal E_T) (1 - e^{-K/9}).
    \]
    Finally, since $\mathbb P(\mathcal A_T) \ge \mathbb P( \mathcal A_T \cap \mathcal E_T )$
    and $\mathbb P(\mathcal E_T) \geq 1 - T^{-\eta}$ by Proposition \ref{prop:cov-est}
    , we have
    \[
        \mathbb P( \mathcal A_T )
        \ge 1 - T^{-\eta} - e^{-K/9},
    \]
    which completes the proof.
\end{proof}

\subsection{Proof of Theorem \ref{thm:erm-pcr-ht}}

\begin{proof}[Proof of Theorem \ref{thm:erm-pcr-ht}]
    We give a sketch of the proof.
    Since $m \delta / 2 > \eta > 0$, fix $\eta' = ( m \delta / 2 + \eta) / 2$,
    so that $\eta < \eta' < m \delta/2$.
    Inspecting the proof of Proposition \ref{prop:cov-est} shows that its claims (i) and (ii)
    go through as before using Propositions \ref{prop:cov-est-u-ht}, \ref{prop:cov-est-p-ht}
    and \ref{prop:cov-est-pu-ht} under Assumption \assref{4}{asm:npk-ht}, that is,
    for all $T$ sufficiently large
    \[
        \hat \lambda_K \ge \frac{c_K}{2}  p^\alpha,
    \]
    holds with probability at least $1 - T^{-\eta'}$.
    
    As Proposition \ref{prop:cov-est-pu-ht} obtains the same rate as Proposition \ref{prop:cov-est-pu},
    adapting Propositions \ref{prop:cov-est-leak}, \ref{prop:leak} and \ref{prop:p-est-oos} yields the same 
    rate as in the thin-tailed case.
    There thus exists a constant $C' > 0$ such that, for all $T$ sufficiently large
    \[
        \mathbb E( \| \hat{\bm P}_{T+1} - \mathbf H \bm P_{T+1} \|_2^2 \mid \mathcal D_T)
        \le C' \frac{1}{p^{2\alpha - 1}} \frac{ K \log(T) }{T},
    \]
    holds with probability at least $1 - T^{-\eta'}$.
    
    Similarly, inspecting the proof of Proposition \ref{prop:p-est-is} 
    and using Proposition \ref{prop:cov-est-u-ht} shows that there exists a constant $C'' > 0$ such that, 
    for all $T$ sufficiently large
    \begin{equation*}
        \frac1T \| \widehat{\mathbf P} - \mathbf P \mathbf H' \|_F^2
        \le C'' \left(1 + \sqrt{ \frac{ p \log(T) }{ T } } 
        + \frac{p \log(T)^3}{ T^{1 - 2 / m - \delta} } \right) 
        \times \frac{1}{p^{2\alpha-1}} \frac{ K \log(T) }{T},
    \end{equation*}
    holds with probability at least $1 - T^{-\eta'}$.
    
    %Proposition \ref{prop:linreg-1} continues to hold using Proposition \ref{prop:cov-est-p-ht}
    %in place of Proposition \ref{prop:cov-est-p}, and 
    Proposition \ref{prop:linreg-conc-ht} adapts Proposition \ref{prop:linreg-conc} to the heavy-tailed setting.
    Using it along with Proposition \ref{prop:cov-est-p-ht} in the proof of Proposition \ref{prop:linreg}
    then shows that there exists a constant $C''' > 0$ such that, for all $T$ sufficiently large
    \[   
	    \| \bm \vartheta^* - \tilde{\bm \vartheta} \|^2_2 \leq C''' {K \log (T) \over T},
    \]
    holds with probability at least $1 - T^{-\eta'}$.
    Finally, Proposition \ref{prop:norm-y-ht} bounds $\| \hat{\bm \vartheta} \|_2^2$ with a constant
    with probability at least $1 - T^{-\eta'}$.
    Taking all of these facts together then implies the claim.
\end{proof}

\begin{prop} \label{prop:cov-est-u-ht} 
    Suppose Assumptions \assref{1}{asm:dist-ht}, \ref{asm:mix}, \ref{asm:eigen} 
    and \assref{4}{asm:npk-ht} are satisfied. 
    Then for every $\kappa / m > \delta > 0$ and every $m \delta / 2  > \eta > 0$
    there exists a constant $C > 0$ such that, for all $T$ sufficiently large
    \[
        \left\| \frac1T \sum_{t=1}^T \bm u_t \bm u_t' - \bm \Sigma_u \right\|_2
        \le C \left( \sqrt{ \frac{ p \log(T) }{ T } }
        + \frac{p \log(T)^3}{ T^{1 - 2 / m - \delta} } \right),
    \]
    holds with probability at least $1 - T^{-\eta}$.
\end{prop}
\begin{proof}
    When $c_{K+1} = 0$ the claim is trivial as $\bm u_t = \bm 0$ almost surely, so we may assume otherwise.
    Note that by part $(iii)$ of Lemma \ref{lem:ht} we have
    $\sup_{\| \bm v \|_2 = 1} \mathbb E[ | \bm v' \bm \Sigma_u^{-1/2} \bm u_t|^m ] \le C_Z$.
    Moreover, $\bm u_t$ inherits the mixing properties of $\bm X_t$ from Assumption \ref{asm:mix}.
    We may thus apply Lemma \ref{lem:ht-cov-conc} to $\bm u_t$.
    Since Assumption \ref{asm:eigen} implies $\| \bm \Sigma_u \|_2 = c_{K+1}$, 
    we have that for every $\nu > (1 + \eta) / m$ there exists a constant $C > 0$ such that, 
    for all sufficiently large $T$
    \[
        \left\| \frac1T \sum_{t=1}^T \bm u_t \bm u_t' - \bm \Sigma_u \right\|_2
        \le C \left( \sqrt{\frac{ p (\log(2p) + \log(T))}{T}}
        + T^{2 \nu} \frac{p \log(T)^2 (\log(2p) + \log(T))}{T} \right), 
    \]
    holds with probability at least $1 - T^{-\eta}$.
    Set $\nu = 1/m + \delta / 2$.
    This is an admissible choice since $m \delta / 2 > \eta > 0$ implies 
    $\nu = 1/m + \delta / 2 > (1 + \eta) / m$.
    We then have that for every $\kappa / m > \delta > 0$ and every $m \delta / 2 > \eta > 0$
    there exists a constant $C_1 > 0$ such that, for all $T$ sufficiently large
    \begin{align*}
        \left\| \frac1T \sum_{t=1}^T \bm u_t \bm u_t' - \bm \Sigma_u \right\|_2
        &\le C_1 \left(  \sqrt{ \frac{p (\log(2p) + \log(T))}{T} }
        + T^{2 / m + \delta} \frac{p \log(T)^2 (\log(2p) + \log(T))}{T} \right) \\
        &\le C \left( \sqrt{ \frac{p \log(T) }{T} } + \frac{p \log(T)^3}{T^{1 - 2/m - \delta}} \right),
    \end{align*}
    which follows for some constant $C > 0$,
    as $\log(2p) + \log(T) \le C_3 \log(T)$ and $\log(T)^2 (\log(2p) + \log(T)) \le C_4 \log(T)^3$,
    for some constants $C_3, C_4 > 0$.
\end{proof}

\begin{prop} \label{prop:cov-est-p-ht} 
    Suppose Assumptions \assref{1}{asm:dist-ht}, \ref{asm:mix}, \ref{asm:eigen} 
    and \assref{4}{asm:npk-ht} are satisfied. 
    Then for every $\kappa / m > \delta > 0$ and every $m \delta / 2  > \eta > 0$
    there exists a constant $C > 0$ such that, for all $T$ sufficiently large
    \begin{align*}
        &(i) \left\| \frac1T \sum_{t=1}^T \bm P_t \bm P_t' - \mathbf I_K \right \|_2
        \le C \left( \sqrt{ \frac{ K \log(T) }{ T } }
        + \frac{K \log(T)^3}{ T^{1 - 2 / m - \delta} } \right) \\
	    &(ii) \lambda_{\min}\left( \frac1T \sum_{t=1}^T \bm P_t \bm P_t' \right) \ge \frac12, 
    \end{align*}
    holds with probability at least $1 - T^{-\eta}$.
\end{prop}
\begin{proof}
    First note that part $(ii)$ of Lemma \ref{lem:ht} implies that 
    $\sup_{\| \bm v \|_2 = 1} \mathbb E[ | \bm v' \bm P_t|^m ] \le C_Z$
    and $\bm P_t$ inherits the mixing properties of $\bm X_t$ from Assumption \ref{asm:mix}.
    We may then apply Lemma \ref{lem:ht-cov-conc} which implies that for every $\nu > (1 + \eta) / m$
    there exists a constant $C > 0$ such that, for all sufficiently large $T$
    \[
        \left\| \frac1T \sum_{t=1}^T \bm P_t \bm P_t' - \mathbf I_K \right\|_2
        \le C \left( \sqrt{\frac{ K (\log(2K) + \log(T))}{T}}
        + T^{2 \nu} \frac{K \log(T)^2 (\log(2K) + \log(T))}{T} \right), 
    \]
    holds with probability at least $1 - T^{-\eta}$.
    We may then follow the same approach as in the proof of Proposition \ref{prop:cov-est-u-ht} 
    to obtain that for every $\kappa / m > \delta > 0$ and every $m \delta / 2 > \eta > 0$, 
    there exists a constant $C > 0$ such that, for all $T$ sufficiently large
    \[
        \left\| \frac1T \sum_{t=1}^T \bm P_t \bm P_t' - \mathbf I_K \right\|_2
        \le C \left( \sqrt{ \frac{ K \log(T) }{T} }
        + \frac{K \log(T)^3}{ T^{1 - 2 / m - \delta} } \right),
    \]
    with probability at least $1 - T^{-\eta}$.
    This establishes the first claim.
    By Weyl's inequality
    \[
        \lambda_{\min}\left( \frac1T \sum_{t=1}^T \bm P_t \bm P_t'\right) 
        \geq 1 - \left \| {1 \over T } \sum_{t=1}^T \bm P_t \bm P_t' - \bm I_K \right \|_2. 
    \]
    The second claim then follows from the first by Assumption \assref{4}{asm:npk-ht}, 
    since for every $m \delta / 2 > \eta > 0$ and for all $T$ sufficiently large, 
    $\| (1/T) \sum_{t=1}^T \bm P_t \bm P_t' - \bm I_K \|_2 \le 1/2$ with probability at least $1-T^{-\eta}$. 
\end{proof}

\begin{prop} \label{prop:cov-est-pu-ht}
    Suppose Assumptions \assref{1}{asm:dist-ht}, \ref{asm:mix}, \ref{asm:eigen} 
    and \assref{4}{asm:npk-ht} are satisfied. 
    Then for every $\kappa / m > \delta > 0$ and every $m \delta / 2  > \eta > 0$
    there exists a constant $C > 0$ such that, for all $T$ sufficiently large
    \[
        \left\| {1 \over T} \sum_{t = 1}^T \bm P_t \bm u_t' \right\|_F 
        \le C \sqrt{ \frac{ Kp \log(T) }{T } },
    \]
    holds with probability at least $1 - T^{-\eta}$.
\end{prop}
\begin{proof}
    We assume that $c_{K+1}$ is positive, since if $c_{K+1}$ is zero the claim is trivial.
    Let $V_t = \operatorname{vec}(\bm P_t \bm u_t')$.
    We begin by showing that the sequence of $(Kp)$-dimensional random vectors 
    $\{\bm V_t\}_{t=1}^T$ satisfies the conditions required by Lemma \ref{lem:ht-conc}.
    First, $\mathbb E[\bm V_t] = \bm 0$.
    Second, we have $\| \bm V_t \|_2^2 = \| \bm P_t \bm u_t' \|_F^2 = \| \bm P_t \|_2^2 \| \bm u_t \|_2^2$.
    Then by Cauchy-Schwarz and parts $(ii)$ and $(iv)$ of Lemma \ref{lem:ht}, 
    there exists a constant $C_1 > 0$ such that
    \[
        \mathbb E[ \| \bm V_t \|_2^{m/2} ] 
        = \mathbb E\left[ \| \bm P_t \|_2^{m/2} \| \bm u_t \|_2^{m/2} \right]
        \le \mathbb E\left[ \| \bm P_t \|_2^m \right]^{1/2} 
        \mathbb E\left[ \| \bm u_t \|_2^m \right]^{1/2}
        \le C_1 (Kp)^{m / 4}.
    \]
    Moreover, $\{ \bm V_t \}_{t=1}^T$ inherits the mixing properties of the sequence 
    $\{(Y_t, \bm X_t)'\}_{t=1}^T$ from Assumption \ref{asm:mix}.

    We may then apply Lemma \ref{lem:ht-conc} with its moment parameter equal to $m / 2$ 
    to the sequence $\{\bm V_t \}_{t=1}^T$.
    The lemma implies that for every $\eta > 0$ and every $\nu > 2 (1 + \eta) / m$ 
    there exist constants $C_1, C_2 > 0$ such that, for all sufficiently large $T$
    \begin{align*}
        \left\| \frac1T \sum_{t=1}^T \bm V_t \right\|_2
        &\le C_1 \sqrt{Kp} \left( \sqrt{\frac{\log(2Kp) + \log(T)}{T}}
        + T^\nu \frac{ \log(T)^2 (\log(2 Kp) + \log(T))}{T} \right) \\
        &\le C_2 \left( \sqrt{ \frac{ Kp \log(T)}{T}}
        + \frac{ \sqrt{Kp} \log(T)^3}{T^{1 - \nu}} \right), 
    \end{align*}
    with probability at least $1 - T^{-\eta}$.
    Notice that the former term dominates if $\nu < 1/2$.
    Take $\nu = 2 / m + \delta$ and notice that by Assumption \assref{4}{asm:npk-ht}
    \[
        \delta 
        < \kappa / m
        < \frac{m - 4}{2m}
        = 1/2 - 2/m.
    \]
    This choice of $\nu$ thus guarantees that the former term dominates.
    Then for every $\kappa / m > \delta > 0$ and every $m \delta / 2 > \eta > 0$
    there exists a constant $C > 0$ such that, for all $T$ sufficiently large
    \[
        \left\| {1 \over T} \sum_{t = 1}^T \bm P_t \bm u_t' \right\|_F
        = \left\| \frac1T \sum_{t=1}^T \bm V_t \right\|_2
        \le C \sqrt{ \frac{ Kp \log(T) }{T} }.
    \]
\end{proof}

\begin{prop} \label{prop:linreg-conc-ht}
    Suppose Assumptions \assref{1}{asm:dist}, \ref{asm:mix}, \ref{asm:eigen}
    and \assref{4}{asm:npk} are satisfied.
    Then for every $\kappa / m > \delta > 0$ and every $m \delta / 2  > \eta > 0$
    there exists a constant $C > 0$ such that, for all $T$ sufficiently large
    \[
        \left\| \frac1T \sum^T_{t = 1} \bm P_t (Y_t - \bm P_t' \bm \vartheta^*) \right\|_2
        \le C \sqrt{ {K \log (T) \over T} },
    \]
    holds with probability at least $1 - T^{-\eta}$.
\end{prop}
\begin{proof}
    As in the proof of Proposition \ref{prop:linreg-conc}, 
    define $\bm W_t = (Y_t - \bm P_t' \bm \vartheta^*) \bm P_t$.
    It is easy to verify that $\bm W_t$ satisfies the conditions of Lemma \ref{lem:ht-conc},
    by Assumptions \assref{1}{asm:dist-ht} and \ref{asm:mix} along with part $(v)$ of Lemma \ref{lem:ht}.
    Taking $\nu = 2 / m + \delta$, 
    the lemma implies that for every $\kappa / m > \delta > 0$ and every $m \delta / 2 > \eta > 0$
    there exists a constant $C_1 > 0$ such that, for all $T$ sufficiently large,
    \[
        \left\| \frac1T \sum_{t=1}^T \bm W_t \right\|_2
        \le C_1 \left( \sqrt{ \frac{ K \log(T) }{T} } + \frac{ \sqrt K \log(T)^3}{T^{1 - 2/m - \delta}} \right),
    \]
    with probability at least $1 - T^{-\eta}$.
    The former term dominates because $\delta < 1/2 - 2/m$ by Assumption \assref{4}{asm:npk},
    which guarantees $1 / 2 < 1 - 2/m - \delta$.
    It thus follows that there exists a constant $C > 0$ such that
    \[
        \left\| \frac1T \sum_{t=1}^T \bm W_t \right\|_2
        \le C \sqrt{\frac{K \log T}{T}}, 
    \]
    with probability at least $1-T^{-\eta}$.
\end{proof}

\begin{prop} \label{prop:norm-y-ht}
    Suppose Assumptions \assref{1}{asm:dist-ht} and \ref{asm:mix} are satisfied. 
    Then for every $1 - 2 / m > \delta > 0$ and every $m \delta / 2  > \eta > 0$
    there exists a constant $C > 0$ such that, for all $T$ sufficiently large
    \[
        \| \hat{\bm \vartheta} \|_2^2
        \le \frac{1}{T} \| \bm Y \|_2^2 
        \le C,
    \]
    holds with probability at least $1 - T^{-\eta}$.
\end{prop}
\begin{proof}
    As in Proposition \ref{prop:norm-y}, we have 
    \[
        \| \hat{\bm \vartheta} \|_2^2 
        \le \frac{1}{T} \| \bm Y \|_2^2
        = \mathbb E[ Y_t^2 ] + \frac{1}{T} \sum_{t = 1}^T (Y_t^2 - \mathbb E[ Y_t^2 ]).
    \]
    We thus wish to apply Lemma \ref{lem:ht-conc} to $U_t = Y_t^2 - \mathbb E[ Y_t^2 ]$.
    By the elementary inequality $|a-b|^q \leq 2^{q-1} (|a|^q + |b|^q )$ for two real numbers $a, b$
    and $q \geq 1$, Assumption \assref{1}{asm:dist-ht} and Jensen's inequality imply that 
    \[
        \mathbb E[ |U_t|^{m/2} ] 
        \le 2^{m/2 - 1} (\mathbb E[ |Y_t|^m ] + | \mathbb E[ Y_t^2] |^{m/2} )
        \le 2^{m/2 - 1} (\mathbb E[ |Y_t|^m ] + \mathbb E[ | Y_t^m|] )
        \leq C_1.
    \]
    for some constant $C_1 > 0$, where we took $q = m / 2 > 2$.
    It is easy to verify that $U_t = Y_t^2 - \mathbb E[ Y_t^2 ]$ satisfies the remaining conditions 
    of Lemma \ref{lem:ht-conc}.
    The lemma then implies that for every $1 - 2 / m > \delta > 0$ and every $m \delta / 2  > \eta > 0$
    there exist constants $C_2, C_3 > 0$ such that, for all $T$ sufficiently large,
    \[
        \left| \frac1T \sum_{t=1}^T ( Y_t^2 - \mathbb E[ Y_t^2 ]) \right|
        \le C_2 \left( \sqrt{\frac{\log T}{T}} + \frac{\log(T)^3}{T^{1-2/m-\delta}} \right)
        \le C_3,
    \]
    with probability at least $1-T^{-\eta}$, where we took $\nu = 2 / m + \delta$.
    The last inequality follows for sufficiently large $T$ because $1 - 2/m - \delta > 0$.
    As $\mathbb E[ Y_t^2 ] \le ( \mathbb E[ |Y_t|^m ])^{2/m} \le C_Z^{2/m}$ by Lyapunov's inequality,
    it follows that for some constant $C > 0$ and all $T$ sufficiently large,
    \[
        \frac1T \|\bm Y \|_2^2 = \frac1T \sum_{t=1}^T Y_t^2 \le C,
    \]
    with probability at least $1 - T^{-\eta}$.
\end{proof}

\subsection{Auxiliary Results}

\subsubsection{Thin-Tailed}

\begin{lem} \label{lem:sub-g} 
    Suppose Assumptions \ref{asm:dist} and \ref{asm:eigen} are satisfied.

    Then we have that 
    $(i)$ there exists a constant $C_1 > 0$ such that $\bm P_t$ is sub-Gaussian with parameter $C_1$; 
    Define $\bm W_t = \mathbf V_R \bm \Lambda_R^{-1/2} \mathbf V_R' \bm u_t$.
    Then $(ii)$ if $c_{K+1} > 0$, there exists a constant $C_2 > 0$ such that $\bm W_t$ is sub-Gaussian 
    with parameter $C_2$, otherwise $\bm W_t$ is degenerate at zero;
    $(iii)$ if $c_{K+1} > 0$ then there exists a constant $C_3 > 0$ such that $\bm u_t$ is sub-Gaussian 
    with parameter $C_3$, otherwise $\bm u_t$ is degenerate at zero;
    $(iv)$ there exists a constant $C_4 > 0$ such that $Y_t - \bm P_t' \bm \vartheta^*$ is sub-Gaussian 
    with parameter $C_4$.
\end{lem}
\begin{proof}
    $(i)$ Despite the fact that $\bm \Sigma$ may be singular, it still holds that 
    $\bm X_t = \bm \Sigma^{1/2} \bm \Sigma^{-1/2} \bm X_t = \bm \Sigma^{1/2} \bm Z_t$ almost surely
    since $\bm X_t \in \operatorname{Range}(\bm \Sigma)$ almost surely.   
    Since $\bm P_t = \bm \Lambda_K^{-1/2} \bm V_K' \bm X_t = \bm V_K' \bm Z_t$ almost surely, 
    it follows that for any $\varepsilon > 0$
    \[
	    \sup_{\bm v: \| \bm v \|_2 = 1} \mathbb P( |\bm v'\bm P_t| > \varepsilon ) 
	    \leq \sup_{\bm v: \| \bm v \|_2 = 1} \mathbb P( |\bm v'\bm Z_t| > \varepsilon ) 
	    \leq C_0 \exp(-C_m \varepsilon^2) ~,
    \]
    so $\bm P_t$ is sub-Gaussian with parameter $C_1 = C_m$. \\
    $(ii)$ As $\bm W_t = \mathbf V_R \bm \Lambda_R^{-1/2} \mathbf V_R' \bm u_t
    = \mathbf V_R \bm \Lambda_R^{-1/2} \bm \Lambda_R^{1/2} \mathbf V_R' \bm Z_t$ almost surely,
    we have that for any $\varepsilon > 0$
    \[
	    \sup_{\bm v: \| \bm v \|_2 = 1} 
        \mathbb P( |\bm v'\mathbf V_R \bm \Lambda_R^{-1/2} \mathbf V_R' \bm u_t| > \varepsilon ) 
	    \leq \sup_{\bm v: \| \bm v \|_2 = 1} \mathbb P( |\bm v'\bm Z_t| > \varepsilon ) 
	    \leq C_0 \exp(-C_m \varepsilon^2),
    \]
    because $\mathbf V_R \bm \Lambda_R^{-1/2} \bm \Lambda_R^{1/2} \mathbf V_R'$ is an orthogonal projector.
    $\mathbf V_R \bm \Lambda_R^{-1/2} \mathbf V_R' \bm u_t$ is thus sub-Gaussian with parameter $C_2 = C_m$.\\
    $(iii)$ The conclusion is obvious if $c_{K+1} = 0$, so assume that $c_{K+1} > 0$.
    Since $\bm u_t = \bm V_R \bm V_R' \bm X_t = \bm V_R \bm \Lambda_R^{1/2} \bm V_R' \bm Z_t$ almost surely, 
    we have that for any $\varepsilon>0$
    \[
    	\sup_{\bm v: \| \bm v \|_2 = 1} \mathbb P( |\bm v'\bm u_t| > \varepsilon ) 
    	\leq \sup_{\bm v: \| \bm v \|_2 = 1} \mathbb P( c_{K+1}^{1/2} |\bm v'\bm Z_t| > \varepsilon ) 
    	\leq C_0 \exp \left(-{C_m \over c_{K+1}} \varepsilon^2 \right),
    \]
    where we have used the fact that $\| \bm V_R \bm \Lambda_R^{1/2} \bm V_R'\|_2 
    \leq \| \bm V_R \|_2  \| \bm \Lambda_R^{1/2} \|_2 \| \bm V_R'\|_2 = c_{K+1}^{1/2}$. 
    As a consequence, $\bm u_t$ is sub-Gaussian with parameter $C_3 = C_m / c_{K+1}$. \\
    $(iv)$ We first show that $\bm P_t' \bm \vartheta^*$ is sub-Gaussian.
    If $\bm \vartheta^* = \bm 0$, the conclusion trivially follows, so we assume otherwise.
    Define $\bm a = \bm\vartheta^* / \| \bm\vartheta^* \|_2$ so that
    $\bm P_t' \bm\vartheta^* = \| \bm\vartheta^* \|_2 \bm a' \bm P_t$.
    As $\bm P_t$ is sub-Gaussian, 
    and because Assumption \ref{asm:dist} implies there exists a constant $C' > 0$ such that
    \begin{align*}
        \mathbb E [Y_t^2]
        &= \int_0^\infty \mathbb P(Y_t^2 > s) ds
        = \int_0^\infty \mathbb P(|Y_t| > \sqrt{s}) ds \\
        &= 2\int_0^\infty \varepsilon \mathbb P(|Y_t| > \varepsilon) d\varepsilon
        \leq 2 C_0 \int_0^\infty \varepsilon\exp(-C_m \varepsilon^2) d\varepsilon
        \leq (C')^2,
    \end{align*}
    so that $\| \bm\vartheta^* \|_2 \leq C'$ by Lemma \ref{lem:pcr},
    we have
    \[
        \mathbb P(|\bm P_t' \bm\vartheta^*| > \varepsilon)
        = \mathbb P\left( |\bm a' \bm P_t| > \frac{\varepsilon}{\|\bm\vartheta^*\|_2} \right)
        \leq C_0 \exp(-C_m \varepsilon^2 / (C')^2),
    \]
    so $\bm P_t' \bm \vartheta^*$ is sub-Gaussian with parameter $C'' = C_m / (C')^2$.
    Finally, we have
    \begin{align*}
        \mathbb P(|Y_t - \bm P_t' \bm\vartheta^*| > \varepsilon)
        &\leq \mathbb P(|Y_t| > \varepsilon / 2) + \mathbb P(|\bm P_t'\bm\vartheta^*| > \varepsilon / 2) \\
        &\leq C_0 \exp(-C_m \varepsilon^2 / 4) + C_0 \exp(- C_m \varepsilon^2 / (4 (C')^2)) \\
        &\leq 2 C_0 \exp(-C_4 \varepsilon^2),
    \end{align*}
    for some constant $C_4 > 0$, which establishes the final claim.
\end{proof}

\begin{lem} \label{lem:conc-bosq}
    Let $\{\bm Z_t \}_{t=1}^T$ be a stationary sequence of $d$-dimensional random vectors
    and $\bm \Sigma_z = \mathbb E[\bm Z_t \bm Z_t']$.
    Suppose
    (i) for any $\varepsilon > 0$, $\sup_{\bm v: \| \bm v \|_2 = 1} 
    \mathbb P(|\bm v' \bm \Sigma_z^{-1/2} \bm Z_t| > \varepsilon) 
    \le 2\exp(-C_m \varepsilon^2)$ for some $C_m > 0$; 
    (ii) the $\alpha$-mixing coefficients of the sequence satisfy $ \alpha(l) < \exp( -C_\alpha l^{r_\alpha} )$
    for some $C_\alpha > 0$ and $r_\alpha > 0$; and
    (iii) $d = \lfloor C_d T^{r_d} \rfloor$ for some $C_d>0$ and $r_d \in [0, r_\alpha)$.

    Then for every $\eta>0$ there exists a positive constant $C_\eta > 0$ such that, 
    for all $T$ sufficiently large
    \[
        \left\| \frac1T \sum_{t=1}^T \bm Z_t \bm Z_t' - \bm \Sigma_z \right\|_2
        \le C_\eta \| \bm \Sigma_z \|_2 \left({ ( d + \log(T) )^{ {r_{\alpha} +1 \over \, r_\alpha}  } \over T }
        + \sqrt{ { ( d + \log(T) )^{ {r_{\alpha} +1 \over \, r_\alpha}  } \over T }} \right),
    \]
    holds with probability at least $1 - T^{-\eta}$.
\end{lem}
\begin{proof}
    If $\bm \Sigma_z = \mathbf 0$, $\bm Z_t = \bm 0$ almost surely and the result follows trivially, 
    so we assume otherwise.
    Let $\bm \Sigma_z = \mathbf V_z \bm \Lambda_z \mathbf V_z'$ be the eigendecomposition of $\bm \Sigma_z$.
    As $\bm \Sigma_z$ may be singular, 
    we have the orthogonal projector $\bm \Pi_z \equiv \bm \Sigma_z^{-1/2} \bm \Sigma_z \bm \Sigma_z^{-1/2}
    = \mathbf V_z \bm \Lambda_z^{-1/2} \bm \Lambda_z \bm \Lambda_z^{-1/2} \mathbf V_z'$,
    with the property that $\bm \Pi_z \bm \Sigma_z^{1/2} = \bm \Sigma_z^{1/2}$.
    Note also that since $\bm Z_t \in \operatorname{Range}(\bm\Sigma_z)$ almost surely,
    we have $\bm Z_t = \bm \Sigma_z^{1/2} \bm \Sigma_z^{-1/2} \bm Z_t$ almost surely.
    Then almost surely
    \begin{align}
        \left\| \frac{1}{T} \sum_{t = 1}^T \bm Z_t \bm Z_t' - \bm \Sigma_z \right\|_2
        &= \left\| \bm \Sigma_z^{1/2} 
        \left( \frac{1}{T} \sum_{t = 1}^T \bm \Sigma_z^{-1/2} \bm Z_t \bm Z_t' \bm \Sigma_z^{-1/2}
        - \bm \Pi_z \right) \bm \Sigma_z^{1/2} \right\|_2 \notag \\
        &\le \| \bm \Sigma_z \|_2 \left\| \frac{1}{T} \sum_{t = 1}^T \bm \Sigma_z^{-1/2} \bm Z_t \bm Z_t' 
        \bm \Sigma_z^{-1/2} - \bm \Pi_z \right\|_2. \label{eq:iso}
    \end{align}
    It is therefore sufficient for us to bound the right hand side.
    Lemma 5.4 of \citet{Vershynin:2012} implies that if $\mathcal N$ is a ${1 \over 4}$-net of 
    $\mathcal S^{d-1}\cap \operatorname{Range}(\bm\Sigma_z)$ then
    \begin{align}
	    \left\| \frac1T \sum_{t=1}^T \bm \Sigma_z^{-1/2} \bm Z_t \bm Z_t' \bm \Sigma_z^{-1/2}
        - \bm \Pi_z \right\|_2 
	    &= \max_{\bm x \in \mathcal S^{d-1} \cap \operatorname{Range}(\bm \Sigma_z)} 
        \left| \frac1T \sum_{t=1}^T (\bm Z_t' \bm \Sigma_z^{-1/2} \bm x)^2-1 \right| \nonumber \\
	    &\leq 2\max_{\bm x \in \mathcal N} 
        \left| \frac1T \sum_{t=1}^T (\bm Z_t' \bm \Sigma_z^{-1/2} \bm x)^2-1 \right| \nonumber \\
	    &\equiv 2\max_{\bm x \in \mathcal N} \left| \frac1T \sum_{t=1}^T W_{\bm x,t} \right|, \label{eq:net}
    \end{align} 
    where we defined $W_{\bm x, t} = (\bm Z_t' \bm \Sigma_z^{-1/2} \bm x)^2 - 1$.
    To establish the claim we first show that for all $T$ sufficiently large \eqref{eq:net} 
    can be bounded with high probability. 

    We begin by establishing a bound for a fixed vector $\bm x \in \mathcal N$.
    Note the following three properties of the sequence $\{ W_{\bm x, t} \}$.
    First, $\mathbb E(W_{\bm x, t}) = 0$ since 
    \[
        \mathbb E[( \bm Z_t' \bm \Sigma_z^{-1/2} \bm x)^2] 
        = \bm x'\bm \Pi_z \bm x
        = \|\bm x \|_2^2 = 1.
    \]
    Second, $W_{\bm x, t}$ is the de-meaned square of a sub-Gaussian random variable with parameter $C_m$ 
    (which is independent of $\bm x$) by (i).
    Third, the sequence $\{W_{\bm x,t}\}_{t=1}^T$ inherits the mixing properties of $\{ \bm Z_t \}_{t=1}^T$ 
    from (ii).
    We may then apply Theorem 1.4 of \citet{Bosq:1998}.\footnote{The $C_r$ inequality and 
    \citet[Theorem 2.1]{BoucheronLugosiMassart:2013} imply that the condition 
    in equation (1.33) of \citet{Bosq:1998} is satisfied.}
    Define
    \[
        \varepsilon_T = { C_1^*( d + \log(T) )^{ {r_{\alpha} +1 \over r_\alpha}  } \over T} 
        +  \left[ { C_1^* ( d + \log(T) )^{ {r_{\alpha} +1 \over \, r_\alpha}  } \over T }\right]^{1/2}
    \]
    and 
    \[
        q_T =  \left\lceil { T \over C^*_2 (d + \log(T) )^{1 \over r_\alpha} } \right\rceil - 1,
    \]
    for positive constants $C_1^*$ and $C_2^*$ to be chosen below.
    Provided that $C_2^* \ge 2$ and $T$ sufficiently large, we have $q_T \in [1, T/2]$ by (iii), 
    as required by the theorem.
    Theorem 1.4 of \citet{Bosq:1998} then implies that for any $r\geq 3$,
    there exists a positive constant $C_1$ that depends on $C_m$ and $r$ such that, 
    for all $T$ sufficiently large
    \begin{equation} \label{eq:bosq} 
	    \mathbb P \left( \left| \frac1T \sum_{t=1}^T W_{\bm x,t} \right| > \varepsilon_T \right) 
	    \leq a_{1T} \exp \left(-{q_T \varepsilon_T^2\over C_1 + C_1 \varepsilon_T} \right) 
        + a_{2T} \alpha\left( \left\lfloor{ T \over q_T + 1 }\right\rfloor \right)^{2r \over 2r+1},
    \end{equation} 
    where
    $a_{1T} = 2 {T / q_T} + 2 [ 1 + {\varepsilon_T^2 / (C_1 + C_1 \varepsilon_T}) ]$ and 
    $a_{2T} = 11 T( 1 + {C_1 / \varepsilon_T})$.
    We proceed by bounding the right hand side of \eqref{eq:bosq}.
    
    For the first term, we have for all $T$ sufficiently large
    \begin{align}
        a_{1T} \exp \left(-{q_T \varepsilon_T^2\over C_1 + C_1 \varepsilon_T} \right)
        &\le \left( 2 T + 2 + 2 {\varepsilon_T \over C_1 } \right) 
        \exp\left(-{ \min( \varepsilon_T, \varepsilon_T^2 ) \over 2 C_1 } q_T \right) \nonumber \\
        &\le \exp\left( \log \left( 3 T + 2 {\varepsilon_T \over C_1 } \right) 
        -{ { C_1^*(d + \log(T) )^{ {r_{\alpha} +1 \over r_\alpha}  } \over 2 T C_1 }} q_T \right) \nonumber \\
        &\le \exp \left( \log\left(3 T + 2 \frac{\varepsilon_T}{C_1} \right) 
        - \frac{C_1^*}{4 C_1 C_2^*} (d + \log(T)) \right) \nonumber \\
        &\le \exp \left( -\left( \frac{C_1^*}{4 C_1 C_2^*} - 1 \right) 
        (d + \log(T)) \right), \label{eq:a1}
    \end{align}
    where the first inequality uses the fact that $x^2 / (1+x) \geq \min(x, x^2) / 2$ for $x > 0$,
    the second uses 
    $\min(\varepsilon_T,\varepsilon_T^2) \ge { C_1^* (d + \log(T) )^{ {r_\alpha +1 \over r_\alpha} } / T}$
    and the third uses the fact that $q_T \ge T / (2 C_2^* (d+\log(T)))^{1/r_\alpha}$ by (iii)
    for sufficiently large $T$.
    For the second term in \eqref{eq:bosq}, we have for all $T$ sufficiently large
    \begin{align}
	    a_{2 T} \alpha\left( \left\lfloor{ T \over q_T +1 }\right\rfloor \right)^{2r \over 2r+1} 
        &\le \exp\left( 2 \log(T ) - C_\alpha {2 r \over 2r+1} 
        \left({ T \over q_T +1 } - 1 \right)^{r_\alpha} \right) \nonumber \\
	    &\le \exp\left(2\log(T) -C_\alpha(C^*_2/4)^{r_\alpha} {2r \over 2r+1} (d +\log(T)) \right) \nonumber\\  
	    &\le \exp\left( -\left( C_\alpha (C^*_2/4)^{r_\alpha} {2r \over 2r+1} - 2 \right) 
        (d + \log(T)) \right), \label{eq:a2}
    \end{align}
    where the first inequality follows from (ii) and the fact that $\varepsilon_T \ge 1/T$ 
    for sufficiently large $T$ and the second follows as 
    $(T / (q_T + 1) - 1)^{r_\alpha} \ge (C_2^*/4)^{r_\alpha} (d + \log(T))$ for all sufficiently large $T$.

    Combining \eqref{eq:a1} and \eqref{eq:a2} into \eqref{eq:bosq} then yields for a given $\bm x$, 
    for all $T$ sufficiently large
    \begin{align}
	    \mathbb P \left( \left| \frac1T \sum_{t=1}^T W_{\bm x, t} \right| > \varepsilon_T \right) 
        &\le \exp \left( -\left( \frac{C_1^*}{4 C_1 C_2^*} - 1 \right) 
        (d + \log T)\right) \nonumber \\ 
	    &+ \exp\left( -\left( C_\alpha (C^*_2/4)^{r_\alpha} {2r \over 2r+1} - 2 \right) 
        (d + \log(T)) \right). \label{eq:pointwise}
    \end{align} 
    Finally, we take the union bound over all $\bm x \in \mathcal N$. 
    The cardinality of a $1 \over 4$-net $\mathcal N$ of the unit sphere $\mathcal S^{d-1}$ is bounded by $9^d$
    by Lemma 5.2 of \citet{Vershynin:2012}.
    Setting 
    \[
        C_2^* = 4 \left({(\eta' + \log(9) + 3) (2r+1) \over 2r C_{\alpha}}\right)^{1 \over r_\alpha},
    \]
    where $\eta' = \eta + \delta$ with $\delta > 0$ and $C_1^* = 4 C_1 C_2^* (\eta' + \log(9) + 2)$, 
    in \eqref{eq:pointwise} we then obtain that, for all $T$ sufficiently large
    \begin{align*}
	    \mathbb P \left( \max_{\bm x \in \mathcal N} 
        \left| {1\over T} \sum_{t=1}^T W_{\bm x,t} \right| > \varepsilon_T \right) 
        &\le 9^d \max_{\bm x \in \mathcal N} 
        \mathbb P \left( \left| {1\over T} \sum_{t=1}^T W_{\bm x,t} \right| > \varepsilon_T \right) \\
	    &\le 2 \exp \left( \log(9) d - (\eta' + \log(9))\left( d + \log(T) \right) \right) \\
        &\le 2 \exp(-\eta' \log(T))
        \le T^{-\eta}.
    \end{align*}
    Combining this with \eqref{eq:net} and \eqref{eq:iso} then yields the result.
\end{proof}

\begin{lem} \label{lem:conc-lieb}
    Let $\{\bm Z_t \}_{t=1}^T$ be a stationary sequence of $d$-dimensional zero-mean random vectors.
    Suppose
    (i) for any $\varepsilon > 0$ and some $C_0, C_m > 0$, 
    $\sup_{1 \leq i \leq d} \mathbb P( |Z_{it} | > \varepsilon ) \leq C_0 \exp( -C_m \varepsilon )$; 
    (ii) the $\alpha$-mixing coefficients of the sequence satisfy $ \alpha(l) < \exp( -C_\alpha l^{r_\alpha} )$
    for some $C_\alpha > 0$ and $r_\alpha > 0$; and
    (iii) $d = \lfloor C_d T^{r_d} \rfloor$ for some $C_d>0$ and $r_d \in [0, 1]$.

    Then for every $\eta > 0$ there exists a positive constant $C_\eta$ such that, 
    for all $T$ sufficiently large
    \[
	    \left\| {1 \over T} \sum_{t=1}^T \bm Z_t \right\|_2 
        \le C_\eta \sqrt{ {d \log(T) \over T } },
    \]
    holds with probability at least $1 - T^{-\eta}$.
\end{lem}
\begin{proof}
    Fix $\eta' = \eta + \delta$ with $\delta > 0$.
    Let $C_*$ denote a positive constant to be chosen below. By the union bound
    \[
    	\mathbb P\left( \left\| {1 \over T} \sum_{t=1}^T \bm Z_t \right\|_2 
        \ge C^* \sqrt{ {d \log (T) \over T } } \right) 
    	\le d \max_{1 \leq i \leq d}  \mathbb P\left( \left| \sum_{t=1}^T Z_{it} \right| 
        \ge  C^* \sqrt{T \log (T) } \right).
    \]
    Define the truncation threshold 
    \[
        b_T = 2 \left(r_d + \frac12 + \eta'\right) \frac{1}{C_m} \log(T),
    \]
    and decompose
    $\sum_{t=1}^T Z_{it} = \sum_{t=1}^T Z_{it}' + \sum_{t=1}^T Z_{it}''$ where
    \begin{align*}
        Z'_{it} &=  Z_{it} \mathbbm 1_{ \{ |Z_{it}| \le b_T \} } 
        - \mathbb E( Z_{it} \mathbbm 1_{ \{ |Z_{it}| \le b_T \} }), \\
        Z''_{it} &=  Z_{it} \mathbbm 1_{ \{ |Z_{it}| > b_T \} } 
        - \mathbb E( Z_{it} \mathbbm 1_{ \{ |Z_{it}| > b_T \} }).
    \end{align*}
    If we define $\varepsilon_T = (C_*/2) \sqrt{T \log (T)}$, we have
    \begin{equation*}
    	\mathbb P\left( \left| \sum_{t=1}^T Z_{it} \right| > C_* \sqrt{T \log(T)} \right)
        \le \mathbb P\left( \left|\sum_{t=1}^T Z'_{it} \right| > \varepsilon_T \right)
        + \mathbb P\left( \left| \sum_{t=1}^T Z''_{it} \right| > \varepsilon_T \right).
    \end{equation*}

    First consider the bounded part.
    The sequence $\{ Z'_{it} \}_{t=1}^T$ has the same mixing properties as $\{ \bm Z_t \}_{t=1}^T$ 
    and $\sup_{1 \le i \le d} \| Z_{it}' \|_\infty < 2 b_T$. 
    Define $M_T = \lfloor b_T^{-1} \sqrt{T / \log (T)} \rfloor$.
    Then, for all $T$ sufficiently large we have $M_T \in [1,T]$ and
    \[
        4 (2 b_T) M_T 
        = 8 \sqrt{\frac{T}{\log(T)}} 
        < \varepsilon_T,
    \]
    so the conditions of Theorem 2.1 of \citet{Liebscher:1996} are satisfied. 
    The theorem gives us
    \begin{equation} \label{eq:b2-lieb}
	    \mathbb P\left( \left| \sum_{t=1}^T Z'_{it}\right| > \varepsilon_T \right) 
	    < 4\exp\left( -{ \varepsilon_T^2 \over 64 (T/M_T) D_T + (16/3) b_T M_T \varepsilon_T } \right) 
        + 4 {T \over M_T} \exp\left( -C_\alpha M_T^{r_\alpha} \right),
    \end{equation}
    with $D_T = \sup_{1\leq i\leq d} \mathbb E[ (\sum_{t=1}^{M_T} Z'_{it})^2]$.
    To bound $D_T$, define $\gamma(l) = \sup_{1 \le i \le d} |\operatorname{Cov}(Z'_{it}, Z'_{it+l})|$ 
    for $l = 0, 1, \ldots$ and note that $D_T \le M_T \sum_{l=-M_T+1}^{M_T-1} \gamma(l)$ by stationarity. 
    For $l = 1, 2, \ldots$ we obtain 
    \[
        \gamma(l) \le 12 \alpha(l)^{1\over 2} \sup_{1 \leq i \leq d} \| Z'_{it} \|^2_{L_4} 
        \le C_1 \alpha(l)^{1 \over 2} \sup_{1 \leq i \leq d} \| Z_{it} \|^2_{L_4},
    \]
    for some constant $C_1 > 0$, where the first inequality is Davydov's inequality 
    \citep[Corollary 1.1]{Bosq:1998} and the second follows from the fact that 
    $\| Z_{it}' \|_{L_4} \le 2 \| Z_{it} \mathbbm 1_{\{ |Z_{it}| \le b_T \}} \|_{L_4} \le 2 \| Z_{it} \|_{L_4}$.
    As $\sup_{1 \le i \le d} \| Z_{it} \|_{L_4}$ is also bounded, 
    Assumption (ii) then implies that $D < C_2 M_T$ where $C_2 > 0$ is a constant.
    The definition of $\varepsilon_T$ and the fact that $b_T M_T \leq \sqrt{T / \log(T)}$ together imply
    $(16/3) b_T M_T \varepsilon_T \leq (8/3) C_* T$.
    As a consequence, \eqref{eq:b2-lieb} becomes
    \begin{equation*}
	    \mathbb P\left( \left| \sum_{t=1}^T Z'_{it}\right| > \varepsilon_T \right) 
	    \le 4\exp\left( -{ C_*^2 \log(T) \over 256 C_2 + (32/3) C_* } \right) 
        + 4 T \exp\left( -C_\alpha M_T^{r_\alpha} \right),
    \end{equation*}
    and using Assumption (iii), we finally obtain
    \begin{align}
	    d \max_{1 \le i \le d} \mathbb P\left( \left| \sum_{t=1}^T Z'_{it}\right| > \varepsilon_T \right) 
	    &\le 4 C_d \left( \exp\left(r_d \log(T) -{ C_*^2 \log(T) \over 256 C_2 + (32/3) C_* } \right) 
        + T^{1 + r_d} \exp\left( -C_\alpha M_T^{r_\alpha} \right) \right) \notag \\
        &\le 4 C_d (\exp( -\eta' \log(T) ) + \exp( 2\log(T) - C_\alpha M_T^{r_\alpha})) \notag \\
	    & \le 5 C_d T^{-\eta'} \le T^{-\eta}, \label{eqn:zp} 
    \end{align}
    where the second inequality follows from choosing $C_*$ large enough to ensure
    $C_*^2 / (256 C_2 + (32/3) C_*) \geq r_d + \eta'$.

    Next we bound the tail component sequence $\{ Z''_{it} \}_{t=1}^T$.
    By Markov's inequality and the triangle inequality
    \[
        S \equiv
	    d \max_{1 \le i \le d} \mathbb P\left( \left| \sum_{t=1}^T Z''_{it} \right| > {\varepsilon_T} \right) 
	    \le {d \over \varepsilon_T} \max_{1 \le i \le d} \mathbb E \left| \sum_{t=1}^T Z''_{it} \right|  
	    \le {d \over \varepsilon_T} \sum_{t=1}^T \max_{1 \le i \le d} \mathbb E | Z''_{it} |. 
    \]
    Using the definition of $Z_{it}''$ and Cauchy-Schwarz we continue with
    \[
        S \le 2{dT \over \varepsilon_T} \max_{1 \le i \le d} 
        \mathbb E(| Z_{it} \mathbbm 1_{ \{ |Z_{it}| > b_T \} } |) 
        \le 2{dT \over \varepsilon_T} \max_{1 \le i \le d} \| Z_{it} \|_{L_2} 
        \| \mathbbm 1_{ \{ |Z_{it}| > b_T \} } \|_{L_2}.
    \]
    If we define $\sigma^2 = \sup_{1 \le i \le d} \| Z_{it} \|_{L_2}$, Assumption (i) implies
    \[
	    S 
        \le 2{dT \over \varepsilon_T} \max_{1 \le i \le d} \| Z_{it} \|_{L_2} \mathbb P( |Z_{it}| > b_T )^{1/2}
        < 2{dT \over \varepsilon_T} \sigma^2 \exp \left( -{  C_m \over 2}b_T  \right).
    \]
    Finally, using Assumption (iii) and the definitions of $\varepsilon_T$ and $b_T$
    \begin{equation}
	    S \le {4 C_d \sigma^2 \over C_* \sqrt{ \log(T)} } 
        \exp\left( (r_d + (1/2)) \log(T) - (r_d + (1/2) + \eta') \log(T) \right)
        \le T^{-\eta}. \label{eqn:zpp}
    \end{equation}
    Equations \eqref{eqn:zp} and \eqref{eqn:zpp} together then imply the claim.
\end{proof}

\subsubsection{Heavy-tailed}

\begin{lem} \label{lem:ht} 
    Suppose Assumptions \assref{1}{asm:dist-ht} and \ref{asm:eigen} are satisfied.
    Then we have that 
    $(i)$ $\mathbb E( \|\bm Z_t\|_2^m) \le C_Z p^{m/2}$;
    $(ii)$ $\sup_{\| \bm v \|_2 = 1} \mathbb E[ | \bm v' \bm P_t|^m ] \le C_Z$ and
    $\mathbb E\left[ \| \bm P_t \|_2^m \right] \le C_Z K^{m/2}$;
    Define $\bm W_t = \mathbf V_R \bm \Lambda_R^{-1/2} \mathbf V_R \bm u_t$.
    Then $(iii)$ if $c_{K+1} > 0$, we have 
    $\sup_{\| \bm v \|_2 = 1} \mathbb E[ | \bm v' \bm W_t|^m ] \le C_Z$ and
    $\mathbb E[ \| \bm W_t \|_2^m] \le C_Z p^{m / 2}$,
    otherwise $\bm W_t$ is degenerate at zero;
    $(iv)$ if $c_{K+1} > 0$ there exists a constant $C_Z' > 0$ such that 
    $\sup_{\| \bm v \|_2 = 1} \mathbb E[ | \bm v' \bm u_t|^m ] \le C_Z'$ and 
    $\mathbb E[ \| \bm u_t \|_2^m] \le C_Z' p^{m / 2}$,
    otherwise $\bm u_t$ is degenerate at zero;
    $(v)$ there exists a constant $C_Z'' > 0 $ such that
    $\sup_{\| \bm v \|_2 = 1} \mathbb E[ | \bm v' (Y_t - \bm P_t' \bm \vartheta^*) \bm P_t |^{m/2}] \le C_Z''$
    and $\mathbb E [ \| (Y_t - \bm P_t' \bm \vartheta^*) \bm P_t \|_2^{m/2} ] \le C_Z'' K^{m/4}$.
\end{lem}
\begin{proof}
    $(i)$ By the standard inequality $\| \bm Z_t \|_2 \le p^{1/2 - 1/m} \| \bm Z_t \|_m$
    we obtain
    \[
        \| \bm Z_t \|_2^m
        \le p^{m/2 - 1} \| \bm Z_t \|_m^m
        = p^{m/2 - 1} \sum_{i = 1}^p | Z_{it} |^m.
    \]
    Taking expectations and using $\sup_{\| \bm v \|_2 = 1} \mathbb E[ | \bm v' \bm Z_t|^m ] \le C_Z$
    from Assumption \assref{1}{asm:dist-ht} then implies
    \[
        \mathbb E\left[ \| \bm Z_t \|_2^m \right]
        \le p^{m/2 - 1} \sum_{i = 1}^p \mathbb E\left[ | Z_{it} |^m \right]
        \le C_Z p^{m/2}.
    \]

    $(ii)$ As $\bm X_t \in \operatorname{Range}(\bm \Sigma)$ almost surely, 
    we have $\bm X_t = \bm \Sigma^{1/2} \bm \Sigma^{-1/2} \bm X_t = \bm \Sigma^{1/2} \bm Z_t$ almost surely.
    We therefore have
    $\bm P_t = \bm \Lambda_K^{-1/2} \mathbf V_K' \bm\Sigma^{1/2} \bm Z_t \equiv \mathbf Q_{\bm P} \bm Z_t$
    almost surely, where
    \[
        \mathbf Q_{\bm P} \mathbf Q_{\bm P}'
        = \bm \Lambda_K^{-1/2} \mathbf V_K' \bm\Sigma \mathbf V_K \bm \Lambda_K^{-1/2}
        = \mathbf I_K,
    \]
    and hence $\| \mathbf Q_{\bm P} \|_2 = 1$.
    For any unit vector $\bm x$ we have $| \bm x' \bm P_t| = | (\mathbf Q_{\bm P}' \bm x)' \bm Z_t|$,
    so Assumption \assref{1}{asm:dist-ht} implies 
    $\sup_{\| \bm v \|_2 = 1} \mathbb E[ | \bm v' \bm P_t|^m ] \le C_Z$.
    By the same arguments as in $(i)$, we also immediately obtain 
    $\mathbb E\left[ \| \bm P_t \|_2^m \right] \le C_Z K^{m/2}$.

    $(iii)$ Let $\bm \Sigma_u = \mathbb E[ \bm u_t \bm u_t']$.
    Similar to before, we have
    $\bm W_t = \bm \Sigma_u^{-1/2} \bm u_t = \bm\Sigma_u^{-1/2} \mathbf V_R\mathbf V_R' \bm\Sigma^{1/2} \bm Z_t
    \equiv \mathbf Q_{\bm W} \bm Z_t$ almost surely, where
    \[
        \mathbf Q_{\bm W} \mathbf Q_{\bm W}'
        = \bm\Sigma_u^{-1/2} \mathbf V_R\mathbf V_R' \bm\Sigma \mathbf V_R\mathbf V_R' \bm\Sigma_u^{-1/2}
        = \bm\Pi_u,
    \]
    where $\bm \Pi_u$ is the orthogonal projector onto the range of $\bm \Sigma_u$, 
    and hence $\| \mathbf Q_{\bm W} \|_2 \le 1$.
    For any unit vector $\bm x$ we have $| \bm x' \bm W_t| = | (\mathbf Q_{\bm W}' \bm x)' \bm Z_t|$,
    where $\| \mathbf Q_{\bm W}' \bm x \|_2 \le 1$.
    We may take $\mathbf Q_{\bm W}' \bm x = \| \mathbf Q_{\bm W}' \bm x \|_2 \bm v$ with $\| \bm v \|_2 = 1$,
    obtaining
    $\mathbb E[ | \bm x' \bm W_t|^m ] = \| \mathbf Q_{\bm W}' \bm x \|_2^m \mathbb E[ | \bm v' \bm Z_t|^m ]$.
    As a consequence, Assumption \assref{1}{asm:dist-ht} implies 
    $\sup_{\| \bm x \|_2 = 1} \mathbb E[ | \bm x' \bm W_t|^m ] \le C_Z$ and
    $\mathbb E[ \| \bm W_t \|_2^m] \le C_Z p^{m / 2}$.
 
    $(iv)$ The conclusion is obvious if $c_{K+1} = 0$, so assume that $c_{K+1} > 0$.
    As $\bm u_t = \bm \Sigma_u^{1/2} \bm W_t$ almost surely,
    we have for any unit vector $\bm x$ that $| \bm x' \bm u_t| = | (\bm \Sigma_u^{1/2} \bm x)' \bm W_t|$.
    Since Assumption \ref{asm:eigen} implies 
    $\| \bm \Sigma_u^{1/2} \bm x \|_2 \le \| \bm \Sigma_u \|_2^{1/2} \le \sqrt{c_{K+1}}$,
    we have by Assumption \assref{1}{asm:dist-ht} that
    $\sup_{\| \bm v \|_2 = 1} \mathbb E[ | \bm v' \bm u_t|^m ] \le C_Z c_{K+1}^{m/2}$
    and $\mathbb E[ \| \bm u_t \|_2^m] \le C_Z c_{K+1}^{m/2} p^{m / 2}$.
 
    $(v)$ We first consider $\bm P_t' \bm \vartheta^*$.
    If $\bm \vartheta^* = \bm 0$, the conclusion trivially follows, so we assume otherwise.
    We can take $\bm v = \bm \vartheta^* / \| \bm \vartheta^* \|_2$ and write 
    $\mathbb E[ | \bm P_t' \bm \vartheta^* |^m ] 
    = \| \bm \vartheta^* \|_2^m \mathbb E[ | \bm v' \bm P_t|^m ]$.
    As Lemma \ref{lem:pcr} and Assumption \assref{1}{asm:dist-ht} imply that 
    $\| \bm \vartheta^* \|_2 \le \sqrt{C_Z}$, we obtain from part (i) that
    $\sup_{\| \bm v \|_2 = 1} \mathbb E[ | \bm P_t' \bm \vartheta^*|^m ] \le C_Z^{1 + m/2}$.
    Next use the elementary convexity inequality $|a - b|^q \le 2^{q-1} (|a|^q + |b|^q)$ for $q \ge 1$ 
    to obtain
    \[
        |Y_t - \bm P_t' \bm \vartheta^*|^m \le 2^{m-1} (|Y_t|^m + |\bm P_t' \bm \vartheta^*|^m).
    \]
    As a consequence
    \begin{equation} \label{eq:ht-pe}
        \mathbb E[|Y_t - \bm P_t' \bm \vartheta^*|^m] 
        \le 2^{m-1} ( \mathbb E[|Y_t|^m] + \mathbb E[|\bm P_t' \bm \vartheta^*|^m])
        \le 2^{m-1} C_Z (1 + C_Z^{m/2}).
    \end{equation}
    Finally, we see that by Cauchy-Schwarz
    \[
        \mathbb E[ | \bm v' (Y_t - \bm P_t' \bm \vartheta^*) \bm P_t |^{m/2} ]
        = \mathbb E[ | Y_t - \bm P_t' \bm \vartheta^* |^{m/2} | \bm v' \bm P_t |^{m/2} ]
        \le \mathbb E[ | Y_t - \bm P_t' \bm \vartheta^* |^m ]^{1/2} \mathbb E[ | \bm v' \bm P_t |^m ]^{1/2}.
    \]
    As a consequence, we obtain from $(i)$ and \eqref{eq:ht-pe}
    \[
        \sup_{\| \bm v \|_2 = 1} \mathbb E[ | \bm v' (Y_t - \bm P_t' \bm \vartheta^*) \bm P_t |^{m/2} ]
        \le 2^{(m - 1) / 2} C_Z \sqrt{1 + C_Z^{m/2}},
    \]
    and then also immediately $\mathbb E [ \| (Y_t - \bm P_t' \bm \vartheta^*) \bm P_t \|_2^{m/2} ] 
    \le 2^{(m - 1) / 2} C_Z \sqrt{1 + C_Z^{m/2}} K^{m/4}$.
\end{proof}

\begin{lem} \label{lem:berbee}
    Let $\{ \bm W_t \}_{t=1}^T$ be a stationary sequence of $d$-dimensional zero-mean random vectors
    with beta-mixing coefficients $\beta(l)$.
    Fix $1 \le t < s \le T$ such that $h = s - t$.
    Then for every $r  > 2$ such that $\| \bm W_t \|_{L_r} < \infty$, we have
    \[
        | \mathbb E[ \bm W_t' \bm W_s] | \le 2 \| \bm W_t \|_{L_r}^2 \beta(h)^{1 - 2/r}.
    \]
\end{lem}
\begin{proof}
    By Berbee's Lemma \citep[Lemma 1.1]{Bosq:1998}, there exists a copy $\bm W_s^*$ that satisfies
    (i) it has the same marginal distribution as $\bm W_s$, (ii) it is independent of $\bm W_t$ and (iii)
    $\mathbb P( \bm W_s^* \neq \bm W_s) \le \beta(h)$.
    As $\mathbb E[\bm W_s^*] = \bm 0$ and $\bm W_s^*$ is independent of $\bm W_t$, 
    we have $\mathbb E[ \bm W_t' \bm W_s^*] = 0$,
    implying $\mathbb E[ \bm W_t' \bm W_s] = \mathbb E[ \bm W_t' (\bm W_s - \bm W_s^*)]$.
    Then note that by Cauchy-Schwarz and the triangle inequality
    \begin{align}
        | \mathbb E[ \bm W_t' (\bm W_s - \bm W_s^*) ] |
        &= | \mathbb E[ \bm W_t' (\bm W_s - \bm W_s^*) \mathbbm 1_{\{\bm W_s \ne \bm W_s^*\}} ] | \notag \\
        &\le \mathbb E[ \| \bm W_t \|_2 ( \| \bm W_s \|_2 + \| \bm W_s^* \|_2 ) 
        \mathbbm 1_{\{\bm W_s \ne \bm W_s^*\}} ] \notag \\
        &= \mathbb E[ \| \bm W_t \|_2 \| \bm W_s \|_2 \mathbbm 1_{\{\bm W_s \ne \bm W_s^*\}} ]
        + \mathbb E[ \| \bm W_t \|_2 \| \bm W_s^* \|_2 \mathbbm 1_{\{\bm W_s \ne \bm W_s^*\}} ]. 
        \label{eq:w}
    \end{align}
    Fix $r > 2$, let $q = r / (r - 2)$ such that $1/r + 1/r + 1/q = 1$, 
    and apply Hölder's inequality to the first term
    \begin{align*}
        \mathbb E[ \| \bm W_t \|_2 \| \bm W_s \|_2 \mathbbm 1_{\{\bm W_s \ne \bm W_s^*\}} ]
        &\le \| \| \bm W_t \|_2 \|_{L_r} \| \| \bm W_s \|_2 \|_{L_r} \| 
        \mathbbm 1_{\{\bm W_s \ne \bm W_s^*\}} \|_{L_q} \\
        &= \| \bm W_t \|_{L_r} \| \bm W_s \|_{L_r} \| 
        \mathbbm 1_{\{\bm W_s \ne \bm W_s^*\}} \|_{L_q} \\
        &\le \| \bm W_t \|_{L_r} \| \bm W_s \|_{L_r} \| 
        \beta(h)^{1 - 2 / r},
    \end{align*}
    where the last inequality follows as 
    \[
        \| \mathbbm 1_{\{\bm W_s \ne \bm W_s^*\}} \|_{L_q} 
        = \mathbb E[ \mathbbm 1_{\{\bm W_s \ne \bm W_s^*\}}^q ]^{1/q}
        = \mathbb P[ \{\bm W_s \ne \bm W_s^*\} ]^{1 - 2/r}
        \le \beta(h)^{1 - 2/r}.
    \]
    Analogous arguments apply to the second term in \eqref{eq:w} as 
    $\| \bm W_s \|_{L_r} = \| \bm W_s^* \|_{L_r}$, yielding
    \begin{align*}
        | \mathbb E[ \bm W_t' \bm W_s] |
        &\le \mathbb E[ \| \bm W_t \|_2 \| \bm W_s \|_2 \mathbbm 1_{\{\bm W_s \ne \bm W_s^*\}} ]
        + \mathbb E[ \| \bm W_t \|_2 \| \bm W_s^* \|_2 \mathbbm 1_{\{\bm W_s \ne \bm W_s^*\}} ] \\
        &\le 2 \| \bm W_t \|_{L_r} \| \bm W_s \|_{L_r} \beta(h)^{1 - 2/r}.
    \end{align*}
    The result then follows from stationarity.
\end{proof}

\begin{lem} \label{lem:ht-cov-conc-b}
    Let $\{\bm Z_t\}_{t=1}^T$ be a stationary sequence of $d$-dimensional random vectors and define
    $\mathbb E(\bm Z_t \bm Z_t') = \bm \Sigma_z$.
    Suppose that $(i)$ there exists a constant $C_0 > 0$ and a deterministic quantity $B \ge 1$ that may depend
    on $T$, such that $\| \bm \Sigma_z \|_2 \le C_0$ and $\|\bm Z_t\|_2\le B \sqrt d$ almost surely; 
    $(ii)$ there exists a constant $m > 4$ and a positive constant $C_Z$ such that 
    $\sup_{\| \bm v \|_2 = 1} \mathbb E( |\bm v' \bm Z_t|^m) \le C_Z$;
    $(iii)$ the beta-mixing coefficients of the sequence satisfy $\beta(l) \le \exp(-C_\beta l^{r_\beta})$
    for some $C_\beta > 0$ and $r_\beta \ge 1$.

    Then for every $\eta > 0$ there exists a positive constant $C_\eta > 0$ such that for any $T \ge 2$
    \[
        \left\| \frac1T \sum_{t=1}^T \bm Z_t \bm Z_t' - \bm \Sigma_z \right\|_2
        \le C_\eta \left( \sqrt{\frac{ d (\log(2d) + \log(T))}{T}}
        + B^2 \frac{d \log(T)^2 (\log(2d) + \log(T))}{T} \right), 
    \]
    holds with probability at least $1 - T^{-\eta}$.
\end{lem}
\begin{proof}
    Define $\mathbf M_t = \bm Z_t \bm Z_t' - \bm \Sigma_z$.
    We wish to apply Theorem 1 of \citet{banna-merlevede-youssef-2016} to this sequence of matrices. 
    We begin by verifying the conditions of the theorem.
    First, $\mathbf M_t$ is symmetric and mean zero.
    Second, we have that
    \begin{equation} \label{eq:m-bound-cov}
        \| \mathbf M_t \|_2 
        \le \|\bm Z_t \bm Z_t'\|_2 + \| \bm \Sigma_z \|_2
        \le B^2 d + C_0
        \le C_1 B^2 d \equiv R,
    \end{equation}
    almost surely by (i), as $\| \bm Z_t \bm Z_t'\|_2 = \| \bm Z_t\|_2^2$,
    and the last step follows for a suitable constant $C_1 > 0$ as $B, d \ge 1$.
    Third, (iii) guarantees that the $\beta$-mixing conditions for the theorem are satisfied,
    since $\beta(l) \le \exp(-C_\beta l^{r_\beta}) \le \exp(-C_\beta(l-1))$ for $l \ge 1$.
    Theorem 1 of \citet{banna-merlevede-youssef-2016} and the union bound then imply that
    there exists a constant $C_2 > 0$ such that for any $x > 0$ we have
    \[
        \mathbb P\left( \left\|\sum_{t=1}^T \mathbf M_t \right\|_2 \le x \right)
        \ge 1 - 2 d \exp\left( -\frac{C_2 x^2}{ v^2 T + C_\beta^{-1} R^2 + x R \gamma(C_\beta, T) } \right),
    \]
    where 
    \begin{equation} \label{eq:bmy-cov-v}
        v^2
        = \sup_{S \subset\{1,\ldots,T\}} \frac1{|S|}
        \lambda_{\max} \left( \mathbb E\left[ \Big(\sum_{t \in S} \mathbf M_t \Big)' 
        \Big(\sum_{t \in S} \mathbf M_t \Big) \right] \right),
    \end{equation}
    and
    \begin{equation} \label{eq:bmy-cov-gamma}
        \gamma(C_\beta, T) = \frac{\log(T)}{\log(2)} \max \left( 2, \frac{32 \log(T)}{C_\beta \log(2)} \right).
    \end{equation}
    We wish to obtain an upper bound that holds with probability at least $1 - T^{-\eta}$.
    To this end, we will choose $x$ large enough to ensure 
    \begin{equation} \label{eq:eta}
        \frac{C_2 x^2}{v^2 T + C_\beta^{-1} R^2 + x R \gamma(C_\beta,T)}
        \ge (1 + \eta) \log(T) + \log(2 d)
        \equiv \eta_T.
    \end{equation}
    It is sufficient to choose 
    \[
        x = A \left( \sqrt{v^2 T \eta_T} + R \sqrt{\eta_T} + R \gamma(C_\beta, T) \eta_T \right),
    \]
    where $A > 0$, which guarantees that $v^2 T \le C_2 x^2 / (3 \eta_T)$, 
    $C_\beta^{-1} R^2 \le C_2 x^2 / (3 \eta_T)$ and $x R \gamma(C_\beta, T) \le C_2 x^2 / (3 \eta_T)$ 
    if we choose $A$ sufficiently large, implying that 
    \[
        v^2 T + C_\beta^{-1} R^2 + x R \gamma(C_\beta, T)
        \le C_2 \frac{x^2}{\eta_T}.
    \]
    With this choice, the theorem implies, after dividing by $T$, that there exists a constant $C_3$ such that
    \begin{align}
        \left\| \frac1T \sum_{t=1}^T \mathbf M_t \right\|_2
        &\le \sqrt{\frac{3}{C_2}} \left( \sqrt{ \frac{v^2 \eta_T}{T}} 
        + R \frac{\sqrt{\eta_T}}{T} 
        + R \frac{\gamma(C_\beta, T) \eta_T}{T} \right) \notag \\
        &\le C_3 \left( \sqrt{ \frac{v^2 \eta_T}{T}} + R \frac{\gamma(C_\beta, T) \eta_T}{T} \right), 
        \label{eq:bmy-cov}
    \end{align}
    with probability at least $1 - 2 d e^{-\eta_T} = 1 - T^{-1 -\eta} \ge 1 - T^{-\eta}$,
    because $\gamma(C_\beta, T) \ge 1$ and $\eta_T \ge 1$.

    It remains to bound $v^2$.
    Consider \eqref{eq:bmy-cov-v}, fix a non-empty set $S \subset \{1, \ldots, T\}$ 
    and fix a unit vector $\bm x$.
    Then
    \begin{align}
        \bm x'\mathbb E \left[ \Big( \sum_{t \in S} \mathbf M_t \Big)'
        \Big( \sum_{t \in S} \mathbf M_t \Big) \right] \bm x
        &= \mathbb E \left( \Big\| \sum_{t \in S} \mathbf M_t \bm x \Big\|_2^2 \right) \notag \\
        &= \sum_{t \in S} \mathbb E[ \| \mathbf M_t \bm x \|_2^2 ]
        + 2 \sum_{t < s \in S} \mathbb E[ \bm x'\mathbf M_t' \mathbf M_s \bm x ]. \label{eq:bmy-cov-v-sum}
    \end{align}
    To bound the diagonal part of \eqref{eq:bmy-cov-v-sum}, first note that
    \[
        \mathbf M_t' \mathbf M_t
        = \bm Z_t \bm Z_t' \bm Z_t \bm Z_t'
        - \bm Z_t \bm Z_t' \bm \Sigma_z
        - \bm \Sigma_z \bm Z_t \bm Z_t'
        + \bm\Sigma_z' \bm \Sigma_z,
    \]
    which implies
    \[ 
        \mathbb E(\mathbf M_t' \mathbf M_t)
        = \mathbb E( \bm Z_t \bm Z_t' \bm Z_t \bm Z_t')
        - \bm\Sigma_z' \bm \Sigma_z
        = \mathbb E( \| \bm Z_t \|_2^2 \bm Z_t \bm Z_t')
        - \bm\Sigma_z' \bm \Sigma_z.
    \]
    We then have
    \[
        \mathbb E[ \|\mathbf M_t \bm x\|_2^2 ]
        = \bm x'\mathbb E[ \mathbf M_t' \mathbf M_t ] \bm x
        = \bm x' \mathbb E( \| \bm Z_t \|_2^2 \bm Z_t \bm Z_t') \bm x - \bm x' \bm\Sigma_z' \bm \Sigma_z \bm x.
    \]
    By Cauchy-Schwarz
    \[
        \bm x' \mathbb E( \| \bm Z_t \|_2^2 \bm Z_t \bm Z_t') \bm x
        = \mathbb E( \| \bm Z_t \|_2^2 (\bm x' \bm Z_t)^2 )
        \le \mathbb E( \| \bm Z_t \|_2^4)^{1/2} \mathbb E( (\bm x' \bm Z_t)^4 )^{1/2}
        \le C_Z d, 
    \]
    because $(ii)$ implies $\mathbb E( \|\bm Z_t\|_2^m) \le C_Z d^{m/2}$
    by the same arguments as in part $(i)$ of Lemma \ref{lem:ht}.
    As a consequence we obtain
    \begin{equation} \label{eq:m-var}
        \mathbb E[ \|\mathbf M_t \bm x\|_2^2 ]
        \le C_Z d + \| \bm \Sigma_z \bm x \|^2
        \le C_4 d,
    \end{equation}
    for a suitable constant $C_4 > 0$, by (i).
    As a consequence, we see that the diagonal part is bounded as
    \begin{equation} \label{eq:bmy-cov-v-diag}
        \sum_{t \in S} \mathbb E[ \| \mathbf M_t \bm x\|_2^2 ]
        \le C_4 | S | d.
    \end{equation}
    Now we turn to the off-diagonal part of \eqref{eq:bmy-cov-v-sum}.
    Define $\bm W_r = \mathbf M_r \bm x$ for all $r \in S$ and consider two values $\bm W_s$ and $\bm W_t$,
    spaced $h = s - t$ time periods apart.
    Note that $\mathbb E[ \bm W_r ] = \mathbb E[ \mathbf M_r \bm x ] = 0$.
    By Lemma \ref{lem:berbee} with $r = m / 2$, we obtain
    \[
        | \mathbb E[ \bm W_t' \bm W_s] |
        \le 2 \| \bm W_t \|_{L_{m/2}}^2 \beta(h)^{1 - 4/m}
        \le C_5 d \beta(h)^{1 - 4/m},
    \]
    for some constant $C_5 > 0$, where the latter inequality follows as
    \begin{align*}
        \| \bm W_t \|_{L_{m/2}}
        = \| \bm Z_t \bm Z_t' \bm x - \bm \Sigma_z \bm x \|_{L_{m/2}}
        &\le \| \| \bm Z_t \|_2 (\bm Z_t' \bm x) \|_{L_{m/2}} + \| \bm \Sigma_z \bm x \|_2 \\
        &\le \| \| \bm Z_t \|_2 \|_{L_m} \| \bm Z_t' \bm x \|_{L_m} + C_0,
    \end{align*}
    by Hölder's inequality and $(i)$, and then applying $(ii)$ to the first term.
    We can then obtain a bound on the off-diagonal part of \eqref{eq:bmy-cov-v-sum} by
    noting that there are at most $|S|$ pairs with $h = s - t$
    \begin{align*}
        \sum_{t < s \in S} \mathbb E[ \bm x'\mathbf M_t' \mathbf M_s \bm x ]
        &\le C_5 |S| d \sum_{h=1}^{\infty}\beta(h)^{1 - 4/m} \\
        &\le C_5 |S| d \sum_{h=1}^{\infty} \exp(- (1 - 4 / m) C_\beta h^{r_\beta}) \\
        &\le C_6 |S| d,
    \end{align*}
    where we used $(iii)$.
    Combining this with the diagonal bound from \eqref{eq:bmy-cov-v-diag} 
    and putting into \eqref{eq:bmy-cov-v-sum}
    \begin{align*}
        \bm x'\mathbb E \left[ \Big( \sum_{t \in S}\mathbf M_t \Big)'
        \Big( \sum_{t \in S}\mathbf M_t \Big) \right] \bm x
        &= \sum_{t \in S} \mathbb E[ \| \mathbf M_t \bm x \|_2^2 ]
        + 2 \sum_{t < s \in S} \mathbb E[ \bm x'\mathbf M_t' \mathbf M_s \bm x ] \\
        &\le C_4 |S| d + 2 C_6 |S| d \\
        &\le C_v |S| d,
    \end{align*}
    with $C_v = C_4 + 2 C_6$.
    Finally, dividing both sides by $|S|$,
    and taking the supremum over all unit vectors $\bm x$ and all non-empty subsets 
    $S \subset \{ 1, \ldots, T \}$, we obtain
    \begin{equation} \label{eq:bmy-cov-v-bound}
        v^2 \le C_v d.
    \end{equation}

    Finally, going back to \eqref{eq:bmy-cov} and inserting \eqref{eq:m-bound-cov} along with 
    \eqref{eq:bmy-cov-v-bound} implies that there exists a constant $C > 0$ such that
    \begin{align*}
        \left\| \frac1T \sum_{t=1}^T \mathbf M_t \right\|_2
        &\le C_3 \left( \sqrt{ \frac{v^2 \eta_T}{T}} + R \frac{\gamma(C_\beta, T) \eta_T}{T} \right) \\
        &\le C_3 \left( \sqrt{C_v} \sqrt{\frac{ d \eta_T}{T}}
        + C_1 B^2 \frac{d \gamma(C_\beta, T) \eta_T}{T} \right) \\
        &\le C \left( \sqrt{\frac{ d (\log(2d) + \log(T))}{T}} 
        + B^2 \frac{d \log(T)^2 (\log(2d) + \log(T))}{T} \right), 
    \end{align*}
    with probability at least $1 - T^{-\eta}$, 
    where we used \eqref{eq:eta} and \eqref{eq:bmy-cov-gamma} in the final step.
\end{proof}

\begin{lem} \label{lem:ht-cov-conc}
    Let $\{\bm Z_t \}_{t=1}^T$ be a stationary sequence of $d$-dimensional random vectors
    and define $\bm \Sigma_z = \mathbb E[\bm Z_t \bm Z_t']$.
    Suppose
    $(i)$ there exists a constant $m > 4$ and a positive constant $C_Z$ such that 
    $\sup_{\| \bm v \|_2 = 1} \mathbb E( |\bm v' \bm \Sigma_z^{-1/2} \bm Z_t|^m) \le C_Z$;
    $(ii)$ the beta-mixing coefficients of the sequence satisfy $\beta(l) \le \exp(-C_\beta l^{r_\beta})$
    for some $C_\beta > 0$ and $r_\beta \ge 1$.

    Then for every $\eta > 0$ and every $\nu > (1 + \eta) / m$
    there exists a constant $C_\eta > 0$ such that, for all sufficiently large $T$
    \[
        \left\| \frac1T \sum_{t=1}^T \bm Z_t \bm Z_t' - \bm \Sigma_z \right\|_2
        \le C_\eta \| \bm \Sigma_z \|_2 \left( \sqrt{\frac{ d (\log(2d) + \log(T))}{T}}
        + T^{2 \nu} \frac{d \log(T)^2 (\log(2d) + \log(T))}{T} \right), 
    \]
    holds with probability at least $1 - T^{-\eta}$.
\end{lem}
\begin{proof}
    If $\bm \Sigma_z = \mathbf 0$, $\bm Z_t = 0$ almost surely and the result follows trivially,
    so we assume otherwise.
    Define $\bm W_t = \bm \Sigma_z^{-1/2} \bm Z_t$ and let $\bm \Pi_z$ be the orthogonal
    projector onto the range of $\bm \Sigma_z$ as in the proof of Lemma \ref{lem:conc-bosq}.
    As $\bm Z_t = \bm \Sigma_z^{1/2} \bm \Sigma_z^{-1/2} \bm Z_t$ almost surely, 
    we have almost surely
    \begin{align}
        \left\| \frac{1}{T} \sum_{t = 1}^T \bm Z_t \bm Z_t' - \bm \Sigma_z \right\|_2
        &= \left\| \bm \Sigma_z^{1/2} 
        \left( \frac{1}{T} \sum_{t = 1}^T \bm W_t \bm W_t' 
        - \bm \Pi_z \right) \bm \Sigma_z^{1/2} \right\|_2 \notag \\
        &\le \| \bm \Sigma_z \|_2 \left\| \frac{1}{T} \sum_{t = 1}^T \bm W_t \bm W_t' 
        - \bm \Pi_z \right\|_2. \label{eq:iso-ht}
    \end{align}
    It is therefore sufficient for us to bound the right hand side.

    Define
    \begin{equation} \label{eq:w-trunc}
        \overline{\bm W}_t
        = \bm W_t \mathbbm 1_{ \{ \| \bm W_t \|_2 \le \sqrt d T^\nu \} },
    \end{equation}
    so that $\| \overline{\bm W}_t \|_2  \le \sqrt d T^\nu$.
    Then define the event $\mathcal E_T = \{ \max_{1 \le t \le T} \| \bm W_t \|_2 \le \sqrt d T^\nu \}$
    and note that the union bound followed by Markov's inequality implies
    \[
        \mathbb P(\mathcal E_T^c) 
        \le \sum_{t = 1}^T \mathbb P( \| \bm W_t \|_2 > \sqrt d T^\nu)
        \le \sum_{t = 1}^T \frac{ \mathbb E[ \| \bm W_t \|_2^m ] }{ d^{m / 2} T^{\nu m} }.
    \]
    Since $\nu m - 1 > \eta$, define $\eta' = (\eta + \nu m - 1)/2$ so that $\eta < \eta' < \nu m - 1$.
    Then, as $(i)$ implies $\mathbb E( \| \bm \Sigma_z^{-1/2} \bm Z_t \|_2^m) \le C_Z d^{m / 2}$
    by the same argument as in part $(i)$ of Lemma \ref{lem:ht}, we have
    \begin{equation} \label{eq:prob-trunc-cov}
        \mathbb P(\mathcal E_T^c) 
        \le \sum_{t = 1}^T C_Z T^{-\nu m} 
        = C_Z T^{1 - \nu m}
        \le T^{-\eta'}.
    \end{equation}

    Next add and subtract to decompose the object of interest as
    \begin{align*}
        \frac{1}{T} \sum_{t = 1}^T \bm W_t \bm W_t' - \bm \Pi_z 
        &= \frac{1}{T} \sum_{t = 1}^T \overline{\bm W}_t \overline{\bm W}_t' 
        - \mathbb E\left[ \overline{\bm W}_t \overline{\bm W}_t' \right] \\
        &+ \mathbb E\left[ \overline{\bm W}_t \overline{\bm W}_t' \right]
        - \mathbb E[ \bm W_t \bm W_t'] \\
        &+\frac{1}{T} \sum_{t = 1}^T \bm W_t \bm W_t' 
        - \frac{1}{T} \sum_{t = 1}^T \overline{\bm W}_t \overline{\bm W}_t',
    \end{align*}
    where $\mathbb E[ \bm W_t \bm W_t' ] = \bm \Pi_z$. 
    Note that on the event $\mathcal E_T$, $\bm W_t = \overline{\bm W}_t$ for all $t$
    so the last difference is zero.
    Hence, on $\mathcal E_T$, we have
    \begin{align}
        \left\| \frac{1}{T} \sum_{t = 1}^T \bm W_t \bm W_t' - \bm \Pi_z \right\|_2
        &\le \left\| \frac{1}{T} \sum_{t = 1}^T \overline{\bm W}_t \overline{\bm W}_t' 
        - \mathbb E\left[ \overline{\bm W}_t \overline{\bm W}_t' \right] \right\|_2 \notag \\
        &+ \left\| \mathbb E\left[ \overline{\bm W}_t \overline{\bm W}_t' \right]
        - \mathbb E( \bm W_t \bm W_t') \right\|_2 \notag \\
        &\equiv \| \mathbf A_T \|_2 + \| \mathbf B_T \|_2. \label{eq:decomp-trunc-cov}
    \end{align}

    We first bound $\| \mathbf B_T \|_2$.
    By \eqref{eq:w-trunc}
    \[
        \bm W_t \bm W_t' - \overline{\bm W}_t \overline{\bm W}_t'
        = \bm W_t \bm W_t' \mathbbm 1_{ \{ \| \bm W_t \|_2 > \sqrt d T^\nu \} },
    \]
    so that
    \[
        \mathbf B_T
        = \mathbb E\left[ \overline{\bm W}_t \overline{\bm W}_t' - \bm W_t \bm W_t' \right]
        = -\mathbb E\left[ \bm W_t \bm W_t' \mathbbm 1_{ \{ \| \bm W_t \|_2 > \sqrt d T^\nu \} } \right]
    \]
    and thus
    \[
        \| \mathbf B_T \|_2
        = \left\| \mathbb E\left[ \bm W_t \bm W_t' 
        \mathbbm 1_{ \{ \| \bm W_t \|_2 > \sqrt d T^\nu \} } \right] \right\|_2
        = \sup_{\| \bm x \|_2 = 1} \mathbb E \left[ (\bm x'\bm W_t)^2 
        \mathbbm 1_{ \{ \| \bm W_t \|_2 > \sqrt d T^\nu \} } \right].
    \]
    For a given $\bm x$ such that $\| \bm x \|_2 = 1$, H\"older's inequality gives
    \begin{align*}
        \mathbb E\left[ (\bm x' \bm W_t)^2 \mathbbm 1_{ \{ \|\bm W_t\|_2 > \sqrt d T^\nu \} } \right]
        &\le \| (\bm x' \bm W_t)^2 \|_{L_r} \| \mathbbm 1_{ \{ \|\bm W_t\|_2 > \sqrt d T^\nu \} } \|_{L_s} \\
        &= ( \mathbb E[ |\bm x' \bm W_t |^m ] )^{2/m} (\mathbb P ( \|\bm W_t\|_2 > \sqrt d T^\nu ) )^{1-2/m},
    \end{align*}
    with $r = m / 2$ and $s = m / (m - 2)$.
    Using Markov's inequality as before followed by (i), we obtain
    \begin{align*}
        \mathbb E\left[ (\bm x' \bm W_t)^2 \mathbbm 1_{ \{ \|\bm W_t\|_2 > \sqrt d T^\nu \} } \right]
        &\le ( \mathbb E[ |\bm x' \bm W_t |^m ] )^{2/m}
        \left( \frac{ \mathbb E[ \| \bm W_t \|_2^m ] }{ d^{m / 2} T^{\nu m} } \right)^{1-2/m} \\
        &\le C_Z^{2/m} \left( \frac{ C_Z }{ T^{\nu m} } \right)^{1-2/m}.
    \end{align*}
    As a consequence, we obtain
    \begin{equation} \label{eq:bt-cov}
        \| \mathbf B_T \|_2
        \le C_Z^{2/m} \left( \frac{ C_Z }{ T^{\nu m} } \right)^{1-2/m}
        = C_Z T^{-\nu (m - 2)}.
    \end{equation}
    
    Next we bound $\| \mathbf A_T \|_2$.
    Before applying Lemma \ref{lem:ht-cov-conc-b} to $\overline{\bm W}_t$, 
    we verify that its conditions are satisfied.
    First, we may note that
    \[
        \left\| \mathbb E\left[ \overline{\bm W}_t \overline{\bm W}_t' \right] \right\|_2
        \le \left\| \mathbb E[ \bm W_t \bm W_t' ] \right\|_2
        = \| \bm \Pi_z \|_2
        \le 1.
    \]
    Second, we saw before that $\| \overline{\bm W}_t \|_2 \le \sqrt d T^\nu$ 
    so that we may take $B = T^\nu$ in the lemma.
    Third, (i) implies the necessary moment conditions for $\overline{\bm W}_t$.
    Finally, the beta-mixing coefficients of $\overline{\bm W}_t$ are bounded by those of $\bm W_t$.
    Recall that $\eta' = (\eta + \nu m - 1)/2$.
    Lemma \ref{lem:ht-cov-conc-b} then implies that there exists a positive constant $C > 0$ such that
    \begin{equation} \label{eq:at-cov}
        \| \mathbf A_T \|_2
        \le C \left( \sqrt{\frac{ d (\log(2d) + \log(T))}{T}}
        + T^{2 \nu} \frac{d \log(T)^2 (\log(2d) + \log(T))}{T} \right), 
    \end{equation}
    holds with probability at least $1 - T^{-\eta'}$.

    Hence, on the intersection of $\mathcal E_T$ and the event of Lemma \ref{lem:ht-cov-conc-b},
    we have by combining \eqref{eq:at-cov} and \eqref{eq:bt-cov} with \eqref{eq:decomp-trunc-cov}
    \begin{align*}
        \left\| \frac{1}{T} \sum_{t = 1}^T \bm W_t \bm W_t' - \bm \Pi_z \right\|_2
        &\le \| \mathbf A_T \|_2 + \| \mathbf B_T \|_2 \notag \\
        &\le C \sqrt{\frac{ d (\log(2d) + \log(T))}{T}} \\
        &+ C T^{2 \nu} \frac{d \log(T)^2 (\log(2d) + \log(T))}{T} \\
        &+ C_Z T^{-\nu (m - 2)}.
    \end{align*}
    Notice that the last term is dominated by the second, 
    since $d \log(T)^2 \ge 1$ and $\log(2d) + \log(T) \ge 1$ for all sufficiently large $T$ and
    \[
        T^{-\nu (m - 2)} \le T^{2 \nu} T^{-1},
    \]
    because $\nu > 1 / m$ implies $\nu (m - 2) + 2 \nu = \nu m > 1$.
    We therefore have, on the same events, 
    that there exists a constant $C_\eta > 0$ such that for all $T$ sufficiently large
    \[
        \left\| \frac{1}{T} \sum_{t = 1}^T \bm W_t \bm W_t' - \bm \Pi_z \right\|_2
        \le C_\eta \left( \sqrt{\frac{ d (\log(2d) + \log(T))}{T}}
        + T^{2 \nu} \frac{d \log(T)^2 (\log(2d) + \log(T))}{T} \right).
    \]
    It only remains to work out the probability of the intersection of the two events.
    The truncation event fails with probability at most $T^{-\eta'}$ from \eqref{eq:prob-trunc-cov},
    and the bound from \eqref{eq:at-cov} fails with probability at most $T^{-\eta'}$.
    Both events thus occur with probability at least $1 - 2 T^{-\eta'} \ge 1 - T^{-\eta}$.
    Combining this result with \eqref{eq:iso-ht} then proves the claim.
\end{proof}

\begin{lem} \label{lem:ht-conc-b}
    Let $\{\bm Z_t\}_{t=1}^T$ be a stationary sequence of $d$-dimensional zero-mean random vectors.
    Suppose that $(i)$ there exists a deterministic quantity $B \ge 1$ that may depend on $T$
    such that $\|\bm Z_t\|_2\le B \sqrt d$ almost surely;
    $(ii)$ there exists a constant $m > 2$ and a positive constant $C_Z$ such that 
    $\mathbb E( \| \bm Z_t \|_2^m) \le C_Z d^{m/2}$;
    $(iii)$ the beta-mixing coefficients of the sequence satisfy $\beta(l) \le \exp(-C_\beta l^{r_\beta})$
    for some $C_\beta > 0$ and $r_\beta \ge 1$.

    Then for every $\eta > 0$ there exists a positive constant $C_\eta > 0$ such that for any $T \ge 2$
    \[
        \left\| \frac1T \sum_{t=1}^T \bm Z_t \right\|_2
        \le C \left( \sqrt{d \frac{(\log(2d) + \log(T))}{T}} 
        + B \sqrt{d} \frac{\log(T)^2 (\log(2d) + \log(T))}{T} \right),
    \]
    holds with probability at least $1 - T^{-\eta}$.
\end{lem}
\begin{proof}
    We follow a similar strategy as in the proof of Lemma \ref{lem:ht-cov-conc-b}.
    Define the $(d+1) \times (d+1)$ matrix
    \[
        \mathbf M_t =
        \begin{pmatrix}
            \mathbf 0_{d \times d} & \bm Z_t \\
            \bm Z_t' & 0
        \end{pmatrix}.
    \]
    We wish to apply Theorem 1 of \citet{banna-merlevede-youssef-2016} to this sequence of matrices. 
    We begin by verifying the conditions of the theorem.
    First, $\mathbf M_t$ is symmetric and mean zero.
    Second, we have that
    \begin{equation} \label{eq:m-bound}
        \| \mathbf M_t \|_2 
        = \|\bm Z_t \|_2
        \le B \sqrt{d} 
        \equiv R,
    \end{equation}
    almost surely by (i).
    Third, (iii) guarantees that the $\beta$-mixing conditions for the theorem are satisfied,
    since $\beta(l) \le \exp(-C_\beta l^{r_\beta}) \le \exp(-C_\beta(l-1))$ for $l \ge 1$.
    Following the proof of Lemma \ref{lem:ht-cov-conc-b} and applying Theorem 1 of 
    \citet{banna-merlevede-youssef-2016} then implies that there exists a constant $C_1 > 0$ such that 
    \begin{equation} \label{eq:bmy}
        \left\| \frac1T \sum_{t=1}^T \mathbf M_t \right\|_2
        \le C_1 \left( \sqrt{ \frac{v^2 (\log(2d) + \log(T))}{T}} 
        + \frac{R \log(T)^2 (\log(2d) + \log(T))}{T} \right), 
    \end{equation}
    with probability at least $1 - T^{-\eta}$,
    where we used $d + 1 \le 2d$ and $v^2$ is as \eqref{eq:bmy-cov-v}.

    We then require a bound on $v^2$.
    Let $\bm Q = \sum_{t \in S} \bm Z_t$ and notice that
    \[
        \Big(\sum_{t \in S} \mathbf M_t \Big)' \Big(\sum_{t \in S} \mathbf M_t \Big) = 
        \begin{pmatrix}
            \bm Q \bm Q'& \bm 0 \\
            \bm 0 & \| \bm Q \|_2^2
        \end{pmatrix}.
    \]
    As a consequence, 
    \[
        \lambda_{\max} \left( \mathbb E\left[ \Big(\sum_{t \in S} \mathbf M_t \Big)' 
        \Big(\sum_{t \in S} \mathbf M_t \Big) \right] \right)
        = \max \lbrace \lambda_{max}( \mathbb E[ \bm Q \bm Q'] ), \mathbb E [ \| \bm Q \|_2^2 ] \rbrace
        = \mathbb E[ \| \bm Q \|_2^2 ],
    \]
    as $\lambda_{max}( \bm Q \bm Q') \le \operatorname{tr}( \bm Q \bm Q') = \| \bm Q \|_2^2$.
    It then follows from \eqref{eq:bmy-cov-v} that
    \begin{equation} \label{eq:bmy-v}
        v^2
        \le \sup_{S \subset\{1,\ldots,T\}} \frac1{|S|}
        \mathbb E\Big[ \Big\| \sum_{t \in S} \bm Z_t \big\|_2^2 \Big].
    \end{equation}
    We thus consider
    \begin{equation} \label{eq:bmy-v-sum}
        \mathbb E\Big[ \Big\| \sum_{t \in S} \bm Z_t \big\|_2^2 \Big]
        = \sum_{t \in S} \mathbb E[ \| \bm Z_t \|_2^2 ] + 2 \sum_{t < s \in S} \mathbb E[ \bm Z_t' \bm Z_s ]
        \le | S | C_Z d + 2 \sum_{t < s \in S} | \mathbb E[ \bm Z_t' \bm Z_s ] |,
    \end{equation}
    by (ii).
    We now turn to the off-diagonal part of \eqref{eq:bmy-v-sum}.
    Fix $h = s - t$.
    By Lemma \ref{lem:berbee} with $r = m$, we obtain
    \[
        | \mathbb E[ \bm Z_t' \bm Z_s] |
        \le 2 \| \bm Z_t \|_{L_m}^2 \beta(h)^{1 - 2 / m}
        \le 2 C_Z^{2/m} d \beta(h)^{1 - 2 / m},
    \]
    as $\| \bm Z_t \|_{L_m} \le C_Z^{1/m} d^{1/2}$ by $(ii)$.
    As there are at most $|S|$ pairs with $h = s - t$,
    we thus obtain by (iii)
    \[
        \sum_{t < s \in S} \mathbb E[ \bm Z_t' \bm Z_s ]
        \le 2 C_Z^{2/m} |S| d \sum_{h=1}^{\infty}\beta(h)^{1 - 2/m}
        \le C_2 |S| d,
    \]
    for some constant $C_2 > 0$.
    We then have by \eqref{eq:bmy-v-sum}
    \[
        \mathbb E\Big[ \Big\| \sum_{t \in S} \bm Z_t \big\|_2^2 \Big] 
        \le | S | C_Z d + 2 C_2 | S | d 
        \le C_v |S| d,
    \]
    with $C_v = C_Z + 2 C_2$.
    Finally, dividing both sides by $|S|$, and taking the supremum over all non-empty subsets 
    $S \subset \{ 1, \ldots, T \}$, we obtain from \eqref{eq:bmy-v}
    \begin{equation} \label{eq:bmy-v-bound}
        v^2 \le C_v d.
    \end{equation}

    Finally, going back to \eqref{eq:bmy} and inserting \eqref{eq:m-bound} along with 
    \eqref{eq:bmy-v-bound} implies that there exists a constant $C > 0$ such that
    \begin{align*}
        \left\| \frac1T \sum_{t=1}^T \mathbf M_t \right\|_2
        &\le C_1 \left( \sqrt{ \frac{v^2 (\log(2d) + \log(T))}{T}} 
        + \frac{R \log(T)^2 (\log(2d) + \log(T))}{T} \right) \\
        &\le C \left( \sqrt{ \frac{d (\log(2d) + \log(T))}{T}} 
        + B \frac{\sqrt{d} \log(T)^2 (\log(2d) + \log(T))}{T} \right),
    \end{align*}
    with probability at least $1 - T^{-\eta}$, which implies the claim.
\end{proof}

\begin{lem} \label{lem:ht-conc}
    Let $\{\bm Z_t \}_{t=1}^T$ be a stationary sequence of $d$-dimensional zero-mean random vectors.
    Suppose
    $(i)$ there exists a constant $m > 2$ and a positive constant $C_Z$ such that 
    $\mathbb E( \| \bm Z_t \|_2^m) \le C_Z d^{m / 2}$;
    $(ii)$ the beta-mixing coefficients of the sequence satisfy $\beta(l) \le \exp(-C_\beta l^{r_\beta})$
    for some $C_\beta > 0$ and $r_\beta \ge 1$.

    Then for every $\eta > 0$ and every $\nu > (1 + \eta) / m$ 
    there exists a constant $C_\eta > 0$ such that, for all sufficiently large $T$
    \[
        \left\| \frac1T \sum_{t=1}^T \bm Z_t \right\|_2
        \le C_\eta \left( \sqrt{\frac{ d (\log(2d) + \log(T))}{T}}
        + T^\nu \frac{\sqrt d \log(T)^2 (\log(2d) + \log(T))}{T} \right), 
    \]
    holds with probability at least $1 - T^{-\eta}$.
\end{lem}
\begin{proof}
    We follow the proof of Lemma \ref{lem:ht-cov-conc} closely.
    Define
    \begin{equation} \label{eq:z-trunc}
        \overline{\bm Z}_t
        = \bm Z_t \mathbbm 1_{ \{ \| \bm Z_t \|_2 \le \sqrt d T^\nu \} },
    \end{equation}
    which implies $\| \overline{\bm Z}_t \|_2  \le \sqrt d T^\nu$.
    Then define the event $\mathcal E_T = \{ \max_{1 \le t \le T} \| \bm Z_t \|_2 \le \sqrt d T^\nu \}$
    and note that
    \[
        \mathbb P(\mathcal E_T^c) 
        \le \sum_{t = 1}^T \mathbb P( \| \bm Z_t \|_2 > \sqrt d T^\nu)
        \le \sum_{t = 1}^T \frac{ \mathbb E[ \| \bm Z_t \|_2^m ] }{ d^{m / 2} T^{\nu m} }.
    \]
    Since $\nu m - 1 > \eta$, define $\eta' = (\eta + \nu m - 1)/2$ so that $\eta < \eta' < \nu m - 1$.
    Then using (i), we obtain
    \begin{equation} \label{eq:prob-trunc}
        \mathbb P(\mathcal E_T^c) 
        \le \sum_{t = 1}^T C_Z T^{-\nu m} 
        = C_Z T^{1 - \nu m}
        \le T^{-\eta'},
    \end{equation}
    for all sufficiently large $T$.

    Next add and subtract to decompose the object of interest as
    \begin{align*}
        \frac1T \sum_{t=1}^T \bm Z_t
        &= \frac{1}{T} \sum_{t = 1}^T \overline{\bm Z}_t - \mathbb E\left[ \overline{\bm Z}_t \right] \\
        &+ \mathbb E\left[ \overline{\bm Z}_t \right] \\
        &+\frac{1}{T} \sum_{t = 1}^T \bm Z_t - \frac{1}{T} \sum_{t = 1}^T \overline{\bm Z}_t.
    \end{align*}
    Note that on the event $\mathcal E_T$, $\bm Z_t = \overline{\bm Z}_t$ for all $t$
    so the last difference is zero.
    Hence, on $\mathcal E_T$, we have
    \begin{equation}
        \left\| \frac{1}{T} \sum_{t = 1}^T \bm Z_t \right\|_2
        \le \left\| \frac{1}{T} \sum_{t = 1}^T \overline{\bm Z}_t 
        - \mathbb E\left[ \overline{\bm Z}_t \right] \right\|_2 
        + \left\| \mathbb E\left[ \overline{\bm Z}_t \right] \right\|_2
        \equiv \| \mathbf A_T \|_2 + \| \mathbf B_T \|_2. \label{eq:decomp-trunc}
    \end{equation}

    We first bound $\| \mathbf B_T \|_2$.
    By \eqref{eq:z-trunc}, 
    $\bm Z_t - \overline{\bm Z}_t = \bm Z_t \mathbbm 1_{ \{ \| \bm Z_t \|_2 > \sqrt d T^\nu \} }$.
    As $\mathbb E[ \bm Z_t ] = 0$, we thus obtain
    \[
        \mathbf B_T
        = \mathbb E\left[ \overline{\bm Z}_t \right]
        = \mathbb E\left[ \overline{\bm Z}_t - \bm Z_t \right]
        = -\mathbb E\left[ \bm Z_t \mathbbm 1_{ \{ \| \bm Z_t \|_2 > \sqrt d T^\nu \} } \right]
    \]
    and H\"older's inequality therefore gives us 
    \begin{align*}
        \| \mathbf B_T \|_2
        = \left\| \mathbb E\left[ \bm Z_t 
        \mathbbm 1_{ \{ \| \bm Z_t \|_2 > \sqrt d T^\nu \} } \right] \right\|_2 
        &\le \left\| \| \bm Z_t \|_{L_r} 
        \| \mathbbm 1_{ \{ \| \bm Z_t \|_2 > \sqrt d T^\nu \} } \|_{L_s} \right\|_2 \\
        &= \mathbb E \left[ \| \bm Z_t \|_2^m \right]^{1/m}
        \mathbb P( \| \bm Z_t \|_2 > \sqrt d T^\nu )^{1 - 1 / m},
    \end{align*}
    with $r = m$ and $s = m / ( m - 1)$.
    Using Markov's inequality as before followed by (i), we obtain
    \begin{align*}
        \left\| \mathbb E\left[ \bm Z_t 
        \mathbbm 1_{ \{ \| \bm Z_t \|_2 > \sqrt d T^\nu \} } \right] \right\|_2 
        &\le \mathbb E \left[ \| \bm Z_t \|_2^m \right]^{1/m}
        \left( \frac{ \mathbb E[ \| \bm Z_t \|_2^m ] }{ d^{m / 2} T^{\nu m} } \right)^{1 - 1/m} \\
        &\le C_Z^{1/m} \sqrt d \left( \frac{ C_Z }{ T^{\nu m} } \right)^{1 - 1/m}.
    \end{align*}
    As a consequence, we obtain
    \begin{equation} \label{eq:bt}
        \| \mathbf B_T \|_2
        \le C_Z^{1/m} \sqrt d \left( \frac{ C_Z }{ T^{\nu m} } \right)^{1 - 1/m}
        = C_Z \sqrt d T^{-\nu (m - 1)}.
    \end{equation}
    
    Next we bound $\| \mathbf A_T \|_2$.
    Let $\bm W_t \equiv \overline{\bm Z}_t - \mathbb E[ \overline{\bm Z}_t]$.
    Before applying Lemma \ref{lem:ht-conc-b} to $\bm W_t$, 
    we verify that its conditions are satisfied.
    First, $\mathbb E[\bm W_t] = \bm 0$.
    Second, we may note that
    \[
        \|\mathbb E[ \overline{\bm Z}_t ] \|_2
        \le \mathbb E\left[ \| \overline{\bm Z}_t \|_2 \right]
        \le \mathbb E\left[ \| \overline{\bm Z}_t \|_2^2 \right]^{1/2}
        \le \mathbb E\left[ \| \bm Z_t \|_2^2 \right]^{1/2}
    \]
    by Jensen's inequality and then Cauchy-Schwarz.
    It then follows by (i) that
    $\mathbb E\left[ \| \bm Z_t \|_2^2 \right]^{1/2} \le \mathbb E\left[ \| \bm Z_t \|_2^m \right]^{1/m}
    \le C_Z^{1/m} \sqrt{d}$.
    We thus obtain
    \[
        \left\| \bm W_t \right\|_2
        \le \| \overline{\bm Z}_t \|_2 + \|\mathbb E[ \overline{\bm Z}_t ] \|_2
        = \sqrt d T^\nu + C_Z^{1/m} \sqrt{d}
        \le C_1 \sqrt d T^\nu,
    \]
    for some constant $C_1 > 0$, because $\| \overline{\bm Z}_t \|_2 \le \sqrt d T^\nu$ as we saw before. 
    Third, (i) implies the necessary moment conditions on $\bm W_t$.
    Finally, the beta-mixing coefficients of $\bm W_t$ follow from those of $\bm Z_t$.
    Recalling our choice of $\eta'$ above and taking $B = C_1 T^\nu$, 
    Lemma \ref{lem:ht-conc-b} then implies that there exists a positive constant $C > 0$ such that
    \begin{equation} \label{eq:at}
        \| \mathbf A_T \|_2
        \le C \left( \sqrt{\frac{ d (\log(2d) + \log(T))}{T}}
        + T^\nu \frac{\sqrt{d} \log(T)^2 (\log(2d) + \log(T))}{T} \right), 
    \end{equation}
    holds with probability at least $1 - T^{-\eta'}$.

    Hence, on the intersection of $\mathcal E_T$ and the event of Lemma \ref{lem:ht-conc-b},
    we have
    \begin{align*}
        \left\| \frac{1}{T} \sum_{t = 1}^T \bm Z_t \right\|_2
        &\le \| \mathbf A_T \|_2 + \| \mathbf B_T \|_2 \notag \\
        &\le C \sqrt d \sqrt{\frac{(\log(2d) + \log(T))}{T}} \\
        &+ C T^\nu \sqrt d \frac{ \log(T)^2 (\log(2d) + \log(T))}{T} \\
        &+ C_Z \sqrt d T^{-\nu (m - 1)}.
    \end{align*}
    Note that the last term is dominated by the second by analogous arguments as in the proof of 
    Lemma \ref{lem:ht-cov-conc}.
    We therefore have, on the same events, 
    that there exists a constant $C_\eta > 0$ such that for all $T$ sufficiently large
    \[
        \left\| \frac{1}{T} \sum_{t = 1}^T \bm Z_t \right\|_2
        \le C_\eta \left( \sqrt{\frac{ d (\log(2d) + \log(T))}{T}}
        + T^\nu \frac{\sqrt d \log(T)^2 (\log(2d) + \log(T))}{T} \right).
    \]
    It only remains to work out the probability of the intersection of the two events.
    The truncation event fails with probability at most $T^{-\eta'}$ from \eqref{eq:prob-trunc},
    and the bound from \eqref{eq:at} fails with probability at most $T^{-\eta'}$.
    Both events thus occur with probability at least $1 - 2 T^{-\eta'} \ge 1 - T^{-\eta}$.
\end{proof}

%{\paragraph{Competing interests statement}
%    We have no competing interests to declare.
%}

{\paragraph{Funding statement}
    Christian Brownlees acknowledges support from the Spanish Ministry of Science and Technology 
    (Grant MTM2012-37195);
    the Ayudas Fundaci\'on BBVA Proyectos de Investigación Cient\'ifica en Matemáticas 2021;
    the Spanish Ministry of Economy and Competitiveness through the Severo Ochoa Programme for 
    Centres of Excellence in R\&D (SEV-2011-0075).

    Gu\dh mundur Stef\'an Gu\dh mundsson acknowledges financial support from the 
    Danish National Research Foundation (DNRF Chair grant number DNRF154) and the 
    Aarhus Center for Econometrics (ACE) funded by the Danish National Research Foundation grant number DNRF186.
}

{\paragraph{Acknowledgements} 
    We have benefited from discussions with Gabor Lugosi, David Rossell and Piotr Zwiernik.
    All errors remain our own.
}

\bibliography{learning}
\bibliographystyle{natbib}

\end{document}